\documentclass[]{aa}  

\usepackage{xcolor}

\usepackage{graphicx}
\usepackage{txfonts}
\usepackage{natbib}
\bibpunct{(}{)}{;}{a}{}{,} 

\usepackage{placeins}

\begin{document}

   \title{A portrait of the Vast Polar Structure as a young phenomenon: hints from its member satellites}

   \subtitle{}

   \author{S.~Taibi\thanks{\email{staibi@aip.de}}
        \and
        M.~S.~Pawlowski
        \and
        S.~Khoperskov
        \and
        M.~Steinmetz
        \and
        N.~I.~Libeskind
          }

   \institute{Leibniz-Institut für Astrophysik Potsdam (AIP), An der Sternwarte 16, D-14482 Potsdam, Germany}

   \date{}

 
  \abstract
   {It has been observed that several Milky Way (MW) satellite dwarf galaxies are distributed along a coherent planar distribution known as the Vast Polar Structure (VPOS). 
}
   {Here we investigate whether MW satellites located on the VPOS have different physical and orbital properties from those not associated with it.
}
   {Using the proper motion measurements of the MW satellites from the \textit{Gaia} mission and literature values for their observational parameters, we first discriminate between systems that may or may not be associated with the VPOS, and then compare their chemical and dynamical properties. 
}
   {Comparing the luminosity distributions of the on-plane and off-plane samples, we find an excess of bright satellites observed on the VPOS. Despite this luminosity gap, we do not observe a significant preference for on-plane and off-plane systems to follow different scaling relations. 
   The on-plane systems also show a striking pattern in their radial velocities and orbital phases: co-orbiting satellites are almost all approaching their pericentre, while both counter-orbiting ones are leaving their last pericentre. This is in contrast to the more random distribution of the off-plane sample. The on-plane systems also tend to have the lowest orbital energies for a given value of angular momentum. 
   These results are robust to the assumed MW potential, even in the case of a potential perturbed by the arrival of a massive LMC. Considering them a significant property of the VPOS, we explore several scenarios, all related to the late accretion of satellite systems, which interpret the VPOS as a young structure.
}
   {From the results obtained, we hypothesise that the VPOS formed as a result of the accretion of a group of dwarf galaxies. More accurate proper motions and dedicated studies in the context of cosmological simulations are needed to confirm this scenario.
}


   \maketitle
%

\section{Introduction}
\label{sec:intro}

Several satellite galaxies and globular clusters around the Milky Way (MW) distribute on a thin extended plane with a polar orientation with respect to the Galactic disc, commonly known as the Vast Polar Structure \citep*[VPOS;][]{Kroupa2005,Pawlowski2012}. This planar distribution shows a coherent motion, with the vast majority of its bright \citep*{Metz2008} and numerous faint systems \citep{Fritz2018,Li2021} co-orbiting within it.

\citet*{Kroupa2005} were the first to show that the VPOS, and its then-known components, is at odds with the more isotropic distribution of dark matter halos in simulations within the framework of the standard cosmological model. Although this argumentation has been challenged (e.g., \citealp[]{Zentner2005,Libeskind2009,Samuel2021}; but see also \citealp{Pawlowski2012a}), it seems extremely rare to find satellite systems with the degree of flattening and kinematic coherence of the VPOS in cosmological simulations ($<1$\% of halos in both dark-matter only and hydro-dynamical simulations; \citealp[see][for a review]{Pawlowski2021}).

Phase-space correlations similar to the VPOS have also been observed around other nearby host galaxies, such as Andromeda \citep{Conn2012,Ibata2013} and Centaurus~A \citep{Tully2015,Muller2018}. At greater distances, flattened distributions were searched for in the MATLAS survey and found in about 25\% of the systems inspected \citep{Heesters2021}.

Several mechanisms, though none satisfactory, have been proposed to explain the origin of the VPOS and planar distributions in general (e.g., infall of satellite groups or satellite accretion along cosmic filaments, in the context of the standard cosmological model; see again \citealp{Pawlowski2021} for a review, but also \citealp{Welker2018}, and \citealp{Heesters2021}). 
Alternative scenarios suggest that satellites in a planar distribution may have formed differently, for instance from gas-rich tidal tails of interacting galaxies \citep[e.g.,][for the case of the Andromeda system]{Hammer2010,Hammer2013}. Although this possibility has its drawbacks (such dwarfs lack dark matter), it can be tested by directly comparing the observed properties of satellite galaxies on and off the inspected plane. \citet{Collins2015} performed such a comparison for the Andromeda satellite system and found no significant differences between the two samples, suggesting similar formation and evolution processes.

Thanks to the \textit{Gaia} mission \citep{Gaia-DR2_2018,Gaia-eDR3_2021}, the number of accurate proper motion measurements available for MW satellites has increased significantly, opening up new opportunities to examine the VPOS \citep[e.g.,][]{Fritz2018,Pawlowski+Kroupa2020,Li2021}. In particular, it has become important to understand its origin taking into account the recent arrival of a massive satellite such as the LMC \citep{Garavito-Camargo2021,Pawlowski2022} and the accretion history of the MW in general \citep{Helmi2020,Hammer2021,Hammer2023}.

Inspired by the work of \citet{Collins2015} and motivated by the possibility of having access to complete information on the three-dimensional position and velocity of MW satellites, in this paper we analyse the observed physical and orbital properties of these systems on and off the VPOS, as determined by their orbital alignment. In particular, we exploit the most recent proper motion measurements obtained by the \textit{Gaia} mission for most MW satellites (especially those of \citealp{Battaglia2022}, but see also \citealp[e.g.,][]{McConnachie+Venn2020a,McConnachie+Venn2020b,Li2021}, and \citealp{Pace2022}).

The article is structured as follows. In Sect.~\ref{sec:data} we present the dataset and the method used to classify MW satellites as either belonging to VPOS or not. Sections~\ref{sec:comp_phy_props} and \ref{sec:comp_orb_props} are dedicated respectively to the analysis and comparison of the physical and orbital properties of the on- and off-plane satellites. In Sect.~\ref{sec:discussion} we discuss our results and finally in Sect.~\ref{sec:conclusions} we present our conclusions.


\section{Data}
\label{sec:data}

\begin{table}
\caption{Adopted global properties for the Magellanic Clouds.}
\label{tab:MC}
\centering
\begin{tabular}{lccc}
\hline
\hline
Parameters & SMC & LMC & References\\
\hline
  R.A. [deg]                        & 13.1866        & 80.8937        & 1, 1 \\
  Dec. [deg]                        & $-72.8286$     & $-69.7561$     & 1, 1 \\
  D.M.                              & $19.03\pm0.12$ & $18.52\pm0.09$ & 1, 1 \\
  $V$                               & $2.2\pm0.2$    & $0.4\pm0.1$    & 1, 1 \\
  $R_e$ [deg]                       & 1.35           & 3.4            & 2, 3 \\
  $v_{\rm los}$ [km\,s$^{-1}$]      & $148.0\pm0.9$  & $262.2\pm3.4$  & 2, 3 \\
  $\sigma_{\rm v}$ [km\,s$^{-1}$]   & $25.6\pm0.2$   & $20.2\pm0.5$   & 2, 3 \\
  $\left<{\rm [Fe/H]}\right>$ [dex] & $-0.99\pm0.01$ & $-0.56\pm0.02$ & 4, 5 \\
  $\mu_{\alpha,*}$ [mas\,yr$^{-1}$] & $0.80\pm0.01$  & $1.85\pm0.01$  & 6, 6 \\
  $\mu_\delta$ [mas\,yr$^{-1}$]     & $-1.22\pm0.01$ & $0.23\pm0.01$  & 6, 6 \\
  \hline  
\end{tabular}
\tablefoot{Rows~1-2 show the central coordinates, while rows~3-8  show the distance modulus, visual magnitude, half-light radius along the projected major axis, heliocentric systemic line-of-sight velocity, velocity dispersion and weighted-average stellar metallicity, respectively for the Small and Large Magellanic Clouds (cols.~2-3). The systemic proper motions are reported in rows~9-10 and, unlike the rest of our sample, are based on the second data release of the \textit{Gaia} catalogue; the systematic error is $0.028$~mas\,yr$^{-1}$ for both galaxies. The corresponding references are: (1) \citet{McConnachie2012}; (2) \citet{DeLeo2020}; (3) \citet{VanDerMarel2002};  (4) \citet{Dobbie2014}; (5) \citet{Olsen2011}; (6) \citet{Gaia-DR2_2018}.}
\end{table}

In this work we are considering 50 confirmed satellite galaxies of the MW. They are selected from the list of \citet{Battaglia2022}, taking into account all those systems as far as Eridanus~II (i.e., with a Galactocentric distance of $D_{\rm GC}\lesssim350$~kpc) with measured systemic radial and transversal velocities. We have included in this list the two Magellanic clouds (MC) that were not considered in the cited work. We have instead not included the Sagittarius dSph, as its observed properties are strongly influenced by the ongoing merger with the MW \citep{Ibata2001}; however, we will take it into account when discussing the luminosity distribution of our sample. 
We note that a similar argument could also apply to Antlia~II and Crater~II, whose peculiar properties (i.e. very low surface brightness, large radii, low velocity dispersion) compared to other MW satellites may have been affected by strong tidal disturbances (\citealp[e.g.][]{Ji2021,Battaglia2022,Pace2022,Vivas2022}, but also \citealp{Borukhovetskaya2022}). However, as they appear to be tidally disturbed before they have lost most of their stars \citep{Ji2021}, we have kept them in our sample.

We refer to \citet[][and references therein]{Battaglia2022} for the physical and orbital parameters adopted in this work, with systemic proper motions obtained using the early-third data release of the \textit{Gaia} catalogue \citep{Gaia-eDR3_2021}.
The V-band luminosity values are adapted from the updated on-line catalogue (January~2021) of \citet{McConnachie2012}. The corresponding values for the MCs with their references are given in Table~\ref{tab:MC}.

We combined the systemic proper motions together with the celestial coordinates, heliocentric distances and systemic radial velocities to calculate the Galactocentric Cartesian positions and velocities of the selected MW satellites \citep[see e.g.,][]{Li2021}. We assumed for the MW, a heliocentric distance of $R_\odot=8.122$~kpc \citep{GravityCo2018}, a solar offset from the Galactic mid-plane of $z_\odot = 25$~pc \citep{Juric2008}, a solar motion $(U,V,W)_\odot = (11.1,12.24,7.25)$~km\,s$^{-1}$ \citep{Schonrich2010} and a circular velocity at Sun's location of $V_{\rm c}(R_\odot)=229$~km\,s$^{-1}$ \citep{Eilers2019}.

\begin{figure*}
    \centering
    \includegraphics[width=\textwidth]{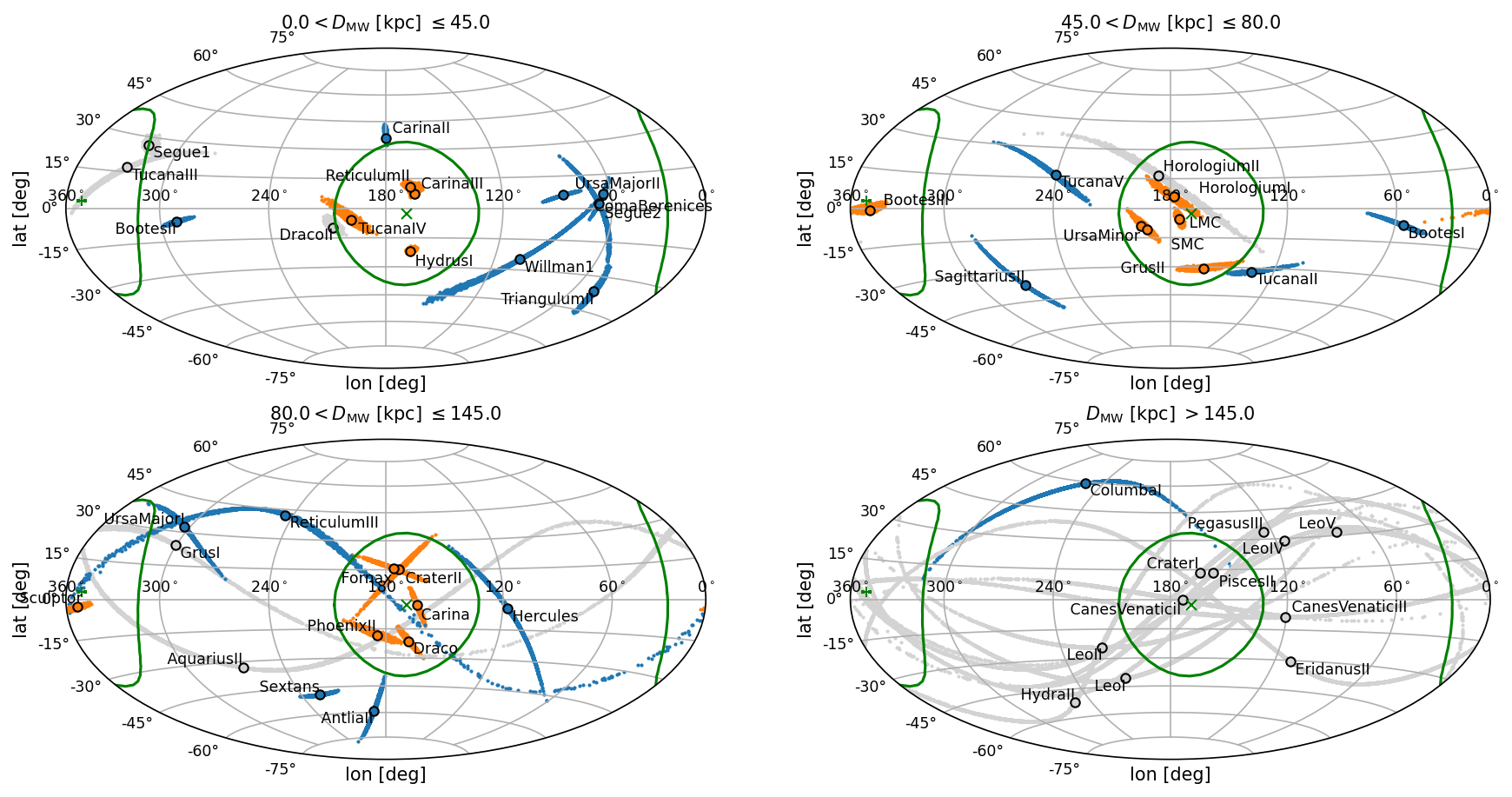}
    \caption{Distribution of orbital poles in Galactic coordinates for the analysed sample of MW satellites. The median orbital poles are shown as open black circles, while the values of 5000 Monte-Carlo realizations drawing on measurement uncertainties are shown as colored dots. The color scheme follows the division of our sample into on-plane (orange dots), off-plane (blue dots) and uncertain systems (gray dots) detailed in Sect.~\ref{sec:data}.}
    \label{fig:orb_poles}
\end{figure*}

\begin{table*}
\caption{Positions, velocities, and orbital poles of the MW satellites considered in this work.}
\label{tab:op}
\centering
\begin{tabular}{lcrrrcccrrr}
\hline
\hline
  \multicolumn{1}{c}{Galaxy} &
  \multicolumn{1}{c}{Class.} &
  \multicolumn{1}{c}{$x$} &
  \multicolumn{1}{c}{$y$} &
  \multicolumn{1}{c}{$z$} &
  \multicolumn{1}{c}{$V_x$} &
  \multicolumn{1}{c}{$V_y$} &
  \multicolumn{1}{c}{$V_z$} &
  \multicolumn{1}{c}{$\theta_{\rm VPOS}^{\rm pred}$} &
  \multicolumn{1}{c}{$\theta_{\rm VPOS}^{\rm meas}$} &
  \multicolumn{1}{c}{$f_{\rm VPOS}$} \\
  \multicolumn{2}{c}{} &
  \multicolumn{3}{c}{[kpc]} &
  \multicolumn{3}{c}{[km\,s$^{-1}$]} &
  \multicolumn{2}{c}{[deg]} &
  \multicolumn{1}{c}{} \\
\hline 
  AntliaII                 & Off & $-19.6 $ & $-128.9$ & $25.8  $  & $-97   \pm 10  $ & $-36  \pm 2   $ & $61   \pm 9   $  & 2.6  & $59 _{-5 }^{+5 }$  & 0.0  \\     
  AquariusII               & ?   & $28.7  $ & $53.2  $ & $-86.1 $  & $103   \pm 88  $ & $7    \pm 77  $ & $-9   \pm 65  $  & 7.7  & $98 _{-62}^{+40}$  & 0.29 \\     
  BootesI                  & Off & $15.2  $ & $-0.8  $ & $62.2  $  & $153   \pm 10  $ & $-97  \pm 21  $ & $58   \pm 4   $  & 16.4 & $113_{-5 }^{+4 }$  & 0.0  \\     
  BootesII                 & Off & $6.9   $ & $-1.6  $ & $38.9  $  & $-336  \pm 21  $ & $-129 \pm 23  $ & $1    \pm 9   $  & 13.2 & $121_{-3 }^{+2 }$  & 0.0  \\     
  BootesIII                & On  & $1.7   $ & $6.9   $ & $45.3  $  & $-10   \pm 6   $ & $-42  \pm 21  $ & $259  \pm 6   $  & 3.2  & $175_{-2 }^{+1 }$  & 1.0  \\     
  CanesVenaticiI           & ?   & $2.6   $ & $35.9  $ & $207.7 $  & $10    \pm 32  $ & $97   \pm 34  $ & $64   \pm 6   $  & 1.6  & $15 _{-10}^{+19}$  & 0.87 \\     
  CanesVenaticiII          & ?   & $-15.8 $ & $18.6  $ & $158.8 $  & $60    \pm 68  $ & $-4   \pm 79  $ & $-92  \pm 8   $  & 4.0  & $75 _{-46}^{+74}$  & 0.4  \\     
  Carina                   & On  & $-25.1 $ & $-96.5 $ & $-39.9 $  & $-29   \pm 10  $ & $-66  \pm 7   $ & $175  \pm 17  $  & 4.6  & $5  _{-1 }^{+2 }$  & 1.0  \\     
  \textbf{CarinaII}        & Off & $-8.2  $ & $-34.6 $ & $-10.6 $  & $122   \pm 7   $ & $-304 \pm 5   $ & $156  \pm 15  $  & 3.3  & $42 _{-1 }^{+2 }$  & 0.0  \\     
  \textbf{CarinaIII}       & On  & $-8.2  $ & $-26.6 $ & $-8.0  $  & $14    \pm 6   $ & $-161 \pm 9   $ & $353  \pm 28  $  & 6.9  & $12 _{-1 }^{+1 }$  & 1.0  \\     
  ColumbaI                 & Off & $-107.8$ & $-125.6$ & $-88.0 $  & $218   \pm 62  $ & $-116 \pm 51  $ & $-36  \pm 53  $  & 27.6 & $91 _{-12}^{+13}$  & 0.0  \\     
  ComaBerenices            & Off & $-10.2 $ & $-4.2  $ & $42.1  $  & $254   \pm 16  $ & $-23  \pm 16  $ & $92   \pm 1   $  & 9.5  & $82 _{-3 }^{+3 }$  & 0.0  \\     
  CraterI                  & ?   & $0.4   $ & $-97.2 $ & $107.7 $  & $37    \pm 87  $ & $79   \pm 47  $ & $59   \pm 43  $  & 9.4  & $34 _{-20}^{+31}$  & 0.54 \\     
  CraterII                 & On  & $11.5  $ & $-84.5 $ & $78.1  $  & $18    \pm 13  $ & $127  \pm 9   $ & $13   \pm 9   $  & 15.4 & $19 _{-3 }^{+5 }$  & 1.0  \\     
  Draco                    & On  & $-3.8  $ & $66.1  $ & $45.9  $  & $68    \pm 9   $ & $4    \pm 4   $ & $-170 \pm 6   $  & 9.8  & $18 _{-3 }^{+3 }$  & 1.0  \\     
  DracoII                  & ?   & $-10.4 $ & $15.7  $ & $14.7  $  & $9     \pm 10  $ & $88   \pm 8   $ & $-332 \pm 9   $  & 31.2 & $38 _{-1 }^{+1 }$  & 0.2  \\     
  EridanusII               & ?   & $-86.9 $ & $-211.6$ & $-284.7$  & $-165  \pm 207 $ & $48   \pm 177 $ & $103  \pm 121 $  & 9.5  & $76 _{-48}^{+47}$  & 0.3  \\     
  Fornax                   & On  & $-39.6 $ & $-48.3 $ & $-126.9$  & $15    \pm 11  $ & $-101 \pm 19  $ & $77   \pm 7   $  & 14.8 & $20 _{-3 }^{+4 }$  & 1.0  \\     
  GrusI                    & ?   & $53.6  $ & $-24.2 $ & $-107.8$  & $-66   \pm 29  $ & $106  \pm 44  $ & $158  \pm 20  $  & 25.0 & $149_{-24}^{+6 }$  & 0.65 \\     
  GrusII                   & On  & $25.2  $ & $-5.2  $ & $-43.4 $  & $-112  \pm 9   $ & $-150 \pm 26  $ & $99   \pm 7   $  & 27.9 & $29 _{-1 }^{+1 }$  & 1.0  \\     
  \textit{Hercules}        & Off & $88.2  $ & $52.7  $ & $81.9  $  & $181   \pm 20  $ & $92   \pm 35  $ & $-22  \pm 22  $  & 38.0 & $52 _{-8 }^{+10}$  & 0.0  \\  
  \textbf{HorologiumI}     & On  & $-7.2  $ & $-45.9 $ & $-64.9 $  & $8     \pm 16  $ & $-143 \pm 32  $ & $141  \pm 23  $  & 1.3  & $12 _{-5 }^{+4 }$  & 1.0  \\     
  HorologiumII             & ?   & $-14.3 $ & $-45.4 $ & $-63.2 $  & $68    \pm 81  $ & $-241 \pm 68  $ & $140  \pm 49  $  & 6.3  & $26 _{-14}^{+15}$  & 0.78 \\     
  HydraII                  & ?   & $48.3  $ & $-117.2$ & $76.3  $  & $-93   \pm 90  $ & $-130 \pm 58  $ & $114  \pm 64  $  & 29.5 & $85 _{-33}^{+25}$  & 0.08 \\     
  \textbf{HydrusI}         & On  & $2.0   $ & $-19.6 $ & $-16.5 $  & $-157  \pm 10  $ & $-190 \pm 19  $ & $282  \pm 16  $  & 10.7 & $19 _{-0 }^{+0 }$  & 1.0  \\     
  LeoI                     & ?   & $-130.0$ & $-126.8$ & $204.1 $  & $-115  \pm 28  $ & $-33  \pm 27  $ & $136  \pm 20  $  & 20.2 & $53 _{-24}^{+30}$  & 0.28 \\     
  LeoII                    & ?   & $-71.7 $ & $-54.2 $ & $200.2 $  & $-55   \pm 35  $ & $56   \pm 33  $ & $21   \pm 13  $  & 13.3 & $51 _{-23}^{+24}$  & 0.27 \\     
  LeoIV                    & ?   & $-14.5 $ & $-84.9 $ & $128.7 $  & $86    \pm 107 $ & $10   \pm 68  $ & $17   \pm 45  $  & 2.9  & $75 _{-40}^{+49}$  & 0.27 \\     
  LeoV                     & ?   & $-20.8 $ & $-92.0 $ & $151.9 $  & $240   \pm 174 $ & $-84  \pm 111 $ & $32   \pm 70  $  & 1.3  & $95 _{-28}^{+38}$  & 0.15 \\     
  PegasusIII$\dagger$      & ?   & $46.6  $ & $150.4 $ & $-143.5$  & / & / & /  & 2.9  & $93 _{-55}^{+55}$  & 0.34 \\     
  \textbf{PhoenixII}       & On  & $25.5  $ & $-24.8 $ & $-72.0 $  & $-21   \pm 17  $ & $-256 \pm 52  $ & $127  \pm 17  $  & 18.9 & $21 _{-1 }^{+2 }$  & 1.0  \\     
  PiscesII                 & ?   & $14.8  $ & $122.4 $ & $-134.1$  & $-277  \pm 360 $ & $-454 \pm 173 $ & $-367 \pm 176 $  & 4.6  & $27 _{-17}^{+20}$  & 0.69 \\     
  \textbf{ReticulumII}     & On  & $-9.5  $ & $-20.4 $ & $-24.1 $  & $21    \pm 4   $ & $-114 \pm 21  $ & $222  \pm 18  $  & 11.8 & $14 _{-1 }^{+1 }$  & 1.0  \\     
  ReticulumIII             & Off & $-4.0  $ & $-64.0 $ & $-65.6 $  & $200   \pm 111 $ & $-110 \pm 53  $ & $-21  \pm 56  $  & 3.0  & $82 _{-17}^{+20}$  & 0.05 \\     
  \textit{SagittariusII}   & Off & $52.2  $ & $20.7  $ & $-27.1 $  & $23    \pm 17  $ & $-145 \pm 32  $ & $192  \pm 16  $  & 47.6 & $97 _{-3 }^{+4 }$  & 0.0  \\   
  Sculptor                 & On  & $-5.3  $ & $-9.5  $ & $-83.4 $  & $17    \pm 8   $ & $155  \pm 8   $ & $-94  \pm 1   $  & 5.2  & $174_{-2 }^{+1 }$  & 1.0  \\     
  Segue1                   & ?   & $-19.2 $ & $-9.5  $ & $17.7  $  & $-133  \pm 7   $ & $-214 \pm 35  $ & $-57  \pm 21  $  & 35.7 & $144_{-1 }^{+1 }$  & 0.71 \\     
  \textit{Segue2}          & Off & $-31.5 $ & $13.8  $ & $-21.1 $  & $-108  \pm 14  $ & $42   \pm 16  $ & $74   \pm 8   $  & 58.7 & $102_{-4 }^{+4 }$  & 0.0  \\  
  Sextans                  & Off & $-35.9 $ & $-56.1 $ & $57.1  $  & $-197  \pm 10  $ & $70   \pm 6   $ & $73   \pm 7   $  & 14.6 & $67 _{-3 }^{+3 }$  & 0.0  \\     
  \textit{TriangulumII}    & Off & $-28.4 $ & $16.4  $ & $-11.4 $  & $227   \pm 6   $ & $-23  \pm 6   $ & $195  \pm 8   $  & 65.0 & $115_{-0 }^{+0 }$  & 0.0  \\  
  TucanaII                 & Off & $21.6  $ & $-18.6 $ & $-45.7 $  & $-239  \pm 18  $ & $-100 \pm 37  $ & $146  \pm 7   $  & 24.6 & $45 _{-3 }^{+4 }$  & 0.0  \\     
  TucanaIII                & ?   & $0.9   $ & $-9.0  $ & $-19.1 $  & $28    \pm 5   $ & $141  \pm 10  $ & $184  \pm 6   $  & 4.4  & $153_{-13}^{+13}$  & 0.77 \\     
  TucanaIV                 & On  & $10.1  $ & $-19.5 $ & $-38.7 $  & $20    \pm 10  $ & $-119 \pm 32  $ & $173  \pm 17  $  & 15.2 & $29 _{-3 }^{+3 }$  & 0.99 \\     
  TucanaV                  & Off & $16.3  $ & $-23.5 $ & $-43.4 $  & $85    \pm 16  $ & $22   \pm 27  $ & $213  \pm 21  $  & 20.5 & $73 _{-7 }^{+8 }$  & 0.0  \\     
  UrsaMajorI               & Off & $-60.9 $ & $20.0  $ & $79.4  $  & $-85   \pm 17  $ & $-88  \pm 27  $ & $-41  \pm 10  $  & 35.7 & $128_{-7 }^{+6 }$  & 0.0  \\     
  \textit{UrsaMajorII}     & Off & $-32.5 $ & $12.7  $ & $21.2  $  & $174   \pm 7   $ & $-108 \pm 23  $ & $215  \pm 21  $  & 54.4 & $105_{-2 }^{+2 }$  & 0.0  \\  
  UrsaMinor                & On  & $-22.0 $ & $52.3  $ & $53.8  $  & $10    \pm 7   $ & $50   \pm 5   $ & $-157 \pm 5   $  & 21.5 & $26 _{-1 }^{+2 }$  & 1.0  \\     
  \textit{Willman1}        & Off & $-27.5 $ & $7.6   $ & $32.0  $  & $88    \pm 17  $ & $64   \pm 35  $ & $80   \pm 17  $  & 38.8 & $65 _{-11}^{+12}$  & 0.0  \\   
  \textbf{SMC}             & On  & $16.6  $ & $-38.5 $ & $-44.8 $  & $9     \pm 9   $ & $-195 \pm 21  $ & $170  \pm 16  $  & 20.2 & $24 _{-1 }^{+1 }$  & 1.0  \\     
  \textbf{LMC}             & On  & $-0.5  $ & $-41.8 $ & $-27.5 $  & $-40   \pm 9   $ & $-227 \pm 16  $ & $222  \pm 23  $  & 6.8  & $7  _{-1 }^{+1 }$  & 1.0  \\     
\hline
\end{tabular}
\tablefoot{Column~2 indicate if the dwarf galaxies listed in col.~1 are classified either as on-plane (On), off-plane (Off) or have an uncertain classification (?); cols.~3-7 show the calculated Galactocentric positions and velocities in Cartesian projection needed to determine the orbital pole directions; cols.~8-9 show the best-possible (``predicted'') and measured angular separation between the satellites orbital plane and the VPOS orientation, while in col.~10 is shown the fraction of Monte Carlo realizations of the orbital poles that fall within the 10\% sky area encircling the VPOS location. 
Galaxy names in italics indicate those systems for which $\theta_{\rm VPOS}^{\rm pred}>\theta_{\rm in VPOS}$ (see Sect.~\ref{sec:data}), while in boldface are marked those systems likely associated with the LMC.
($\dagger$) The systemic proper motion of Pegasus~III is too uncertain to obtain a velocity vector with meaningful values.}
\end{table*}

For each dwarf satellite galaxy, we generated 5000 Monte Carlo realizations sampling from the parameter error distributions, assumed to be Gaussian. For the proper motions we considered a bi-variate distribution, taking into account the correlation between parameters and the systematic errors. We then obtained the specific angular momenta for the MW satellites as the cross-product of the 3D Galactocentric positions and velocities. The resulting orbital poles (i.e., the direction of the angular momentum vectors) are shown in Fig.~\ref{fig:orb_poles}, after being transformed into Galactic latitudes and longitudes. Also plotted are the position of the VPOS normal vector pointing to $(l,b)=(169.3^{\circ},-2.8^{\circ})$ \citep{Pawlowski+Kroupa2013, Fritz2018}, and the VPOS area enclosing the 10\% of the sky around such position (i.e., with an aperture angle $\theta_{\rm inVPOS} = 36.87^{\circ}$) and the opposite pole. For each system, we calculated the median angular difference $\theta_{\rm VPOS}^{\rm meas}$ between their orbital poles and the VPOS normal, as well as the best possible alignment $\theta_{\rm VPOS}^{\rm pred}$ as defined solely by the satellite current position \citep[see details in][]{Pawlowski+Kroupa2013}.
We also calculated the fraction $f_{\rm VPOS}$ of Monte Carlo realization that are found within the VPOS area, quantifying then the degree of alignment of the orbital poles with the VPOS normal vector. The calculated values are reported in Table~\ref{tab:op}.

\begin{figure*}
    \centering
    \includegraphics[width=.49\textwidth]{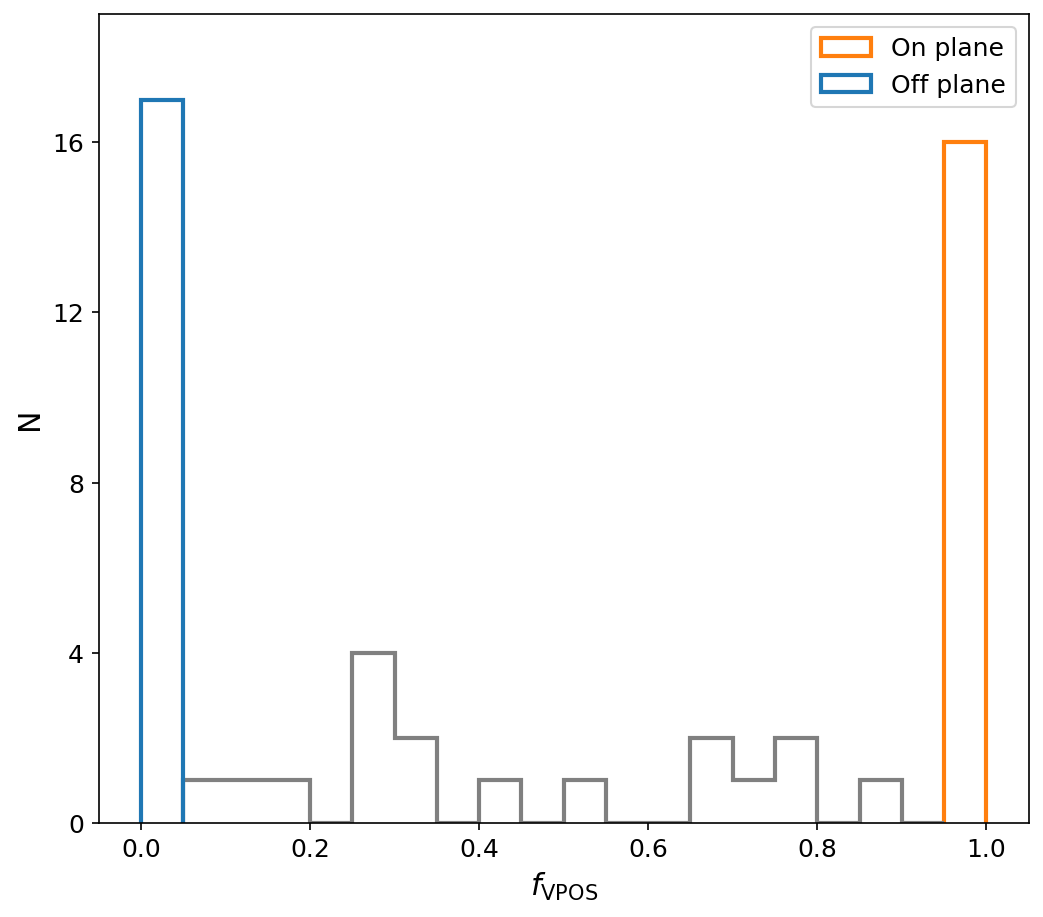}
    \includegraphics[width=.49\textwidth]{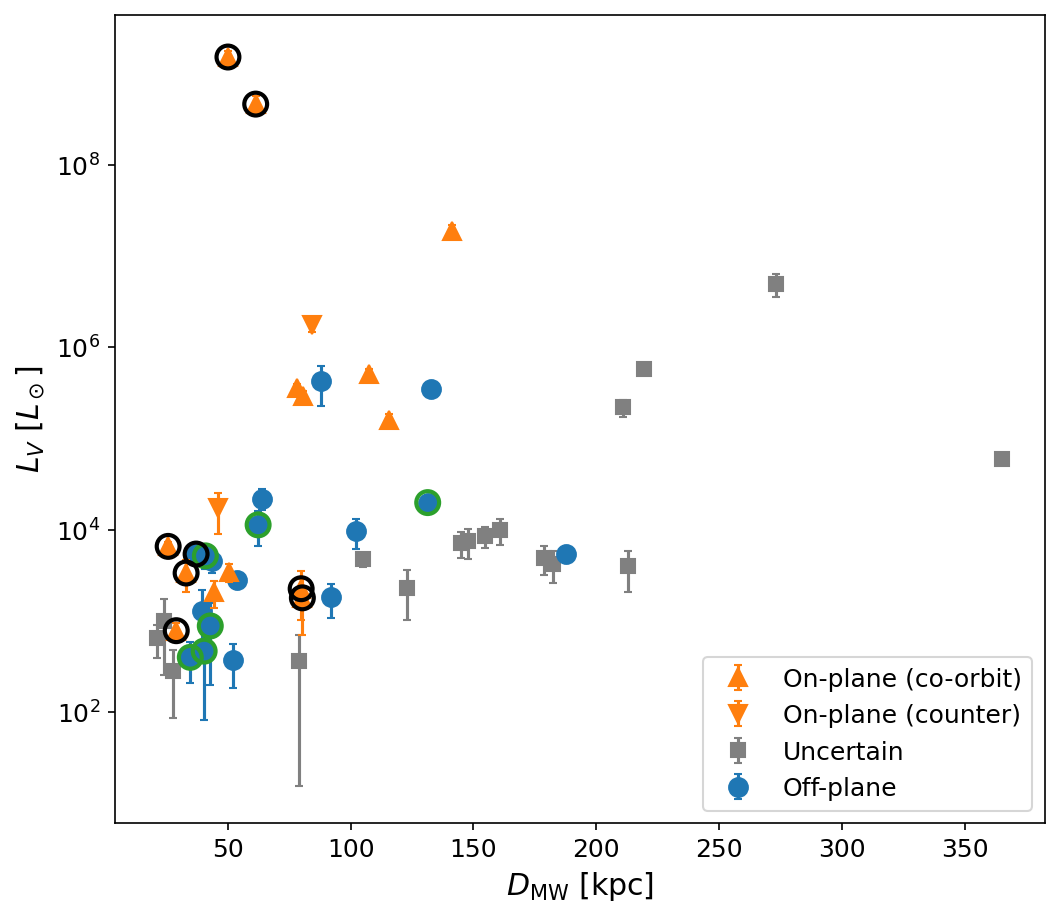}
    \caption{Sub-sample selection. 
    \textit{Left:} distribution of the fraction $f_{\rm VPOS}$ of Monte Carlo realization for the orbital poles of our systems that are found within the VPOS area (i.e., within $\theta_{\rm VPOS} = 36.87^{\circ}$ from the orbital pole of the VPOS normal).
    \textit{Right:} distribution of the V-band luminosity values for our systems as a function of their Galactocentric distance. 
    In both panels, the on-plane systems are plotted in orange, the off-plane ones in blue, while systems with intermediate $f_{\rm VPOS}$ are marked in grey. In the right panel, they are plotted using filled triangles (pointing up or down depending on whether a system is co- or counter-orbiting within the VPOS), circles and squares, respectively. In addition, black circles indicate the MCs and likely associated satellites, while green circles mark those systems that are in the off-plane sample independently of their proper motions.
    }
    \label{fig:f_VPOS}
\end{figure*}

We define as "on-plane" those systems with a $f_{\rm VPOS}>0.95$ (i.e., with a large fraction of their orbital poles distribution aligned with the VPOS normal vector). This definition includes both co- and counter-orbiting systems along the VPOS. The "off-plane" systems have $f_{\rm VPOS}<0.05$, while the remaining galaxies have intermediate $f_{\rm VPOS}$ values, making their association with the VPOS uncertain. This is mainly due to the highly uncertain orbital pole directions because of the relatively high uncertainties in proper motion (see Fig.~\ref{fig:orb_poles}).

The distribution of $f_{\rm VPOS}$ values, shown in Fig.~\ref{fig:f_VPOS} (left panel), is bi-modal with 16 systems resulting on-plane and 17 off-plane. We note that satellites located outside the VPOS area only on the basis of their spatial position (i.e., having $\theta_{\rm VPOS}^{\rm pred}>\theta_{\rm inVPOS}$) are assigned to the off-plane sample independently of their proper motion accuracy. These are: Hercules, Sagittarius~II, Segue~2, Triangulum~II, Ursa Major~II, and Willman~1. Thus, the off-plane sample is by definition more complete, since it can include systems having less constrained proper motions and thus poorly determined orbital properties. 
We also note that the orbital poles of the on-plane sample are confined on a much smaller area ($\sim20\%$ of the sky) with respect to the off-plane sample, and yet both contain a similar number of satellites. This highlights the overdensity of the VPOS-aligned objects; if they were distributed uniformly, there would be four times more in the non-VPOS area.

It is also instructive to consider the distribution of luminosity values of our systems as a function of their Galactocentric distances, as shown in Fig.~\ref{fig:f_VPOS} (right panel). This figure is complementary to the distribution of orbital poles shown in Fig.~\ref{fig:orb_poles}. We can see that systems with intermediate values of $f_{\rm VPOS}$ (and therefore labelled as uncertain) are either low-luminosity or distant dwarfs, which has an impact on their proper motion determination (their associated errors are on average a factor of 10 larger than those of the on- and off-plane sub-samples), and thus on their orbital pole distributions. On- and off-plane systems have proper motion errors that translate into comparable average tangential velocity errors $\sim30$~km\,s$^{-1}$, with only Columba~I and Reticulum~III (both among the off-planes) having large errors $>70$~km\,s$^{-1}$. These two systems are faint satellites whose exclusion from the off-plane sample would not impact our results.

We further note that there are systems such as Carina~II (and marginally Draco~II, Segue~1, but also Leo~I and II at greater distances) which, based on their $\theta_{\rm VPOS}^{\rm meas}$, are located just outside the VPOS area (see again Fig.~\ref{fig:orb_poles}). In particular, Carina~II is with high probability a former satellite of the LMC \citep{Battaglia2022}, which instead is safely on the VPOS. Indeed, the LMC has probably brought in several dwarf satellites in its fall into the MW potential (we consider here the most likely associated systems listed in \citealp{Battaglia2022}, but see also \citealp{Correa-Magnus+Vasiliev2022}). As they might follow similar orbits, they would not be an independent sample and could increase the presence of dwarf galaxies on the VPOS \citep[][]{Pawlowski2022}. Therefore, in the following section we also take into account the contribution of the LMC satellite system.

\section{Comparison of the physical properties}
\label{sec:comp_phy_props}

\begin{table}
\caption{Average distribution values and statistical test results.}
\label{tab:params}
\centering
\begin{tabular}{lrrr}
\hline
\hline
Parameters & On-plane & Off-plane & KS-test\\
\hline
  $M_V$ [dex]                       & $ -7.0 \,(4.7)$ & $ -4.4 \,(2.2)$ & 0.15 (0.05) \\
  $R_e$ [pc]                        & $ 264  \,(304)$ & $ 91   \,(88)$  & 0.15 (0.15) \\
  $\sigma_{\rm v}$ [km\,s$^{-1}$]   & $ 7.3  \,(5.8)$ & $ 4.0  \,(5.3)$ & 0.15 (0.02) \\
  $\left<{\rm [Fe/H]}\right>$ [dex] & $ -2.0 \,(0.6)$ & $ -2.2 \,(0.2)$ & 0.13 (0.08) \\
  Peri (L) [kpc]                & $ 44   \,(25)$  & $ 41   \,(22)$  & 0.98 \\
  Apo  (L) [kpc]                & $ 147  \,(60)$  & $ 161  \,(87)$  & 0.84 \\
  Ecc  (L)                      & $ 0.6  \,(0.1)$ & $ 0.7  \,(0.2)$ & 0.37 \\
  $P$ (L) [Gyr]                 & $ 3.4  \,(1.4)$ & $ 3.6  \,(2.0)$ & 0.86 \\
  $T_{\rm last-peri}$ (L) [Gyr] & $ -1.8 \,(1.3)$ & $ -0.6 \,(0.7)$ & 0.25 \\
  $\phi$ (L)                    & $ -0.7 \,(0.2)$ & $ -0.4 \,(0.3)$ & 0.01 (0.07) \\
  Peri (H) [kpc]                & $ 32   \,(10)$  & $ 39   \,(15)$  & 0.71 \\
  Apo  (H) [kpc]                & $ 99   \,(34)$  & $ 102  \,(60)$  & 0.53 \\
  Ecc  (H)                      & $ 0.6  \,(0.2)$ & $ 0.6  \,(0.1)$ & 0.90 \\
  $P$ (H) [Gyr]                 & $ 1.6  \,(0.6)$ & $ 1.6  \,(0.8)$ & 0.58 \\
  $T_{\rm last-peri}$ (H) [Gyr] & $ -1.0 \,(0.5)$ & $ -0.8 \,(0.7)$ & 0.53 \\
  $\phi$ (H)                    & $ -0.7 \,(0.2)$ & $ -0.4 \,(0.5)$ & 0.44 \\

\hline  
\end{tabular}
\tablefoot{In col.~1 the inspected physical (luminosity, half-light radius, velocity dispersion, average metallicity) and orbital (pericentre, apocentre, eccentricity, time since the last pericentric passage, orbital period and phase; for both the low-mass - L, and high-mass - H, potentials) parameters of our sample of satellites. Columns 2 and 3 show the median (and median absolute deviation) of the distribution of values for the on- and off-plane systems, respectively, as well as col.~4 show the p-value results of the KS-test (and AD-test when the p-value was $<0.25$) between these two sub-samples.
}
\end{table}

\begin{figure*}
    \centering
    \includegraphics[width=.95\textwidth]{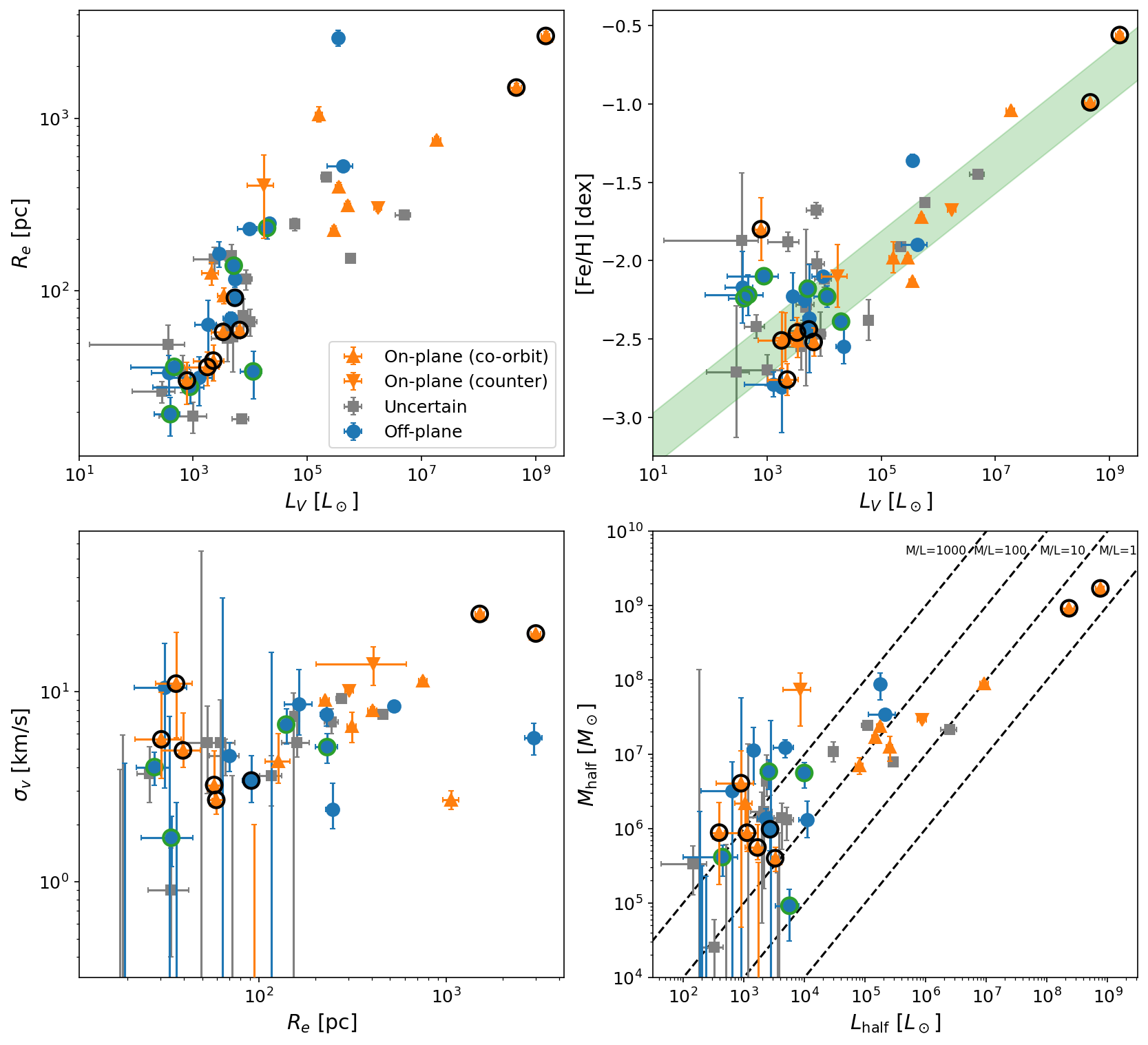}
    \caption{Comparison of the physical properties for our sample of dwarf galaxies. \textit{Top left:} V-band luminosity $L_V$ versus the half-light radius $R_e$. \textit{Top right:} $L_V$ versus the global average metallicity [Fe/H]; the green band is the $L_V$-[Fe/H] relation from \citet{Kirby2013}. \textit{Bottom left:} $R_e$ versus the velocity dispersion $\sigma_v$. \textit{Bottom right:} luminosity $L_{\rm half}$ at the half-light radius versus the calculated dynamical mass $M_{\rm half}$ at the same radius; dashed black lines indicate constant $M/L$ ratios of 1000, 100, 10 and 1, from left to right. In all panels, on- and off-plane systems are indicated respectively with orange triangles, and blue circles, while uncertain systems are plotted with gray squares. The MCs and probable LMC satellites are marked with black circles, while green circles mark systems being in the off-plane sample independently of their proper motions.}
    \label{fig:VPOS_ph}
\end{figure*}

\begin{figure*}
    \centering
    \includegraphics[width=.49\textwidth]{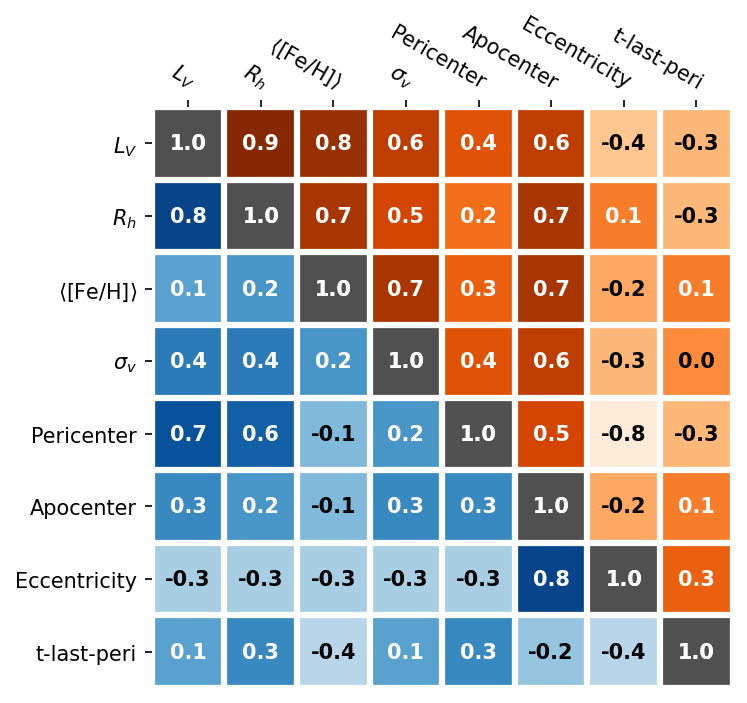}
    \includegraphics[width=.49\textwidth]{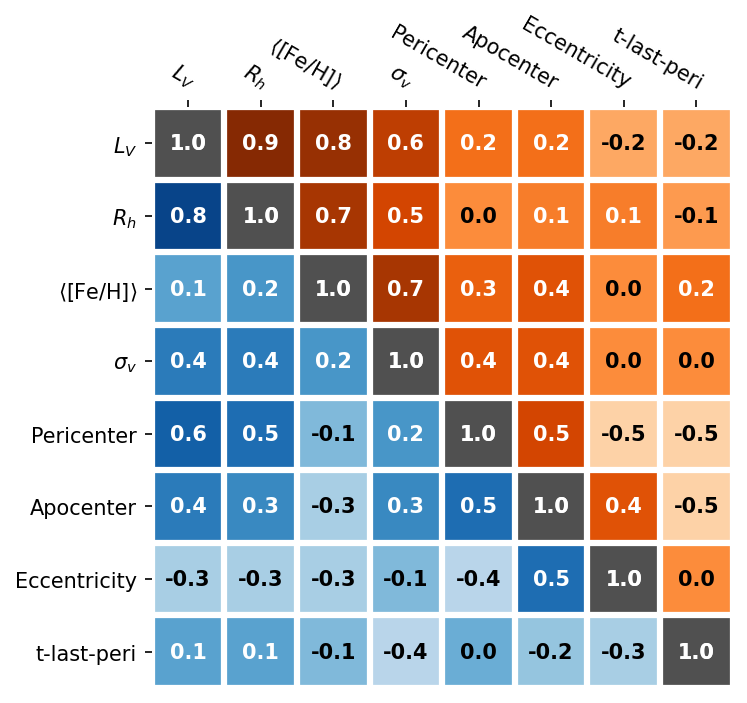}
    \caption{Correlation matrixes showing Spearman's rank coefficients between the inspected physical (luminosity, half-light radius, global average metallicity, velocity dispersion) and orbital parameters (pericentre, apocentre, eccentricity, time since the last pericentric passage) in the case of a low (left panel) and high (right panel) mass MW potential. In both cases, the upper right part corresponds to values for the on-plane sample, while the lower left part is for the off-plane sample; orange and blue tones correspond respectively to the numerical values indicated, so that darker tones correspond to a stronger correlation.
    }
    \label{fig:VPOS_corr}
\end{figure*}

In Fig.~\ref{fig:VPOS_ph} we compare the photometric (i.e., visual luminosity $L_V$; semi-major axis half-light radius $R_e$; weighted-average metallicity $\left<{\rm [Fe/H]}\right>$) and dynamical (i.e., velocity dispersion $\sigma_v$; dynamical mass $M_{1/2}$, as in \citealp[]{Wolf2010}) properties of our sample, obtained using values from Table~B.1 of \citet[][and references therein]{Battaglia2022}. 
Plots are made in order to highlight well-known scaling relations between LG dwarf galaxies, such as luminosity-size \citep{Brasseur2011}, luminosity-metallicity \citep{Kirby2013}, luminosity-dynamical mass \citep{Walker2009c,Wolf2010,McConnachie2012} and related parameters (i.e., $\sigma_v$ and $R_e$). 

We used the Spearman's rank correlation coefficient $\rho$ to measure the level of correlation between the inspected physical parameters, recovering expectedly high correlation values (i.e., $\left| \rho \right| \gtrsim0.7$) when considering the full sample. Restricting the analysis to the on- and off-plane systems, we found instead a marked difference between them in terms of $\rho$ values. As shown in the correlation matrix of Fig.~\ref{fig:VPOS_corr}, on-plane systems have higher correlation values between the physical parameters than the off-plane ones. In the figure, we also provide $\rho$ values for the orbital parameters described in the next section, for the two MW potentials we considered. We note that in both cases, the upper right part of the correlation matrix shows values for the on-plane sample, while the lower left part is for the off-plane sample.

\begin{figure}
    \centering
    \includegraphics[width=.49\textwidth]{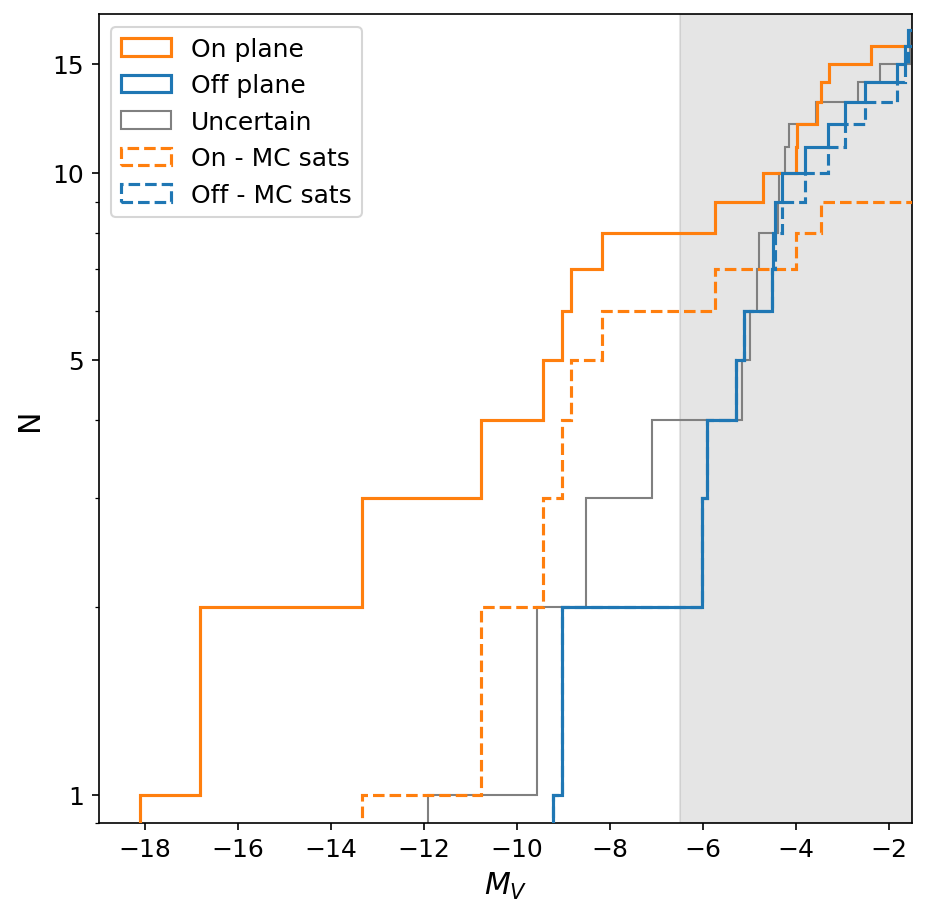}
    \caption{Cumulative luminosity distribution of the satellites on-plane (orange solid line) compared to those off-plane (blue solid line); the distribution of the uncertain systems is reported too for comparison (grey solid line); dashed lines show the same comparison but removing the MC system. The gray band indicates the search completeness on the SDSS footprint \citep{Koposov2008,Walsh2009,Simon2018}.}
    \label{fig:VPOS_hist}
\end{figure}

Considering that for the full sample the relations between the inspected physical parameters are mostly linear, the differences observed when dividing into on- and off-plane systems are mainly due to the lack of luminous systems (i.e., with $L_V \gtrsim 10^5 L_\odot$) among the off-plane sample (2 vs. 8, respectively). This is best shown in Fig.~\ref{fig:VPOS_hist}, where the cumulative luminosity distributions of the two sub-samples are compared. We have used the absolute magnitude $M_V$ in the figure to highlight the observed difference at the bright-end of their distributions. The observed luminosity gap between the two samples is for $M_V \sim-8$. 

Before examining the significance of this result, we first note that systems being in the off-plane sample independently of their proper motions (because they have $\theta_{\rm VPOS}^{\rm pred}>\theta_{\rm inVPOS}$) do not contribute to the observed luminosity gap, since they all have a luminosity $L_V<10^5\,L_\odot$ (i.e., $M_V>-6$, see Fig.~\ref{fig:f_VPOS} right). We also note that on-plane systems are brighter than the off-plane ones even as a function of the Galactocentric distance (see again Fig.~\ref{fig:f_VPOS} right). 

We compared the luminosity distribution of the on- and off-plane samples using two-sided Kolmogorov-Smirnov (KS) and Anderson-Darling (AD) statistical tests, recovering p-values of 0.15 and 0.05, respectively. Similar results are also obtained for the $R_e$, $\left<{\rm [Fe/H]}\right>$, and $\sigma_v$ distributions due to the existing scaling relations, as shown in Table~\ref{tab:params}, where we also report the average distribution values of the inspected parameters. Therefore, we cannot reject with high formal significance the hypothesis that the two samples come from the same parent distribution. Even though the observed luminosity gap is highly suggestive of this, we thus cannot firmly establish it, mostly due to the relatively low total number of dwarf galaxies involved.

The above results are linked to two main sources of uncertainty that could influence them. These are the incompleteness of the MW luminosity function, and the inclusion of satellites with an uncertain orbital pole in one of the two samples.

Regarding the first point, we observe that the largest difference between the inspected distributions is in a luminosity range where the MW luminosity function is considered to be mostly complete, while yet-to-be-discovered satellites are expected to be ultra-faint dwarfs with $M_V \gtrsim -7$ (\citealp[]{Simon2018}, but also \citealp[]{Koposov2008,Walsh2009}). It is true, however, that the discovery of the Antlia~II and Crater~II dSphs \citep{Torrealba2016,Torrealba2019}, suggests that even at the bright-end of the MW luminosity function there might still be some systems missing. In fact, the inclusion of a number $\geq1$ of Antlia~II-like systems with $M_V = -9$ in the off-plane (on-plane) sample would significantly reduce (increase) the statistical significance of the observed luminosity gap. A similar argument would also apply if the Sagittarius dSph ($M_V = -13.5$) were included in the off-plane sample \citep[see e.g.][for details on its orbital properties]{Pawlowski+Kroupa2020}. We recall that we have excluded this galaxy from our analysis because its observed properties are strongly influenced by the ongoing tidal disruption caused by the MW. 

Regarding the second point, there are a number of bright systems (i.e., $M_V \leq -7$) with an uncertain orbital pole determination based on \textit{Gaia} proper motions, due to their large Galactocentric distances: Leo~I, Leo~II, Canes Venatici~I, and Eridanus~II. Canes Venatici~I, despite the uncertainties, has its median orbital pole value almost perfectly aligned with the VPOS, while the median values of Leo~I and II are close to the $\theta_{\rm inVPOS}$ area (see Fig.~\ref{fig:orb_poles}). For the latter two systems, studies have shown that their orbital poles are much better aligned with the VPOS combining Gaia's astrometric measurements with those of the Hubble Space Telescope, in order to reduce the uncertainty on proper motions \citep{Pawlowski+Kroupa2020,Casetti-Dinescu2022}.
Adopting these additional proper motion measurements thus results in including these three galaxies in the on-plane sample, while Eridanus~II is added to the off-plane sample. Even including Sagittarius in the off-plane sample then not only maintains the large gap at the bright-end between the luminosity distributions of the on- and off-plane samples, but in fact does so with high statistical significance (KS and AD tests returning p-values $<0.05$).

The above points show that it is crucial not only to improve the proper motions of distant bright systems to firmly establish their association with the VPOS, but also to have a uniform coverage of the sky to find potentially missing bright systems. A further concern indeed is that the footprint of current surveys may influence the number of satellites that end up on the VPOS. \citet{Pawlowski2016}, analysing the SDSS footprint, ruled out this possibility by showing that the SDSS satellites found then increased  the significance of the VPOS \citep*[see also][]{Metz2009}.

Considering the contribution of the MC system, the LMC in its infall into the MW-potential has probably brought in, together with the SMC, a number of faint satellites (\citealp[we consider here those identified in][]{Battaglia2022} and marked in Table~\ref{tab:op}, \citealp[but see also e.g.][]{Patel2020}). The LMC associated systems have orbital poles mostly aligned with the VPOS (except for Carina~II, which however is just outside the $\theta_{\rm inVPOS}$ area; see Fig.~\ref{fig:orb_poles}). By removing the LMC and associated systems, the observed gap between the luminosity distributions of the on- and off-plane samples is still present at $M_V \sim -8$, as shown in Fig.~\ref{fig:VPOS_hist}, although less severe (6 counts compared to the previous 8). We note that in this case we are taking a conservative approach since the LMC, being an independent object, could be left in the count of on-plane systems.

\section{Comparison of orbital dynamics}
\label{sec:comp_orb_props}

The analysis of the orbital properties of our sample depends necessarily on the assumed MW potential for calculating the orbital parameters. Therefore, we first inspect the velocity vector of our systems, which is independent of such assumptions, focusing on their radial and tangential velocity values. Then, we continue inspecting the orbital parameters (i.e., pericentre, apocentre, eccentricity, period, time since the last pericentric passage, together with other parameters we derived from them) made publicly available by \citet{Battaglia2022}, obtained assuming two different MW potentials. We note that from \citet{Battaglia2022} we have only used tabulated values of the orbital parameters, whereas we explicitly state when we performed our orbital integration.

\subsection{Kinematic properties}
\label{subsec:comp_orb_props-kin}

\begin{figure*}
    \centering
    \includegraphics[width=.49\textwidth]{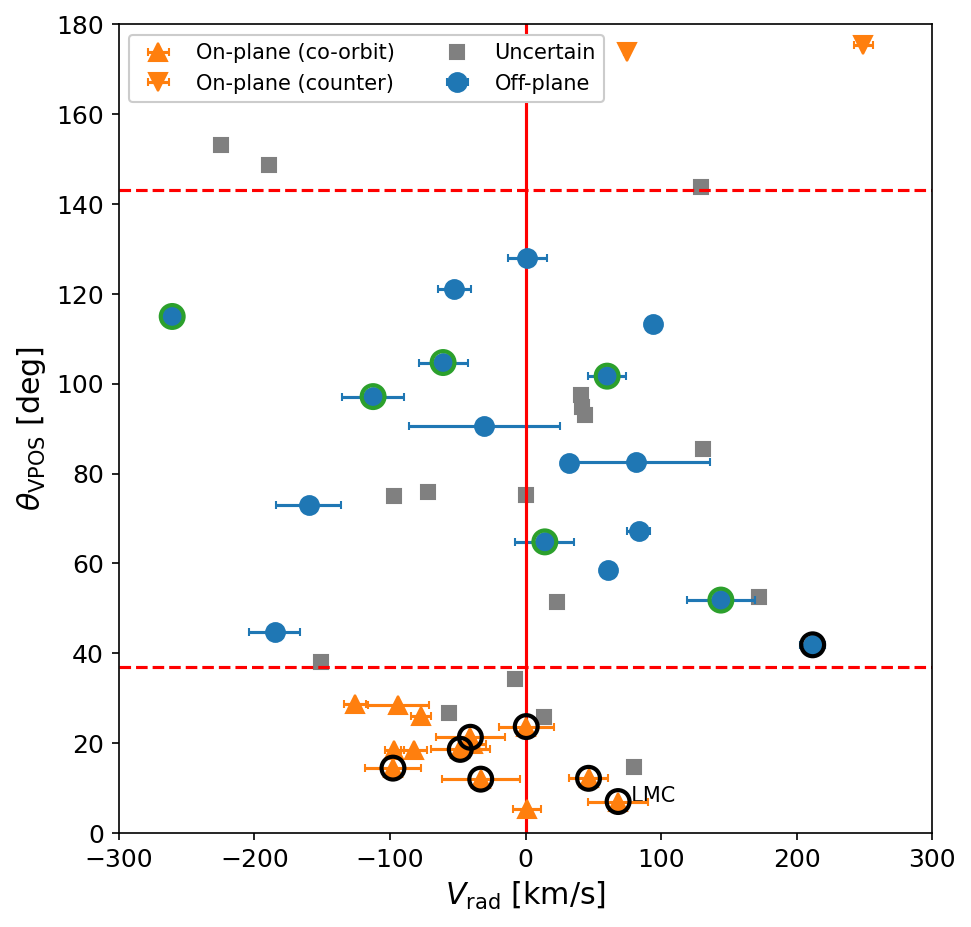}
    \includegraphics[width=.49\textwidth]{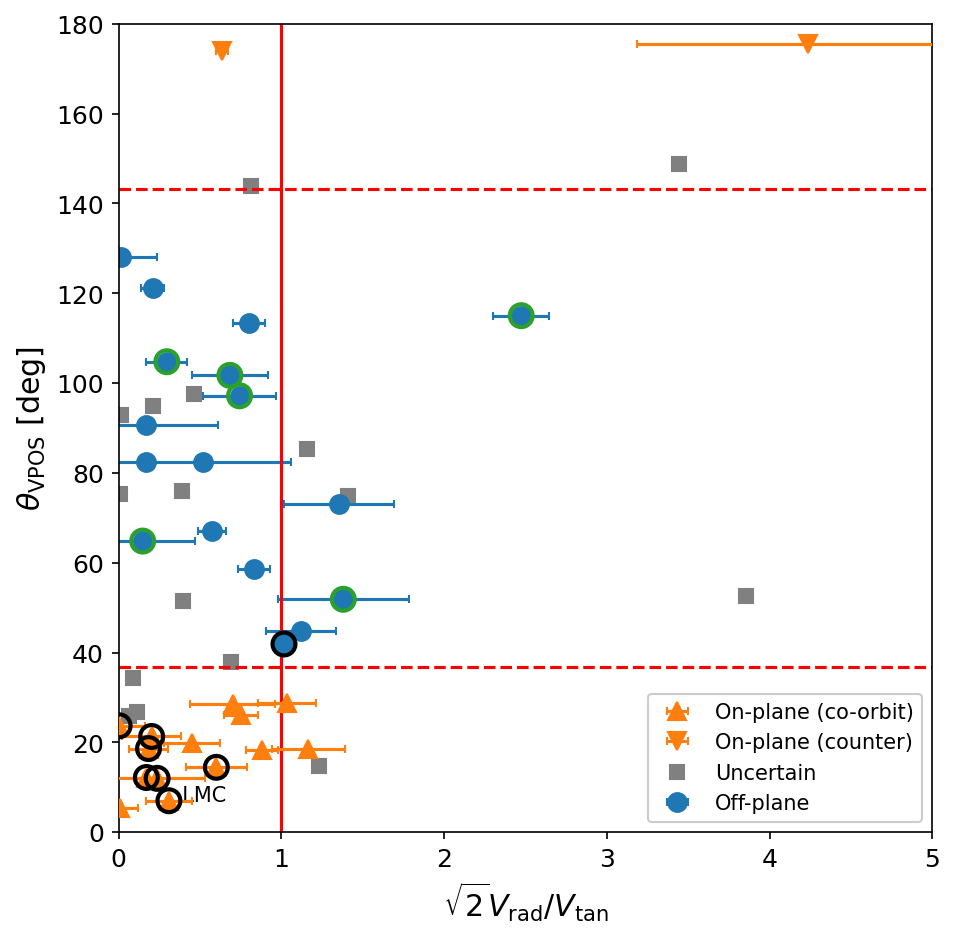}
    \caption{
    \emph{Left:} Median angular difference $\theta_{\rm VPOS}$ between the orbital poles and the VPOS normal for our sample of MW satellites, as a function of their Galactocentric radial velocity $V_{\rm rad}$; the red solid line is for reference only, since it helps to identify satellites that are approaching the MW ($V_{\rm rad}$[km\,s$^{-1}$]~$<0$) or receding from it ($V_{\rm rad}$[km\,s$^{-1}$]~$>0$); the red dashed lines represent the aperture angles around the two opposite poles of the VPOS. 
    \emph{Right:} $\theta_{\rm VPOS}$ as a function of the $\sqrt{2}V_{\rm rad}/V_{\rm tan}$ ratio; systems with $\sqrt{2}V_{\rm rad}/V_{\rm tan}<1$, marked by the red solid line, show a tangential velocity excess.
    Symbols and colors in both panels as in Fig.~\ref{fig:VPOS_ph}; error bars for the uncertain systems, which are large, are not shown for clarity.}
    \label{fig:thetaVPOS_Vrad}
\end{figure*}

Our on-plane and off-plane samples were selected on the basis of the orbital poles of the satellite galaxies. The categorisation thus depends only on the accuracy of their tangential velocity direction (since the orbital pole is the normalised cross product of the Galactocentric position and velocity vector) and is therefore independent of the radial component and the absolute value of the tangential component of the velocity vector. Do the on-plane and off-plane samples differ in these velocity components with respect to the MW?

In the left panel of Fig.~\ref{fig:thetaVPOS_Vrad}, we show the Galactocentric radial velocity component $V_{\rm rad}$ of our systems\footnote{Radial and tangential velocities obtained by transforming the Galactocentric Cartesian coordinate system $(x,y,z; V_x,V_y,V_z)$ into a spherical one $(r,\theta,\phi;V_{\rm rad},V_\theta,V_\phi)$, where the tangential velocity is defined as $V_{\rm tan}=\sqrt{V_\theta^2+V_\phi^2}$.}, which mostly depends on their line-of-sight velocity, compared to the median angular difference between their orbital poles and the VPOS normal, $\theta_{\rm VPOS}$. This reveals a striking feature: the majority of the co-orbiting on-plane systems are currently approaching the MW ($V_{\rm rad} < 0$), while both counter-orbiting on-plane systems are receding from it. Only four of the 14 co-orbiting satellites do not contribute to this pattern: two have clearly receding velocities (i.e. Carina~III and the LMC), while the other two have almost zero radial velocity (i.e. Carina and the SMC). This is in contrast to the radial velocity distribution of the off-plane (as well as the uncertain) systems, which distribute evenly around $V_{\rm rad}=0$~km\,s$^{-1}$. Indeed, the median and median absolute deviation of the co-orbiting systems are $-44$~km\,s$^{-1}$ and $67$~km\,s$^{-1}$ ($-80$~km\,s$^{-1}$ and $43$~km\,s$^{-1}$, if we remove the LMC satellites), while for the off-plane systems we have $14$~km\,s$^{-1}$ and $103$~km\,s$^{-1}$, respectively. 
Performing a KS and AD test between the radial velocity distributions of the co-orbiting and off-plane samples, we recover p-values $<0.15$, for both tests. The same conclusions are obtained by removing the LMC satellites.

We note that the radial velocity pattern is robust and remains unchanged even when we consider the effect of the MW's reflex motion caused by the infall of a massive LMC \citep[see e.g.][]{Pawlowski2022}. This pattern implies that almost all co-orbiting VPOS members are currently approaching the MW and thus their respective orbits' pericenters, while the two counter-orbiting VPOS members are receding and thus had experienced their last pericenter recently. In contrast, the off-plane systems are spread more randomly between approaching or receding, and thus between before or after, their orbits' nearest pericenter.

Regarding the Galactocentric tangential velocity $V_{\rm tan}$ properties of MW satellites, it has been reported in the literature that most ($\sim70-80\%$) of those with measured proper motion show an excess of tangential velocity, meaning that their orbital kinetic energy is dominated by tangential motion \citep{Cautun+Frenk2017,Riley2019,Hammer2021}. This seems to be in sharp contrast with the low fraction ($\sim1.5\%$) expected from simulations \citep{Cautun+Frenk2017}. We confirm this finding also with the proper motion data used in this work. As shown in the right panel of Fig.~\ref{fig:thetaVPOS_Vrad}, we recover a tangential velocity excess for $\sim75\%$ of the satellites (i.e., having $\sqrt{2}V_{\rm rad}/V_{\rm tan} <1$), both considering each sub-sample or taking them as a whole. Furthermore, we do not find particular differences between the on- and off-plane samples in this regard, although the LMC satellites are more tangentially biased than the other on-plane systems. It can be argued that the observed tangential velocity can be a biased estimator of the true value, but as already noted in \citet{Battaglia2022}, such biases are minimised for measurement uncertainties $<70$~km\,s$^{-1}$. As reported in Sect.~\ref{sec:data}, our on- and off-plane systems are safely within this limit. 

\subsection{Orbital properties}
\label{subsec:comp_orb_props-orb}

We inspect now the orbital parameters (i.e., pericentre, apocentre, eccentricity, period, and time since the last pericentric passage) made publicly available by \citet{Battaglia2022}, obtained assuming either a low-mass ($M_{\rm vir}=8.8\times10^{11}M_\odot$ within $r_{\rm vir}=251$~kpc) or a high-mass potential ($M_{\rm vir}=1.6\times10^{12}M_\odot$ within $r_{\rm vir}=307$~kpc) for an isolated MW. The former is a triaxial potential published originally in \citet{Vasiliev2021}, while the latter is a scaled and more massive version of the \texttt{MWPotential2014} from \citet{Bovy2015}. Such potentials bracket the range of likely MW masses, which distribute around $M_{\rm vir}\sim10^{12}M_\odot$ \citep[e.g.][and references therein]{Wang2020}.
We note that for some systems only the lower limits of their orbital parameters are available, since they do not reach their apocentre during the orbital integration period\footnote{Orbits were integrated 6~Gyr backward, but also forward in order to calculate the pericentre of the most distant systems \citep[see][]{Battaglia2022}.}.
We also warn that these orbital parameters are not free of biases, particularly with regard to tangential velocity, but as already mentioned, we are within safe limits. For more details on the MW potentials and how the orbital parameters were derived, we refer to \citet{Battaglia2022}. 

Comparison of the orbital parameter distributions between on- and off-plane samples shows no significant differences using the KS and AD statistical tests, for both MW potentials (see again Table~\ref{tab:params}). 
Similarly, analysing the level of correlation between the orbital parameters, we found low values (i.e. $\left| \rho \right| \leq 0.5$) for both samples considering the high-mass potential case, while we found stronger values among some parameters in the low-mass potential case (see again the correlation matrix in Fig.~\ref{fig:VPOS_corr}, but also Figs.~\ref{fig:VPOS_orb_iso} and \ref{fig:VPOS_orb_heavy}). 
We caution, however, that in the latter case stronger $\left| \rho \right|$ values may arise because fewer targets have a complete orbital solution, which may lead to possible spurious correlations between parameters.

\begin{figure*}
    \centering
    \includegraphics[width=.49\textwidth]{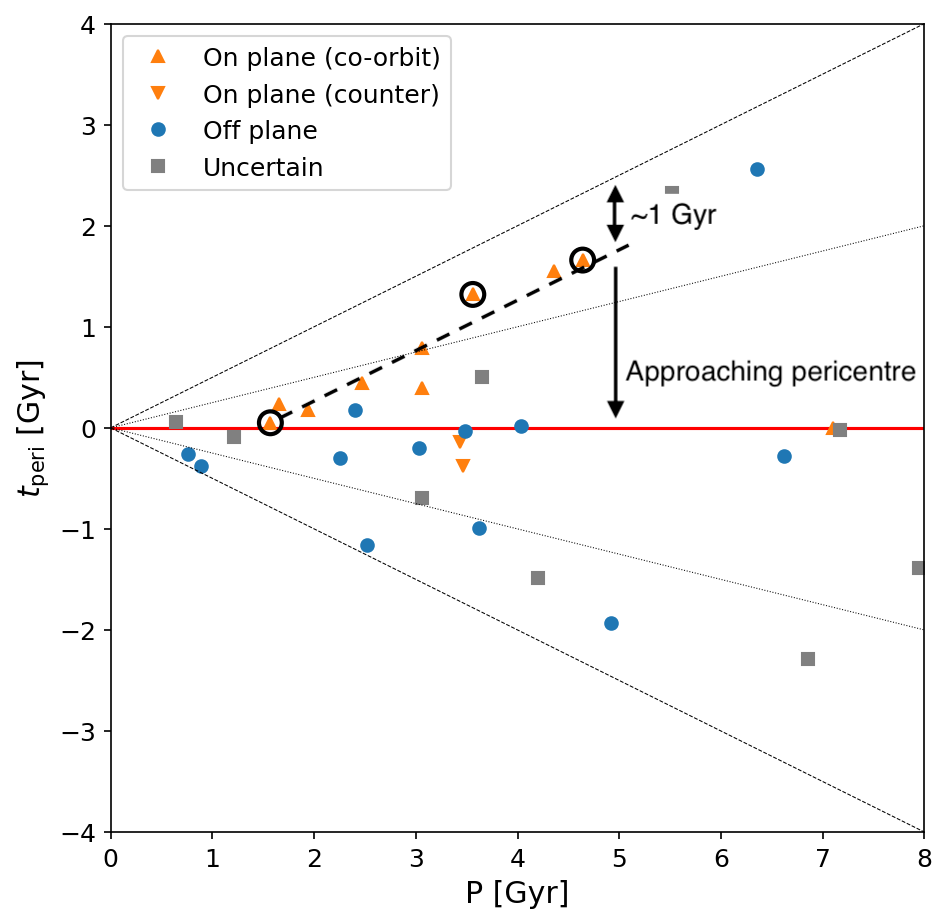}
    \includegraphics[width=.49\textwidth]{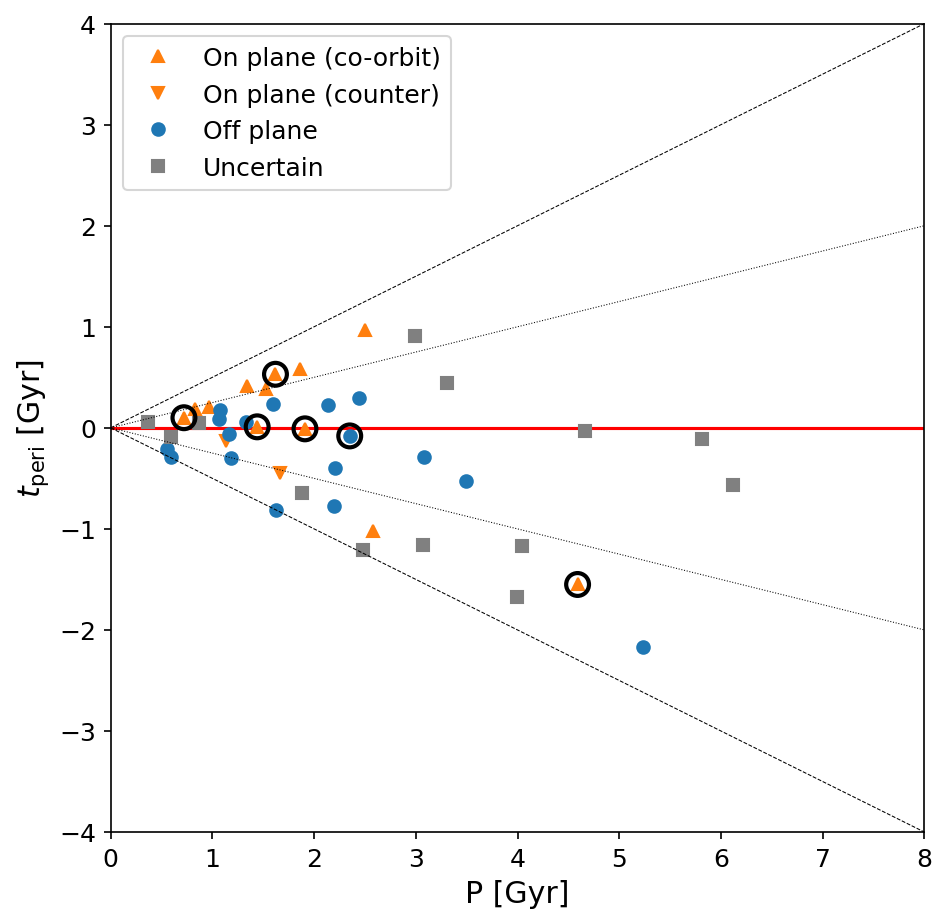}\\
    \includegraphics[width=.497\textwidth]{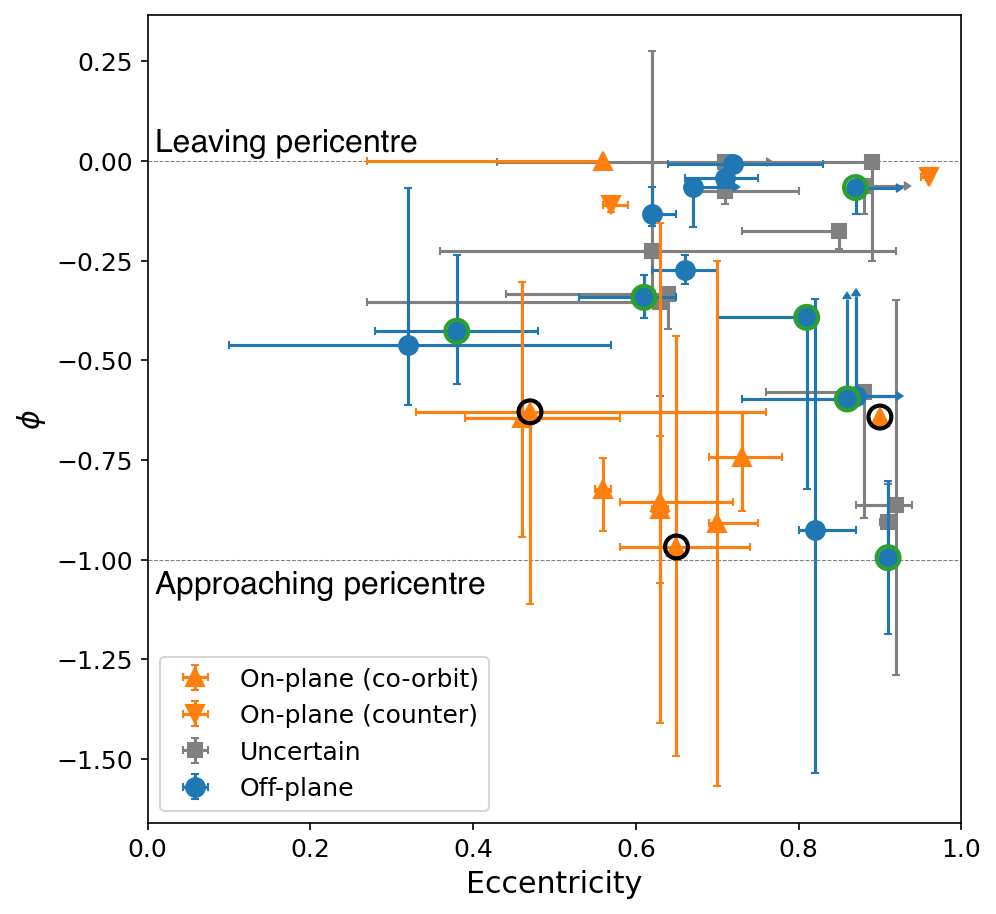}
    \includegraphics[width=.49\textwidth]{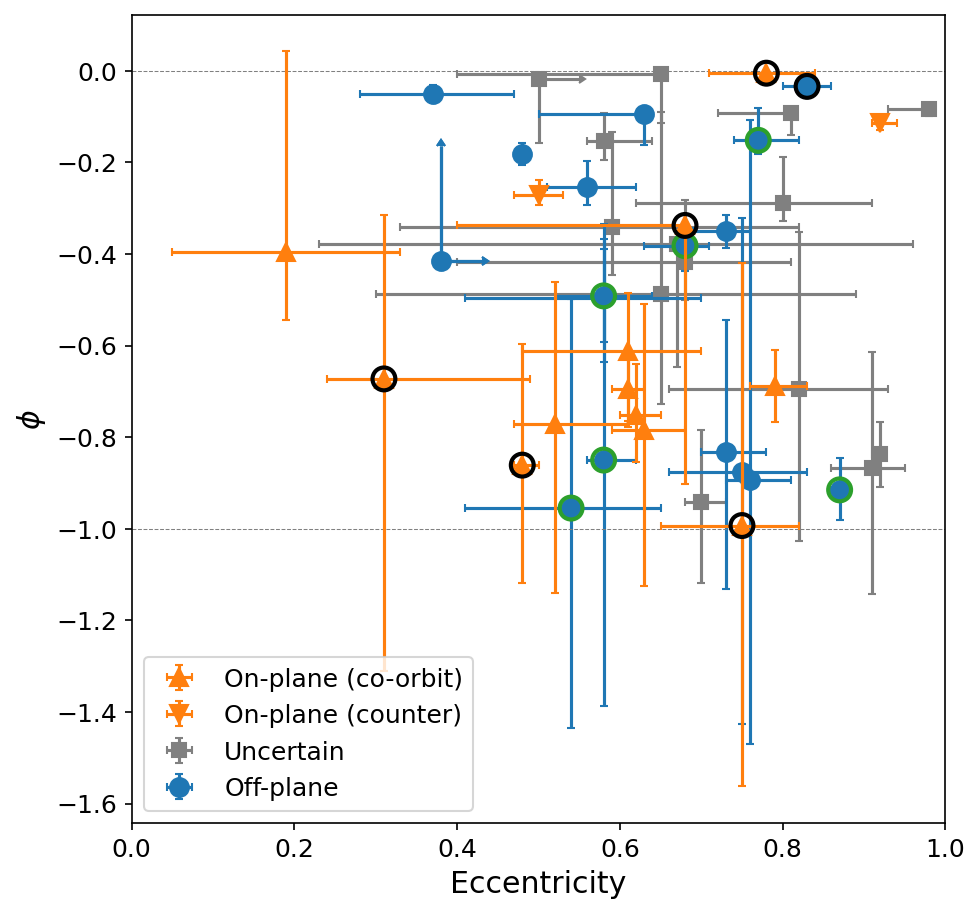}
    \caption{\textit{Top:} time to the next or since the last pericentric passage, $t_{\rm peri}$, as a function of the satellites' orbital period, $P$, both in units of Gyr; the red solid line at $t_{\rm peri}=0$ indicate when satellites are at pericentre, the thin dashed lines indicate the apocenter limits, while the dotted lines mark the half-orbit; the thick dashed line, shown for illustrative purposes, is parallel to the upper apocentre line, shifted by $\sim1$~Gyr. \textit{Bottom:} satellite orbital phases, $\phi$, as a function of the orbital eccentricity; thin dotted lines represent the limits when a satellite has just left its pericentre ($\phi$ close to 0), or is approaching it ($\phi$ close to $-1$). Panels on the left show values obtained with the low-mass MW potential, while those on the right with the high-mass MW potential. In all panels, symbols and colors are as described in Fig.~\ref{fig:VPOS_ph}.}
    \label{fig:VPOS_orb_Tperi}
\end{figure*}

An additional aspect we examined is related to the excess of MW satellites found near their orbital pericentre, first reported using \textit{Gaia}-DR2 data \citep[see][]{Fritz2018,Simon2018}. This is a potential problem because, as MW satellites are on eccentric orbits, one would expect to find them close to their apocentre, where they spend more time. \citet{Pawlowski2021} confirmed the significance of the pericentre excess by also analysing the orbital parameters of \citet{Battaglia2022}, finding that such an effect is more prominent when the low-mass MW potential is assumed, in agreement with previous results (\citealp[e.g.][]{Li2021}, \citealp[but see also][]{Correa-Magnus+Vasiliev2022}). This problem may be related to the tangential velocity excess described in the previous section, since the radial velocity is minimal around the pericentre. As in that case, we checked for possible differences between the on-plane and off-plane samples in the distribution of $f_{\rm peri}$, the fractional distance in time of a satellites to be found between pericentre ($f_{\rm peri}=0$) and apocentre ($f_{\rm peri}=1$), finding none significant (see further details in the Appendix). 

The preferred negative radial velocities of the co-orbiting on-plane systems have already indicated a clear preference of this sample of satellites to move towards their pericentre. We now investigate this in more detail using the full orbit integration, taking into account whether a satellite is moving towards or away from the pericentre and in what orbital phase it is in. Again, we find that the co-orbiting on-plane systems preferentially approach their pericentre, but they also show a remarkable pattern in their time since pericentre. This is illustrated in the top panels of Fig.~\ref{fig:VPOS_orb_Tperi}, where we inspected $t_{\rm peri}$, the time to the next ($t_{\rm peri} > 0$) or since the last ($t_{\rm peri} < 0$) pericentric passage as a function of $P$, the satellites' orbital period. 
In both figures, the solid red line indicates that a satellite is at its pericentre, the dashed black lines represent when it is at the apocentre (i.e., at $\pm P/2$), while the dotted lines represent when it is halfway between these two points (at $\pm P/4$). Therefore, a satellite leaving its apocentre, passing through its pericentre and then returning to the apocentre, will evolve on the diagram going from positive to negative $t_{\rm peri}$ values. If there is a group of satellites with different \textit{P} leaving their apocentre at the same time, they will move towards their pericentre parallel to the upper dashed line shown in the figure. \textit{This seems to be the case for the on-plane systems: the majority of the co-orbiting satellites are moving towards their pericentres after having left their apocentres $\sim1$~Gyr ago}, while the counter-orbiting systems have just passed their pericentres. 

Our result holds true for both galactic potentials adopted, although it is more striking for the low-mass potential, which is more realistic in light of recent MW mass estimates \citep[e.g.][and references therein]{Bird2022}. The systems that deviate most from this pattern are the probable LMC satellites, which however have more uncertain orbital parameters and should not be considered as independent satellites. Therefore, if we exclude them, we are left with only one system that does not follow the pattern in either case, the Carina dSph. We note that this system has a radial velocity with respect to the MW that is consistent with zero (see Fig.~\ref{fig:thetaVPOS_Vrad}), so its orbital properties depend strongly on its tangential motion and the assumed MW potential and it can either be close to peri- or apocenter, or on a circular orbit. It follows that Carina has an uncertain apocentre and eccentricity in the case of the low-mass potential, while for the high-mass potential its orbit is mostly circular, which leads to large uncertainties in the orbital period in this case as well. In both cases, its position within the errors is consistent with approaching the pericentre and can therefore be associated with the rest of the co-orbiting systems. 

A complementary way to highlight the coordinated motion of the on-plane systems is by inspecting the satellites' orbital phase $\phi$ (i.e. the ratio between the time since the last pericentric passage and the orbital period). In our definition, a satellite with $\phi$ close to 0 has just left its pericentre, whereas if $\phi$ is close to $-1$ it is approaching the pericentre.
Similarly to the previous case, we find that almost all systems co-orbiting (counter-orbiting) within the plane have comparable $\phi$ values, within errors, and move towards (away from) their pericenters for both potentials, as shown in the bottom panels of Fig.~\ref{fig:VPOS_orb_Tperi}, where we plot $\phi$ values as a function of orbital eccentricity. Such a result is again more pronounced in the case of the low-mass potential, where the $\phi$ distribution of the on-plane systems differs significantly from that of the off-plane systems (see Table~\ref{tab:params}). 
Performing KS and AD tests to assess that the $\phi$ distribution of on-plane satellites is not drawn from a uniform distribution yields significant p-values $<0.05$, considering only co-orbiting satellites. We can thus reject the hypothesis that the co-orbiting satellites are found in random orbital phases. This result is valid for both potentials, whereas the inclusion of counter-orbiting systems would cause it to lose significance only in the case of the less realistic high-mass potential. This result is also not dominated by the LMC satellites (for which we can be confident that their orbital phase is not random but linked to that of the LMC) and is still present for co-orbiting on-plane satellites not linked to the LMC (leading to p-values of $\sim0.04$ and $\sim0.06$ for the low and high mass potentials respectively). 
We also note that the co-orbiting systems share similar eccentricities values, with a small scatter.
High p-values are instead obtained by analysing whether the $\phi$ distribution of the off-plane systems is drawn from a uniform distribution, indicating that they are consistent with random orbital phases.

Although the large errors associated with $\phi$, especially for the low-mass potential case (see Fig.~\ref{fig:VPOS_orb_Tperi} again), limit our interpretation of the orbital motion of the on-plane systems, the radial velocity of the co-(counter)orbiting systems alone already clearly indicates that they are approaching (moving away from) the MW. 
The expected phase mixing of the on-plane systems approaching the MW depends on their orbital periods and that the relevant timeline to wash out the observed coherence is of the order of $\sim1.5$~Gyr (i.e. half of the difference between the minimum and maximum of their $P$ values). Therefore, the observed $\phi$ distribution mixes relatively quickly. If we evolve the positions of the satellites on the $t_{\rm peri}-P$ plane, we observe that the current configuration of the on-plane systems does not repeat itself for at least a Hubble time. Does this indicate that the on-plane systems have only recently arrived in the MW potential, possibly as a group of dwarf galaxies, or is it just a chance occurrence? And what is the possible role played by the LMC?

\subsection{The role of the LMC}
\label{subsec:comp_orb_props_LMC}

\begin{figure}
    \centering
    \includegraphics[width=.49\textwidth]{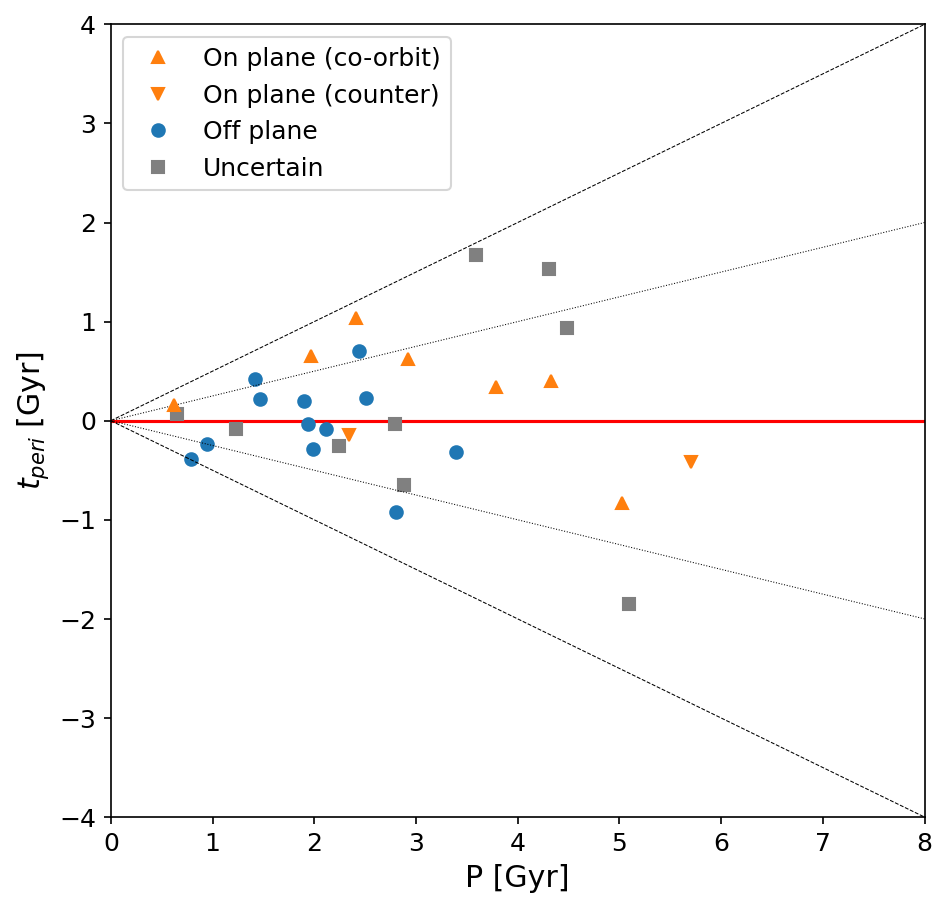}\\
    \includegraphics[width=.49\textwidth]{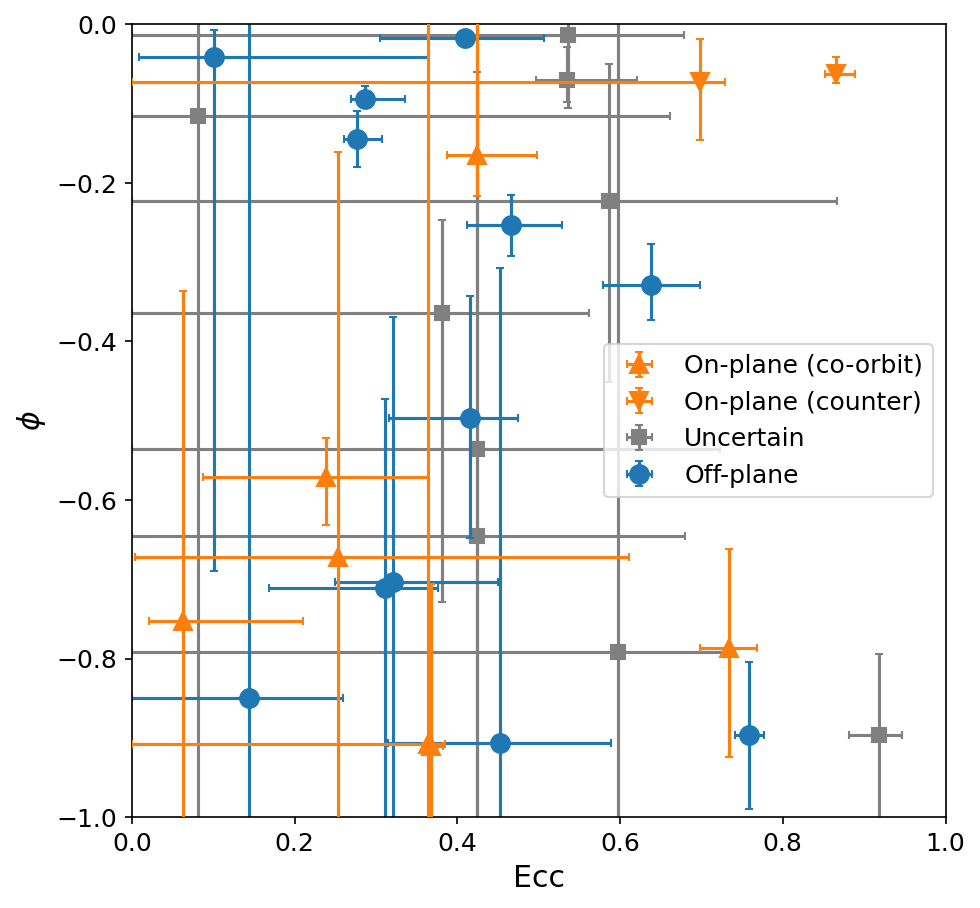}
    \caption{As in Fig.~\ref{fig:VPOS_orb_Tperi}, but considering results obtained with a MW potential perturbed by a massive LMC (see details in Sect.~\ref{subsec:comp_orb_props_LMC}), in the \textit{top} panel is shown the time to the next or since the last pericentric passage, $t_{\rm peri}$, versus the satellites' orbital period, $P$, while in the  \textit{bottom} panel the satellites' orbital phases, $\phi$, versus the orbital eccentricity. We note that while some systems have a measured $t_{\rm peri}$, they may not have a closed orbit and therefore have $P=0$ (like the LMC and its satellites).}
    \label{fig:VPOS_orb_Tperi_evolvPot}
\end{figure}

Several lines of evidence have been presented in the literature in support of a massive LMC (i.e. between $1-2\times10^{11}M_\odot$) that has just passed its pericentre, probably for the first time \citep[see][for a review]{Vasiliev2023}. Such a massive halo can significantly affect the orbits of the MW's satellites \citep{Patel2020,Battaglia2022,Correa-Magnus+Vasiliev2022,Makarov2023}, but it can also cause changes in the MW itself, such as the reflex motion of the MW's centre caused relative to its more extended dark halo by the gravitational pull of the LMC \citep{Garavito-Camargo2019,Garavito-Camargo2021,Petersen+Penarrubia2020,Petersen+Penarrubia2021,Vasiliev2021}.
In particular, \citet{Garavito-Camargo2021} showed the ability of the LMC to produce an excess of orbital poles of dark matter halo particles towards its own pole direction, although \citet{Pawlowski2022} showed that this might be insufficient to explain the presence of MW satellites co-orbiting along the VPOS.
Here we rather discuss the impact of the LMC on the orbital analysis results shown above.

In this regard, we have further investigated the orbital properties of MW satellites by considering the low-mass MW potential perturbed by a $1.5\times10^{11}M_\odot$ LMC, first introduced in \citet{Vasiliev2021} and then used in \citet{Battaglia2022} to study the effects of such a massive LMC on the orbital properties of MW satellites. They found that, as far as the VPOS satellites are concerned, especially the bright ones, Sculptor would be on its first infall, while Carina and Fornax would have a fraction ($25\%$ and $4\%$ respectively) of their Monte Carlo orbit realisation linked to the LMC, the former possibly being a past satellite. On the other hand, the former LMC satellites would not yet be in a closed orbit (or have reached their pericentre). These results depend on the assumed masses of the MW and LMC, but they already underline that the effect of the LMC on the orbits of other MW satellites may not have been negligible \citep[see also][]{Correa-Magnus+Vasiliev2022,Makarov2023}.

In our case, we checked the possible influence of the LMC in comparison with the results in Sect.~\ref{sec:comp_orb_props}. 
We used the code \texttt{Agama} \citep{Vasiliev2019} to integrate the orbits of our sample of satellites assuming the aforementioned perturbed potential of \citet{Vasiliev2021}.
This is a time-dependent potential of the MW that takes into account the trajectory of a massive LMC, which entered the dark matter halo of the MW about 1.6~Gyr ago, and the acceleration induced by the reflex motion of the MW. We integrated the orbits backwards for 5 Gyr from 100 Monte-Carlo realisations of the position and velocity of our systems generated from their present values (see Table~\ref{tab:op}). The orbital parameters (time since last pericentre, period, eccentricity and phase) were calculated for each realisation, from whose distributions we derived their median and, as uncertainties, the 16th and 84th quantiles. We note that for the LMC and its satellites, which in this case are at their first infall, orbital parameters have not been recovered, a part for $t_{\rm peri}$ when available.

As shown in the top panel of Fig.~\ref{fig:VPOS_orb_Tperi_evolvPot}, we find that the co-orbiting on-plane systems are still approaching their pericenters (again excluding Carina), but with changes in their $t_{\rm peri}$ and $P$, here to smaller values compared to the low-mass potential case. This is expected due to the higher total mass of the combined MW and LMC. 
Looking at the bottom panel, we again see that the co-orbiting on-plane systems have $\phi<-0.5$, meaning that they are approaching the pericentre, but on less eccentric orbits, as is also the case for the other satellites. The associated errors are, however, larger since the influence of the LMC makes the determination of orbital parameters more uncertain. 

Despite the significant influence of the LMC and the uncertainties it introduces, we confirm our previous finding that co-orbiting on-plane systems approach the pericentre, although it is uncertain whether they do so in a coordinated manner. It therefore appears to be a signal independent of the presence of the LMC and thus not caused by it.

\subsection{Energy and angular momentum properties}
\label{subsec:comp_orb_props-E_L}

\begin{figure*}
    \centering
    \includegraphics[width=.49\textwidth]{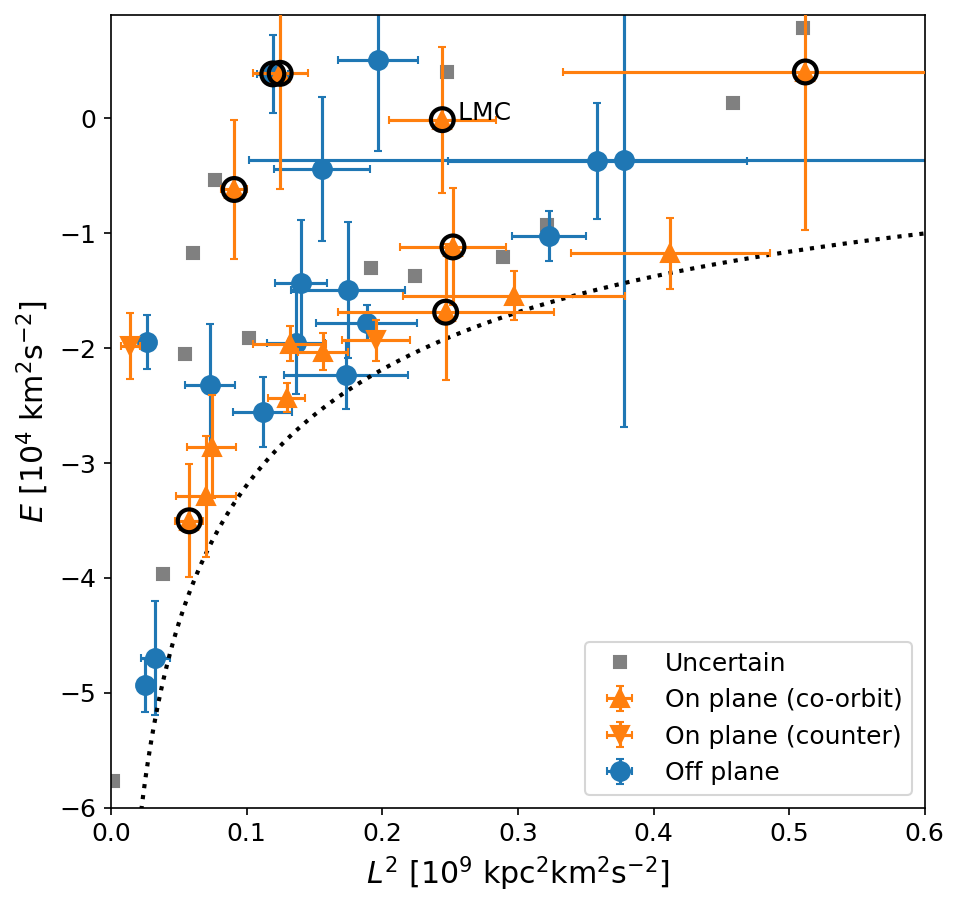}
    \includegraphics[width=.49\textwidth]{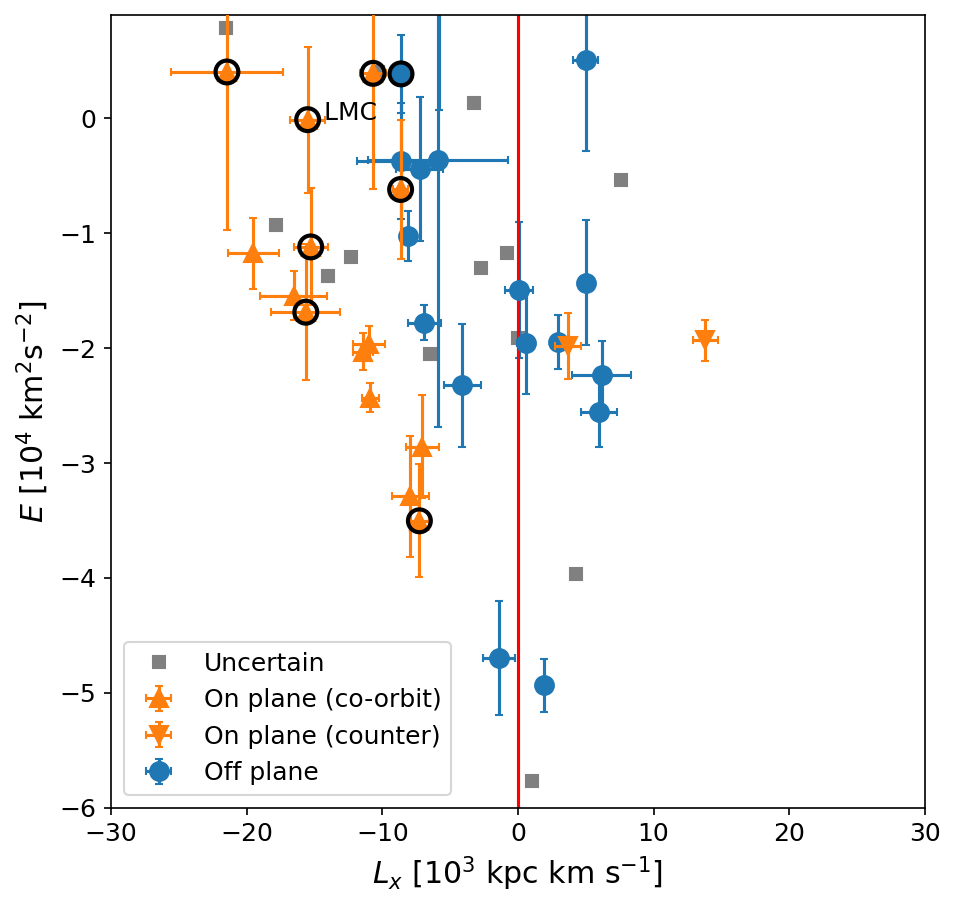}
    \caption{\textit{Left:} total specific energy $E$ (in units of $10^4$~km\,s$^{-1}$) versus the square of the specific angular momentum $L$ (in units of $10^9$~kpc$^2$\,km$^2$\,s$^{-2}$) at the present time; dotted line represents the relation for a particle of negligible mass on a circular orbit. \textit{Right:} $E$ against the specific angular momentum in the Cartesian x-direction of a Galactocentric reference frame, $L_x$ (in units of $10^3$~kpc\,km\,s$^{-1}$); the red solid line at $L_x=0$ is for reference only. In both panels, symbols and colors as in Fig.~\ref{fig:VPOS_ph}; error bars for the uncertain systems, which are large, are not shown for clarity.}
    \label{fig:VPOS_E_L}
\end{figure*}

Before discussing the implications of our results on possible scenarios for the formation of the VPOS, we first examine the properties of our satellite sample in terms of specific energy $E$ and angular momentum $L$ at the present time. As already noted by \cite{Hammer2023}, satellites associated with the VPOS have minimal specific energy $E$ values for a given angular momentum $L$, closely following the relation for circular motion on the $E-L$ plane. We confirm this result, as shown in the left panel of Fig.~\ref{fig:VPOS_E_L}, where we see that the on-plane systems generally tend to have the lowest $E$ for a given $L^2$, in contrast to the other satellites which show a larger scatter. In the right panel, we instead plot $E$ versus the specific angular momentum in the Cartesian x-direction of a Galactocentric reference frame, $L_x$, and again find that the on-plane systems tend to have the lowest $E$ for a given $L_x$ value\footnote{We choose $L_x$ because the VPOS is a planar structure perpendicular to the MW disc and distributed mainly in the Cartesian yz-plane.}. In both panels we only show the $E$ values obtained for the low-mass potential, but similar conclusions are drawn for the high-mass potential, and also for the LMC perturbed potential considered in the previous section. These results seem to indicate that the VPOS has a cold dynamics, rather than being formed by satellites on their first arrival. A comparison can be made with the LMC system (which arrived $<2$~Gyr ago), whose satellites (including the LMC) have on average higher $E$ for a given $L$. Combined with the previous results, what are the implications on the longevity of the VPOS?

\section{Discussion}
\label{sec:discussion}

Our comparison of the satellite galaxies associated with the VPOS and those that are not has revealed several distinct differences in their properties: the known on-plane satellite galaxies tend to be more luminous; they show a strong pattern in their orbital phases, with the majority of the co-orbiting satellites currently approaching their pericenters, while the two counter-orbiting satellites have both recently passed their pericenters. In addition, the on-plane satellites not associated with the LMC tend to have the lowest orbital energies for a given angular momentum value. We shall discuss then what are the implications for the longevity of the VPOS in light of our results.

The possibility that the LMC system is on its first pericentric passage \citep[][and references therein]{Vasiliev2023} and the fact that most of its components are part of the on-plane sample seems to indicate that the VPOS is a dynamic structure that could be currently forming. The question is therefore how long the other on-plane systems have been part of the VPOS and whether they have been associated with the LMC in the past. 
We can see two possible interpretations: \textit{(i)} the on-plane systems not currently associated with the LMC are long-time satellites of the MW, having completed several pericentric passages, and thus they are old members of the VPOS; \textit{(ii)} they may or may not have previously been associated with the LMC, but have recently entered the MW potential and have completed at most a few pericentric passages.

Assuming that the VPOS is an old structure, as advanced in \textit{(i)}, would imply that the orbital features reported in Sect.~\ref{sec:comp_orb_props} are a mere coincidence. We have already seen that it is difficult to interpret the dynamic properties and phase alignment of the on-plane systems in this way, but it is also true that these aspects have not been yet investigated in the context of cosmological simulations. In fact, such comparisons have focused rather on the spatial flattening and clustering of the orbital poles of the simulated systems. We leave this topic for a future study, while we focus here on further investigating point \textit{(ii)}, which interprets the observed orbital features as a significant property of the VPOS from which clues about its origin can be deduced (which a chance alignment does not allow).

The interpretation that the VPOS may be a rather young structure, as proposed in \textit{(ii)}, mainly concerns three aspects, all related to the late accretion of satellite systems: \textit{(a)} VPOS members are tidal dwarf galaxies (TDGs) formed after a major merger or close interaction between the MW and another galaxy; \textit{(b)} the VPOS may be the result of a late merger experienced by the MW, but in this case its components are the former satellites stripped from the merged galaxy; \textit{(c)} VPOS members were part of a bound group of dwarf galaxies that was tidally stripped by the MW and has not yet had time to fully phase mix.

The proposed scenarios underline the complexity of understanding the origin of the VPOS. As each of them deserves an in-depth study that would go beyond the scope of this article, we limit ourselves in the following sections to exploring their implications given our findings.

\subsection{Do on-plane systems have a tidal origin?}
\label{subsec:comp_phy_props_TDG}

\begin{figure}
    \centering
    \includegraphics[width=.49\textwidth]{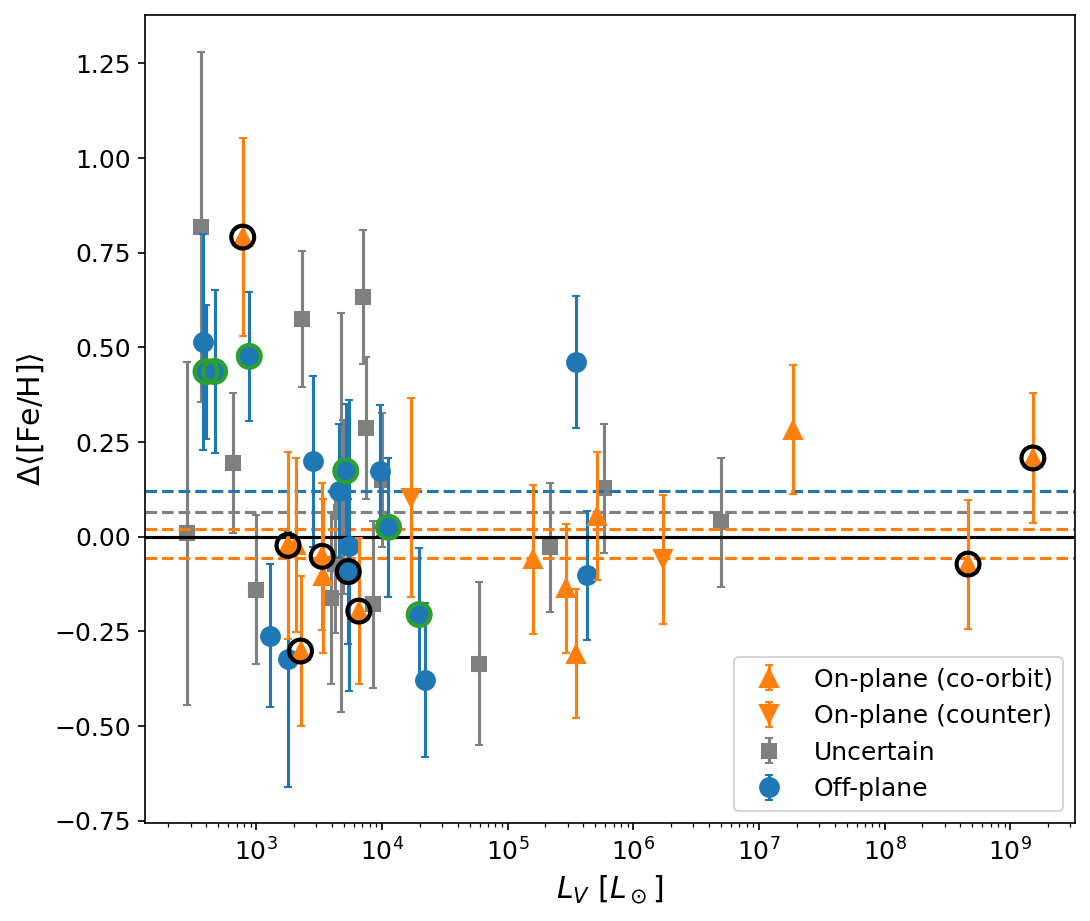}
    \caption{Difference between the observed weighted-average $\left<{\rm [Fe/H]}\right>$ of our galaxies and the expected values at their luminosity obtained from the \citet{Kirby2013} relation; symbols and colors as in Fig.~\ref{fig:VPOS_ph}. Error bars obtained adding in quadrature the errors on $\left<{\rm [Fe/H]}\right>$ and the $rms=0.17$~dex of the \citeauthor{Kirby2013} relation for the MW satellites. Dashed lines are the median offsets of each sub-sample.}
    \label{fig:VPOS_Lv_FeH}
\end{figure}  

The possibility that the VPOS is populated by TDGs was one of the first proposals put forward on its origin \citep{Lynden-Bell1976,Kunkel1979}. Indeed, TDGs are second-generation galaxies that form in a common tidal tail from material recycled from a previous collision, thus sharing its orbital plane orientation. 
Being dark-matter-free systems with a high gas content at the time of formation, TDGs are also more likely to experience the effects of ram-pressure stripping and tidal disruption during their orbit around the host galaxy than dark-matter-dominated systems (\citealp{Klessen+Kroupa1998}, \citealp{Yang2014}, but see also \citealp{Ploeckinger2015}). 
Therefore, if we take into account that TDGs could disperse after approaching the MW, we could assume that the ones currently observed are on a first pericentric passage and interpret this as consistent with the strongly skewed orbital phase distribution we found, where the majority of co-orbiting VPOS members are currently located ahead of their respective pericenters. 

Interestingly, the peculiar observed orbital phase distribution is reminiscent of a proposal by \citet*{Pawlowski2011} to explain the presence of counter-orbiting satellite galaxies aligned with the VPOS. In this model, tidal debris from a past galaxy collision accumulates on orbits around the MW. As the tidal debris tail, in which TDGs are formed, passes over the centre of the MW, its orbital direction changes and the later accreted TDGs move in the opposite orbital direction to the earlier accreted ones. This is consistent with one orbital direction (the two counter-orbiting VPOS members) having recently passed their pericenters and thus being further along their orbits than the co-orbiting satellite galaxies, which are still approaching their pericenters. This would imply that the TDG members of the VPOS only arrived in the last few Gyrs around the MW \citep[as envisaged in e.g.][]{Fouquet2012}. 

Based on the analysis of \textit{Gaia} data, \cite{Hammer2021} suggested that VPOS systems may have entered the MW potential $<3$~Gyr ago as gas-rich dwarf galaxies that lost their gaseous content due to ram pressure stripping, which also caused a deceleration along the radial direction of their orbits. This mechanism is most efficient for luminous dwarfs, which in fact dominate the VPOS population (see Sect.~\ref{sec:comp_phy_props}), but also assumes that they are TDGs formed in a past merger event between the MW and Andromeda \citep[see e.g.][]{Yang2014}.

In the scenario in which the VPOS systems are late-arriving TDGs, we would expect them to be different from other MW satellites, which are instead dark-matter dominated dwarf galaxies that have followed different formation and evolutionary paths. In particular, TDGs should be more metal-rich than the general population of dwarf galaxies because they would have formed from the recycled enriched material released by the merging parent galaxies. 

In Fig.~\ref{fig:VPOS_Lv_FeH}, we show the difference between the observed weighted-average $\left<{\rm [Fe/H]}\right>$ and the expected values from the \citet{Kirby2013} relation for the selected sub-samples. The on-plane satellites show a median offset of $-0.06$~dex, while the off-planes of 0.12~dex (0.06~dex for the uncertain sample). The $\sim0.2$~dex difference between the on- and off-plane samples is mainly driven by the faint satellites with $L_V<10^3\,L_\odot$, which tend to no longer follow the observed luminosity-metallicity relation, but rather a flat trend with $L_V$ characterised by significant scatter \citep[also][]{Simon2018}. If we limit the analysis to $L_V>10^3\,L_\odot$, we find that the on- and off-plane samples show similar offsets of $-0.06$~dex and $-0.02$~dex, respectively, which are well within the rms scatter of the \citeauthor{Kirby2013} relation. 
We note that the probable LMC satellites tend to be more metal-poor than the \citeauthor{Kirby2013} relation, with a median offset of $\sim -0.1$~dex \citep[excluding Carina~III, which is at the low-luminosity end, due to its uncertain metallicity value obtained from an isochrone fitting,][]{Torrealba2018}. It is difficult, however, to establish the significance of such result, considering that the metallicity values come from heterogeneous sources \citep[see references within][]{Battaglia2022}. 
Therefore, the faint end of the luminosity-metallicity relation deserves further investigation, as it is still unclear what are the causes of the large scatter observed, whether it is mainly due to observational errors or an actual physical cause, such as tidal and ram-pressure stripping \citep[see e.g. discussion in][]{Simon2018}.

On a broader perspective, our results constitutes a tight constraint on the hypothesis that the on-plane systems might be TDGs, while the off-plane systems are not. 
In fact, for them to follow the observed luminosity-metallicity relation, they should have formed from low-metallicity material ejected during an early merger event ($z\gtrsim1$) between the MW and another galaxy (\citealp*[see][]{Recchi2015}, but also \citealp{Ploeckinger2018} and \citealp{Dumont+Martel2021}), probably Andromeda \citep[e.g.][]{Hammer2010,Hammer2013,Yang2014}. The TDGs thus formed should then have evolved for several Gyrs as regular dwarf galaxies, maintaining similar size and dynamical mass to the off-plane systems for several orders of magnitude (although they are expected to follow a different mass-size relation, \citealp[e.g.][]{Haslbauer2019}). We recall, however, that TDGs are more likely to be significantly affected by tidal and ram-pressure stripping. Therefore, if they were long-lived satellites of the MW, it would be difficult to understand how they could have survived for so long and exhibit properties similar to those of the off-plane systems. 

Our conclusions are in agreement with those reached on a similar basis by \citet{Collins2015}, who argued against a tidal origin for the dwarf galaxies on the Great Plane of Andromeda. 
We finally note that the scenario in which VPOS systems are TDGs has other alternatives that abandon the dark matter paradigm in favour of models based on modified Newtonian dynamics (MOND; \citealp[see][]{Banik2022} for an extensive introduction on TDGs in a MOND scenario and an application to the MW-M31 case), although there are limitations here too \citep[see][for a review]{Pawlowski2018}.

\subsection{A link between the VPOS and the MW's merger history?}
\label{subsec:comp_orb_props_merge_hist}

\begin{figure*}
    \centering
    \includegraphics[width=\textwidth]{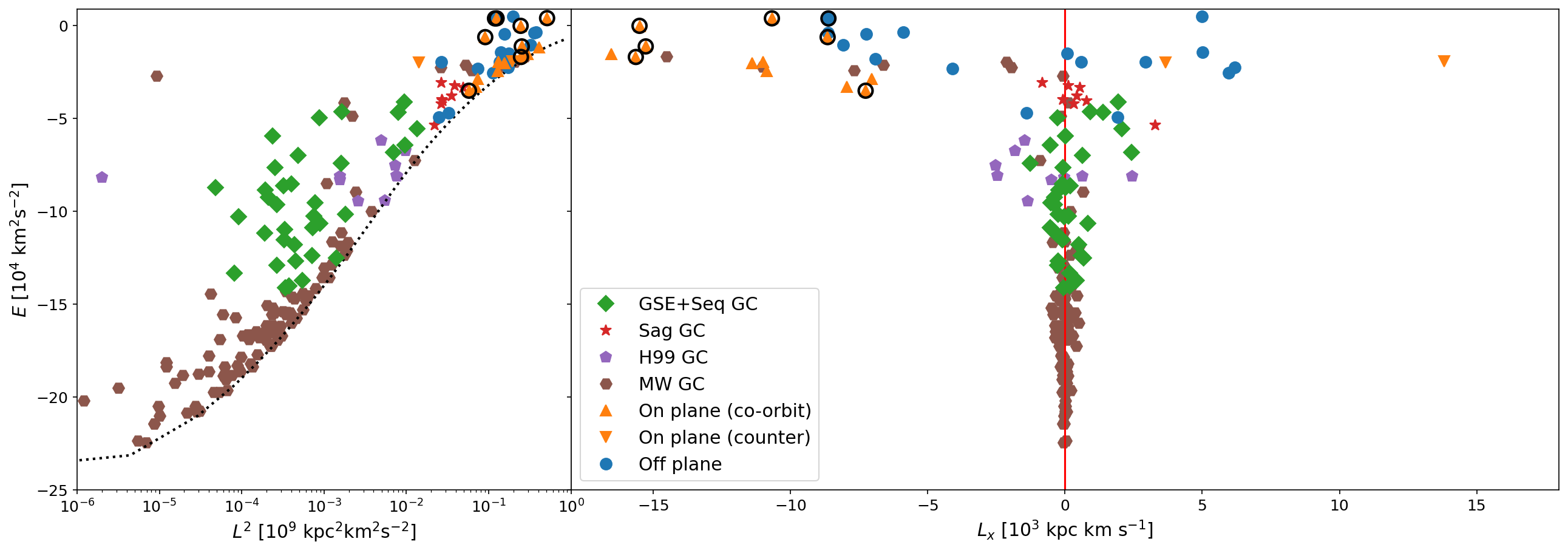}
    \caption{As in Fig.~\ref{fig:VPOS_E_L}, but here including the MW's globular clusters (GCs), which probably formed in-situ (brown hexagons) or were accreted, associated with GSE/Sequoia (green diamonds), Sagittarius (red stars) or Helmi (purple pentagons) streams, according to the compilation of \citet{Massari2019}.}
    \label{fig:VPOS_GSE}
\end{figure*}  

We introduced the possibility that the VPOS could be the result of a merger event in which a galaxy brings its own satellite population when it merges with the MW. \cite{Smith2016} showed that, under certain conditions, the acquired satellite population can reorganise itself into an extended planar structure with coherent motion around the merger remnant.
This raises the question of whether there are specific merger events in the MW's past that could be responsible for the emergence of the VPOS. One possible association would be with the Gaia-Sausage-Enceladus (GSE), the most recent significant merger identified to date~\citep{Belokurov2018,Haywood2018,Helmi2018N}. 

The GSE progenitor is thought to have been accreted about $8-11$~Gyr ago~\citep{DiMatteo2019,Naidu2021,Montalban2021}. 
Before its infall, the GSE progenitor had a total mass of about $\sim10^{11} M_\odot$~\citep{Helmi2018N,Mackereth2019,Kruijssen2020,Lane2023}, so it is reasonable to assume that it hosted a population of bound satellites, like the present-day LMC. 
The exact number and parameters of these satellites are difficult to determine from the available data. However, if we assume that $9$ VPOS satellites (i.e. excluding the LMC group) were associated with the GSE, the number of ``genuine'' MW satellites would be about $34$ (where some, but fewer than the GSE's, could have been accreted earlier in other massive accretion events). 
This gives a GSE-MW merger mass ratio of $\approx 1:4$, which is remarkably consistent with the constraints on the GSE merger parameters. In terms of chemical abundance, the average metallicity of the VPOS satellites is lower than the maximum metallicity of the GSE merger debris~($\approx-0.5$~dex), so their formation around the GSE progenitor is not precluded.

However, it is not easy to determine whether the VPOS systems were associated with GSE on the basis of their orbital properties. 
For instance, by analysing stellar debris from GSE, \citet{Chandra2023} found evidence for an associated stream that extends into the MW halo and moves in the plane defined by the Sagittarius stream, but in the opposite direction. We note that the Sagittarius stream defines a plane perpendicular to the VPOS \citep[e.g.][]{Pawlowski+Kroupa2020}, while \citet{Hammer2021} noted the presence of a possible planar structure associated with the Sagittarius dwarf, in which its components are counter-rotating with respect to it.
On the other hand, the reconstruction of the infall orbit of the GSE progenitor is hindered by various dynamical effects \citep{Vasiliev2022,Dillamore2022} that together with the growth of the MW potential \citep{Panithanpaisal2021,Khoperskov2023} affect stellar debris differently from the globular clusters and satellite galaxies associated with them \citep{Pagnini2023}.

Nevertheless, we investigated whether the VPOS systems share orbital properties with the globular clusters (GCs) likely to be associated with GSE or other MW streams.
We used the compilation of GCs from \citet{Massari2019} and in particular we refer to their classification to distinguish between in-situ and accreted GCs (probably associated with GSE/Sequoia, Sagittarius and Helmi streams, but see \citealp{Pagnini2023} for possible caveats in the classification).
Examining their position on the $E-L$ diagrams shown in Fig.~\ref{fig:VPOS_GSE}, we find no particular orbital features shared between the VPOS systems and any of the considered GC populations.
This result should not be surprising, since accreted satellite galaxies are less gravitationally bound to their host (GSE-progenitor in the considered case) than its GC and stars, so they would have undergone stripping earlier and would not necessarily end up sharing the orbital properties of the host galaxy's stars and GCs. It is also true that by remaining mostly outside the tidal influence of the MW disk, they would be more likely to retain the infall orbital parameters of their progenitor, but whether this applies to the GSE case remains to be demonstrated. Indeed, we would rather expect the orbits of former GSE satellites to be already phase-mixed if their accretion occurred $8-11$~Gyr ago.

In summary, it is unclear whether the satellites currently comprising the VPOS were initially associated with the GSE progenitor and were stripped away during its infall into the MW. Further efforts are needed to verify the probability of such processes, especially in the context of the orbital phase coherence of VPOS satellites. 
The analysis of a statistical sample of satellite systems from the IllustrisTNG simulation, however, seems already to indicate that major mergers (i.e. with a mass ratio $>1:3$) play a negligible role in the formation of correlated planar structures such as the VPOS \citep*{Kanehisa2023}.

\subsection{Are the on-plane systems part of a group infall?}
\label{subsec:comp_orb_props_infall}

\begin{figure}
    \centering
    \includegraphics[width=.49\textwidth]{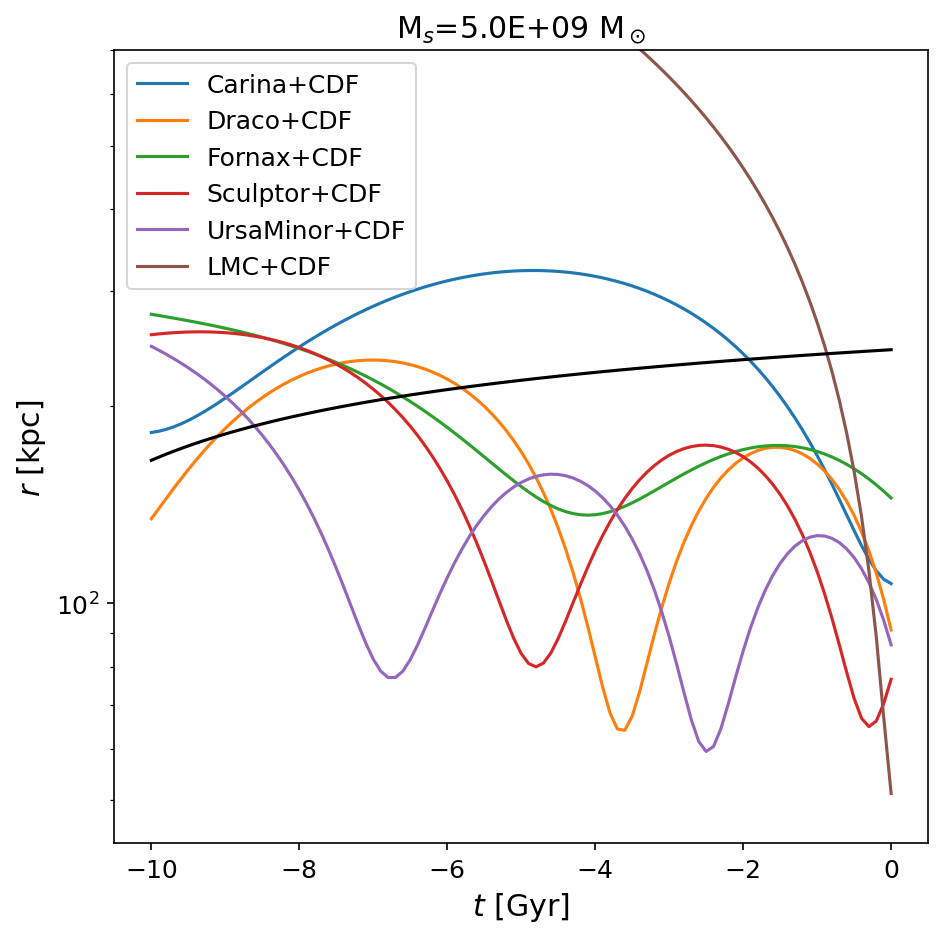}\\
    \includegraphics[width=.49\textwidth]{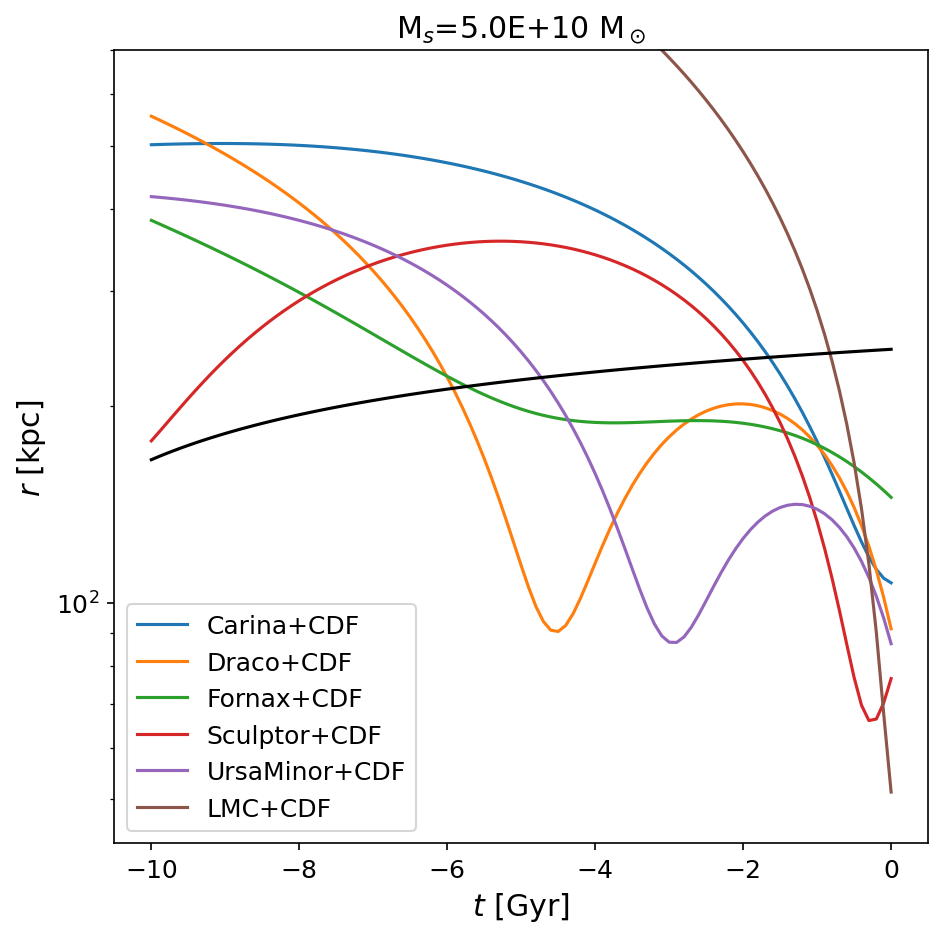}
    \caption{Orbital evolution of the bright dwarf of the VPOS integrating backwards in time for 10~Gyr assuming an evolving MW potential and taking into account the effects of dynamical friction, assuming that the satellites have all a total mass of $M=5\times10^{10}\,M_\odot$. The black solid line represent the time evolution of the virial radius $r_{\rm vir}$.}
    \label{fig:VPOS_orb_dyn_frict+evolvMW}
\end{figure}

Group infall is a promising way to explain the origin of a coherent planar structure such as the VPOS \citep[e.g.][]{Lynden-Bell1995,Li+Helmi2008,dOnghia+Lake2008,Klimentowski2010}. The evidence that the LMC arrived with its own group of satellites \citep[e.g.][]{Patel2020} also seems to support this hypothesis \citep[see also][]{Samuel2021}. However, it is difficult to determine from orbital integration alone whether the LMC system was part of a single group comprising the other VPOS satellites, or whether they were part of separate groups. For instance, only Carina and Fornax have been suggested to be dynamically associated with the LMC \citep[e.g.][]{Pardy2020}, although this association does not seem to be conclusive \citep[e.g.][]{Patel2020,Battaglia2022}. 

We further note that the accretion of groups of dwarf galaxies is a mechanism already present in cosmological simulations, but the predicted fraction of satellites arriving in groups and the number of dwarf galaxies forming such groups seems insufficient at the moment to explain the formation of a structure such as the VPOS (\citealp[see][for a review]{Pawlowski2021}, but also \citealp{Li+Helmi2008,Wetzel2015,Shao2018}). 
It is true that much remains to be understood about the dependence of the environment in which dwarf galaxy groups form. This is important not only to determine the brightness and compactness of such groups, but also to establish whether they could possibly form a coherent and spatially thin structure such as the VPOS once being accreted. In this context the accretion of (groups of) satellites along cosmic filaments is particularly relevant, a mechanism that provides both a dense environment in which groups can form and a preferential direction for their infall onto a host galaxy (\citealp[e.g.][]{Libeskind2011,Libeskind2015}, but see also \citealp[]{Pawlowski2012a}).

The fact that the on-plane systems (excluding the LMC and its satellites) are approaching or moving away from their pericentre showing a strong pattern in their orbital phases can be interpreted as a sign that they have recently arrived in the MW potential together. 
However, in order to reconcile this scenario with their $E-L$ properties shown in Sect.~\ref{subsec:comp_orb_props-E_L}, a mechanism is needed to cause them to lose orbital energy.
\citet{Hammer2021} proposed ram-pressure stripping as the main energy loss mechanism, capable of reducing radial velocity and circularising orbits. This process, however, is only effective for dark-matter-free galaxies, such as TDGs. We here explore a different approach which instead relies on the dwarfs being heavily dark matter dominated by instead considering a toy model of dynamical friction. Our choice is justified by the fact that the VPOS is dominated by bright satellites on which the effect of dynamical friction should be maximal. 

We applied the Chandrasekhar prescription for the induced acceleration by dynamical friction \citep{Chandrasekhar1943}. This formula has provided a good approximation for reproducing the orbital evolution of low-mass objects such as satellite dwarf galaxies (\citealp{Binney+Tremaine2008}, but also \citealp{Tremaine1976,Hashimoto2003,VanDerMarel2012,Patel2020}). One of its terms is the Coulomb logarithm ${\rm ln}(\Lambda)={\rm ln}(b_{\rm max}/b_{\rm min})$, where $b_{\rm max}$ and $b_{\rm min}$ are the maximum and minimum impact factors in an encounter. These terms usually indicate the distance to the centre of the host galaxy, and a deflection radius equal to the half-mass size of the satellite (hence $b_{\rm max} \gg b_{\rm min}$). However, ${\rm ln}(\Lambda)$ can be quite uncertain due to the difficulties in accurately measuring its terms, and while a constant value has usually been assumed, a term varying with distance from the centre of the host galaxy offers better results \citep{Just+Penarrubia2005,Petts2016}.

We used the \texttt{galpy} code \citep{Bovy2015} to integrate the orbits of the luminous dwarfs on the VPOS: Carina, Draco, Fornax, Sculptor, Ursa Minor (i.e., part of the classic MW dwarf spheroidals), and the LMC for comparison. We assumed the \texttt{MWPotential2014} ($M_{\rm vir}=0.8\times10^{12}M_\odot$ and $r_{\rm vir}=245$~kpc, similar to the low-mass potential) and integrated backwards for 10~Gyr. 
As total masses for the satellites we assumed four values of $M_S=(5\times10^{9};10^{10};5\times10^{10};10^{11})\,M_\odot$, compatible with abundance matching expectations for such luminous systems \citep[e.g.,][]{Bullock+Boylan-Kolchin2017}. 
We have kept the total mass of the satellites constant during the integration process, since they are expected to have already assembled more than half of their mass at $z\sim2$ \citep[e.g.][]{Sawala2016,Libeskind2020,Deason2022}.
We also neglect tidal stripping of their dark-matter halos in order to maximise the effects of dynamical friction in our toy model.
For the dynamical friction term\footnote{See \url{https://docs.galpy.org/en/v1.8.3/reference/potentialchandrasekhardynfric.html}.}, we assumed a varying ${\rm ln}(\Lambda)$, with the satellite's half-mass radius set at 5 kpc \citep[e.g.][]{Iorio2019}. 
We note that we would reach qualitatively similar conclusions also assuming a constant ${\rm ln}(\Lambda)=3$ \citep[e.g.][]{Vasiliev2021}.

We found that the effects of dynamical friction begin to significantly affect orbit integration only from $M_S\geq5\times10^{10}\,M_\odot$ (see Fig.~\ref{fig:VPOS_orb_dyn_frict}). This is a rather high mass for classic dwarf spheroidals, which are instead estimated to be currently around $M_S\sim5\times10^{9}\,M_\odot$. In this case we found the effects of dynamical friction to be negligible, in agreement with previous studies investigating the effects of tides on Fornax and Sculptor \citep{Battaglia2015,Iorio2019}.

The previous exercise did not take into account the evolution of the MW potential, which we included as an additional element. We followed the prescription of \citet[][in particular their Eq.~2]{Miyoshi+Chiba2020}, but using the halo component of the \texttt{MWPotential2014} and considering for the time evolution of its concentration parameter $c(M_{\rm vir},z)$ the relation provided by \citet{Dutton14}.
We integrated the orbits in steps of 10~Myr, assuming two values for the total mass of the individual satellites: $M_S=5\times10^{9}\,M_\odot$ and $M_S=5\times10^{10}\,M_\odot$ (chosen as a representative total mass for a dwarf spheroidal and as the minimum mass at which we saw the orbits start to be significantly affected).

As shown in Fig.~\ref{fig:VPOS_orb_dyn_frict+evolvMW}, the inclusion of an evolving potential does indeed have a significant impact on orbital recovery. At the assumed high-mass limit, three satellites cross the MW virial radius as early as 6~Gyr ago, while the other three at much recent times ($<2$~Gyr ago), none making more than a single pericentric passage. For the more plausible low-mass limit, the majority of satellites cross the virial radius around 8~Gyr ago, but in this case the satellites make as many as two pericentric passages. This is in contrast to the previous case with a static MW potential, where satellites at this mass stayed on average within the MW virial radius during the integration period (see Figs.~\ref{fig:VPOS_orb_dyn_frict_sats_1} and~\ref{fig:VPOS_orb_dyn_frict_sats_2}).

Although our approach of estimating infall times by orbit integration is simplistic, due to the inherent difficulty of knowing the MW potential and its evolution over time accurately \citep[e.g.][]{DSouza+Bell2022,Santistevan2023arXiv}, it is nevertheless helpful to draw some qualitative conclusions. The first is that in order for the classic dwarf spheroidals to have a late arrival into the MW potential, we would have to assume too high masses for them. Even allowing for an evolving MW potential, at the low-mass limit they still tend to make some pericentric passages over several Gyr. Therefore, if we consider them as independent systems, it is difficult to see how they can now exhibit an orbital coherence such as that described in Sect.~\ref{sec:comp_orb_props}, considering that orbital phases tend to mix in a few Gyr.
It may also be that such systems were significantly more massive at the time of first infall and were severely tidally stripped by the MW, although their current stellar populations do not appear to show significant signs of this \citep[e.g.][]{Battaglia2022}.

On the other hand, a natural hypothesis would be to consider the satellites in question having been part of a group of dwarf galaxies. We have already seen that a late arrival in the MW potential requires masses (i.e. $M_S\ge5\times10^{10}\,M_\odot$) that are too high for individual systems (excluding the LMC), but compatible with that of a group comprising the classic dwarf spheroidals considered here. We verified that in this mass range it is possible that they shared a similar spatial position when they were crossing the MW's virial radius, $\sim6$~Gyr ago, but mainly because the uncertainties became too large at this point in time (see Fig.~\ref{fig:VPOS_orb_dyn_frict_YZ}). Therefore, to test the hypothesis of group infall it would be necessary to move forward from simple orbital integration and set up a dedicated study involving cosmological simulations, which we reserve for future work.

If group accretion were the case, further aspects to consider would be the formation environment of this group of dwarf galaxies \citep[e.g.][]{Libeskind2011} and the subsequent tidal stripping by the MW after an initial pericentric passage that would have contributed to its disruption \citep[e.g.][]{Li+Helmi2008}. More recently, \citet{Vasiliev2023a}, exploring the implications that the LMC may be on its second pericentric passage around the MW, have speculated on the origin of the VPOS and the possibility that the present on-plane systems may have formed together with the LMC a primordial group from which the luminous satellites were stripped off in a first passage occurred some 6~Gyr ago. As noted by the author, the scenario proposed by \citet*{Pawlowski2011} and reported above would still be valid in this context. Instead of accreted TDGs, we would have a group of dwarf galaxies tidally stripped by the MW, eventually forming the VPOS. 

The possibility that the VPOS is the result of a single group accretion is intriguing, but as already noted by \citet{Vasiliev2023a} is a complex matter that deserve further investigation, in particular to establish whether or not the LMC and its satellites belonged to such a group.
The accretion of groups of dwarf galaxies on MW-like hosts has been poorly studied in the literature \citep[see][]{Pawlowski2021}, in particular when related to the formation of plane of satellites. Our results place emphasis on this particular mechanism that deserves further exploration.

\section{Conclusions}
\label{sec:conclusions}

We compared the observed properties of a sample of 50 satellite galaxies of the MW, looking for differences between systems on and off the VPOS. In particular, we compared their physical and orbital properties.

The comparison of the physical properties using different parameters (i.e. $L_V$, $R_e$, $\left<{\rm [Fe/H]}\right>$, $M_{1/2}$) showed that any differences found between the on-plane and off-plane samples are mainly driven by the presence of the brightest MW satellites in the former sample (i.e. with $L_V \gtrsim 10^5 L_\odot$). The statistical significance of such a luminosity gap between the on-plane and off-plane sample ranged from moderate to strong, depending on whether the bright systems with more uncertain proper motions in Gaia are also included via other proper motion measurements. 

The analysis of the orbital properties of our sample of galaxies depended on the MW potential assumed. Therefore, we first inspected the velocity vector of our systems, independent of these assumptions, focusing on the radial and tangential components. We found a striking result: the majority of the on-plane co-orbiting systems are currently approaching the MW ($V_{\rm rad} < 0$~km\,s$^{-1}$), while both the on-plane counter-orbiting systems are moving away from it. The radial velocity distribution of the off-plane (as well as the uncertain) systems instead distribute evenly around $V_{\rm rad} = 0$~km\,s$^{-1}$. 
Regarding the tangential velocity properties of our systems, we confirm the results reported in the literature about an excess of galaxies with orbital kinetic energy dominated by tangential motion, with no particular differences between on-plane and off-plane satellites.

Inspection of the orbital parameters, on the other hand, revealed no significant differences between on-plane and off-plane systems (i.e. taking into account their pericentre, apocentre, orbital eccentricity, time since the last pericentric passage), both in terms of distributions and correlations between parameters. However, when we examined their orbital phases (i.e. the ratio between the time since the last pericentric passage and the orbital period), we again found another striking feature: almost all on-plane systems that are co-orbiting (counter-orbiting) are moving towards (away from) their pericenters within errors. In particular, co-orbiting systems seem to do so after leaving their apocentre at the same time (about 1~Gyr ago). This is a high degree of phase coherence that is not expected. In fact, the satellites have quite different orbital periods, which should lead to rather rapid phase mixing that would cancel out such coherence. This could indicate that the feature, and hence the VPOS, is young. We then hypothesised that the on-plane satellites may have recently arrived into the MW potential. We also showed that on-plane systems tend to occupy the lowest orbital energy values for a given angular momentum, suggesting instead that the VPOS may be a rather long-lived structure.

We have also examined the role of a massive LMC and its probable former satellites in relation to the VPOS. Since they were accreted by the MW as a group, they do not form an independent sample of dwarfs and their presence on the VPOS could bias our results. On one hand, we showed that by removing the LMC satellites we still recover the observed luminosity gap. On the other hand, accounting for the passage of a massive LMC, we confirm our previous findings that co-orbiting on-plane systems are approaching pericentre, although it is uncertain whether they do so in a coordinated fashion.

We briefly discussed the possible implications of our results in light of the longevity of the VPOS. If the VPOS were a long-lived structure, the results of the orbital analysis could be interpreted as a mere coincidence. Instead, interpreting them as a significant property of the VPOS from which clues about its origin can be deduced, allowed us to explore three possible formation mechanisms for such a structure, all related to the late and/or joint accretion of satellite systems. In particular, we discussed whether the VPOS consists of tidal dwarf galaxies, whether it is the result of a major merger event or of the accretion of a bound group of dwarf galaxies.

Regarding the first hypothesis, and similar to what was found for the plane of Andromeda by \citet{Collins2015}, we did not observe a significant preference for on-plane and off-plane systems to follow different scaling relations, in particular by examining their values on the luminosity-metallicity relation, as would be expected if they were tidal dwarf galaxies. These results place a constraint on this scenario, making it less plausible.

We also examined the hypothesis that VPOS systems are a population of satellites acquired after a major merger event experienced by the MW, in particular whether they could be associated with GSE. In principle, the GSE progenitor could have been massive enough to host a population of satellites coinciding in number with the VPOS systems (excluding the LMC group). However, the orbital properties of the stellar remnants of GSE and those of the globular clusters probably associated with it make the possibility of an association with the VPOS unclear, for which further tests are needed.

Regarding the possibility of a group infall behind the origin of VPOS, we tested an indirect approach based on an orbital integration toy-model. Since the VPOS is dominated by bright satellites, we explored the role of dynamical friction as an energy-loss mechanism to understand whether a late arrival of the on-plane systems is possible despite their comparably low orbital energies. 
We showed that it would require rather high total masses for the considered bright satellites to cross the MW virial radius no earlier than 6~Gyr ago, experiencing then as much as a single pericentric passage. Such a condition is unrealistic if we consider them as independent systems, but it could be plausible if they were accreted as a group.

In summary, to explain the results obtained we put forward the hypothesis that the on-plane systems were part of a group of galaxies before their arrival in the MW potential. This could help explain the observed luminosity gap, assuming that many of the on-plane systems were part of a single group that brought in the bright MW satellites. Dynamical friction, in this regard, could be an effective energy-loss mechanism to allow a late arrival, which could explain why the co-orbiting on-plane systems are approaching pericentre. However, it is difficult to prove this scenario by orbital integration alone, and a dedicated simulation should be set up.
In particular, more work needs to be done on the formation environment of such a group and/or the possible link with the merger history of the MW (e.g.~an association with the GSE progenitor).

The accretion of groups of dwarf galaxies on MW-type hosts has been little studied in the literature, as has their role in the formation of phase-space correlations \citep{Pawlowski2021}. As in the case of the LMC, we expect that future improvements in the measurements of the proper motion of MW satellites, together with the study of their properties in the context of cosmological simulation, will help to shed further light on the formation of the VPOS.

\begin{acknowledgements}
We wish to thank the anonymous referee for constructive comments that improved the manuscript. We thank E.~Vasiliev and Y.~Revaz for the useful discussion and comments.
M.~S.~P.~acknowledges funding of a Leibniz-Junior Research Group (project number J94/2020) and thanks the German Scholars Organization and Klaus Tschira Stiftung for support via a KT Boost Fund.
\end{acknowledgements}

%
%

\bibliographystyle{bibtex/aa} 
\bibliography{bibtex/VPOS_props_arXiv.bib} 

\FloatBarrier
\begin{appendix}
    
\section{Supplementary material}

We provide here additional material to support results shown in the main text.
In Sect.~\ref{sec:comp_orb_props} we examined the issue of the excess of MW satellites found near their orbital pericentre. 

\begin{figure}
    \centering
    \includegraphics[width=.49\textwidth]{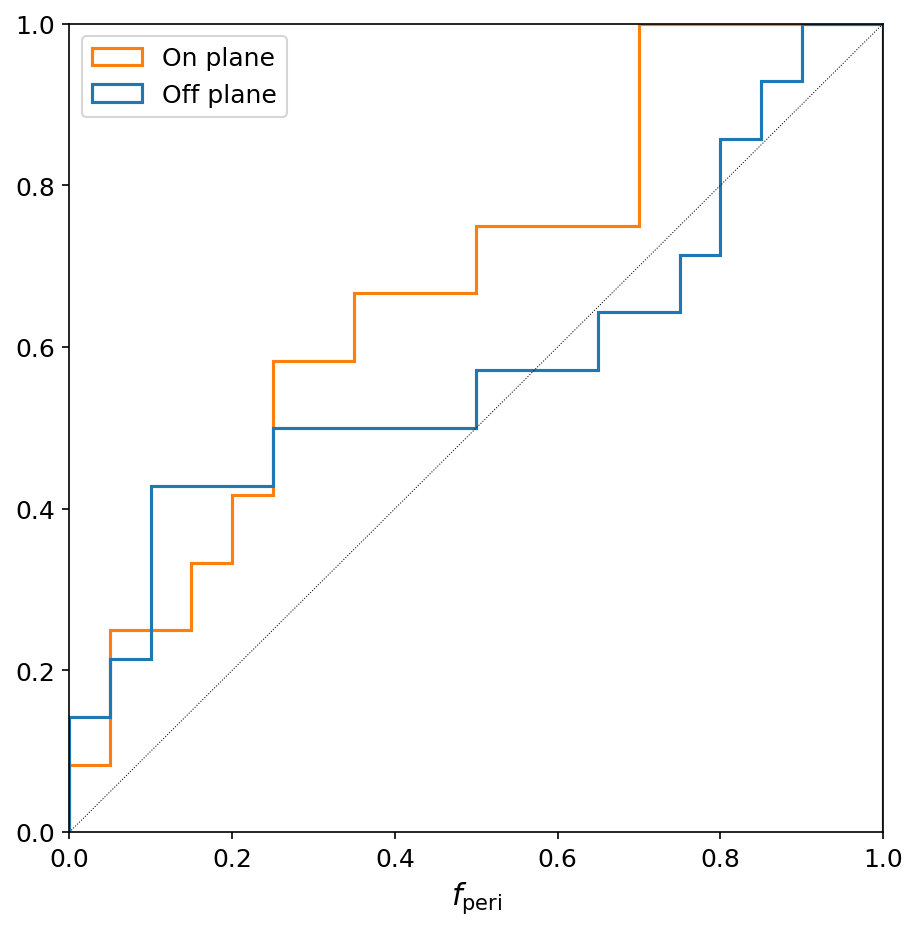}\\
    \includegraphics[width=.49\textwidth]{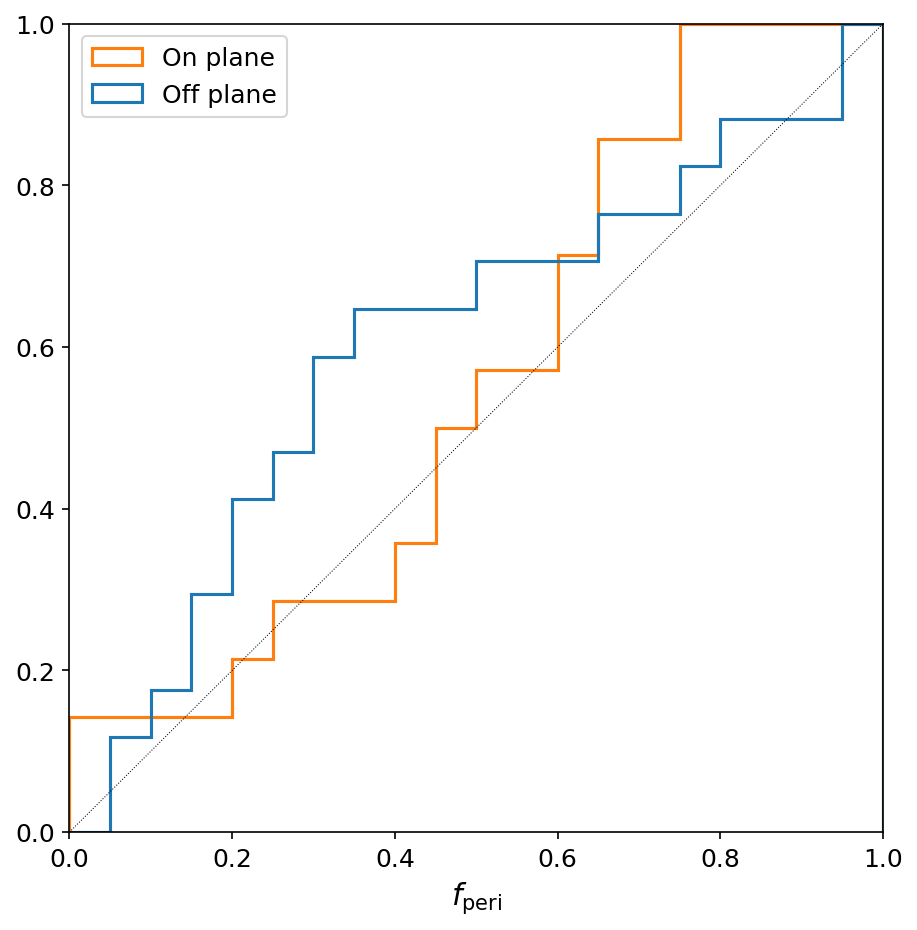}
    \caption{Cumulative distributions of $f_{\rm peri}$, the fractional distance in time between the pericentre ($f_{\rm peri}=0$) and the apocentre ($f_{\rm peri}=1$), for the on-plane (orange solid lines) and off-plane (blue solid lines) samples, considering values obtained with the low-mass (\textit{top} panel) and high-mass (\textit{bottom} panel) MW potentials. Dashed lines represent the expectation of a uniform distribution for each sub-sample.}
    \label{fig:VPOS_orb_fperi}
\end{figure}

In Fig.~\ref{fig:VPOS_orb_fperi} we show whether there are differences between the on- and off-plane samples for both potentials in their $f_{\rm peri}$, the fractional time distance between the pericentre ($f_{\rm peri}=0$) and the apocentre ($f_{\rm peri}=1$), defined as the ratio between the absolute value of $t_{\rm peri}$ (the time since/to pericentre) and half the orbital period $P$. Examining the cumulative distribution of $f_{\rm peri}$, we find no significant differences between the on-plane and off-plane samples, for both potentials. In fact, by performing a 2-sample KS-test, we find that they are likely to be drawn from the same parent distribution.

\begin{figure*}
    \centering
    \includegraphics[width=.95\textwidth]{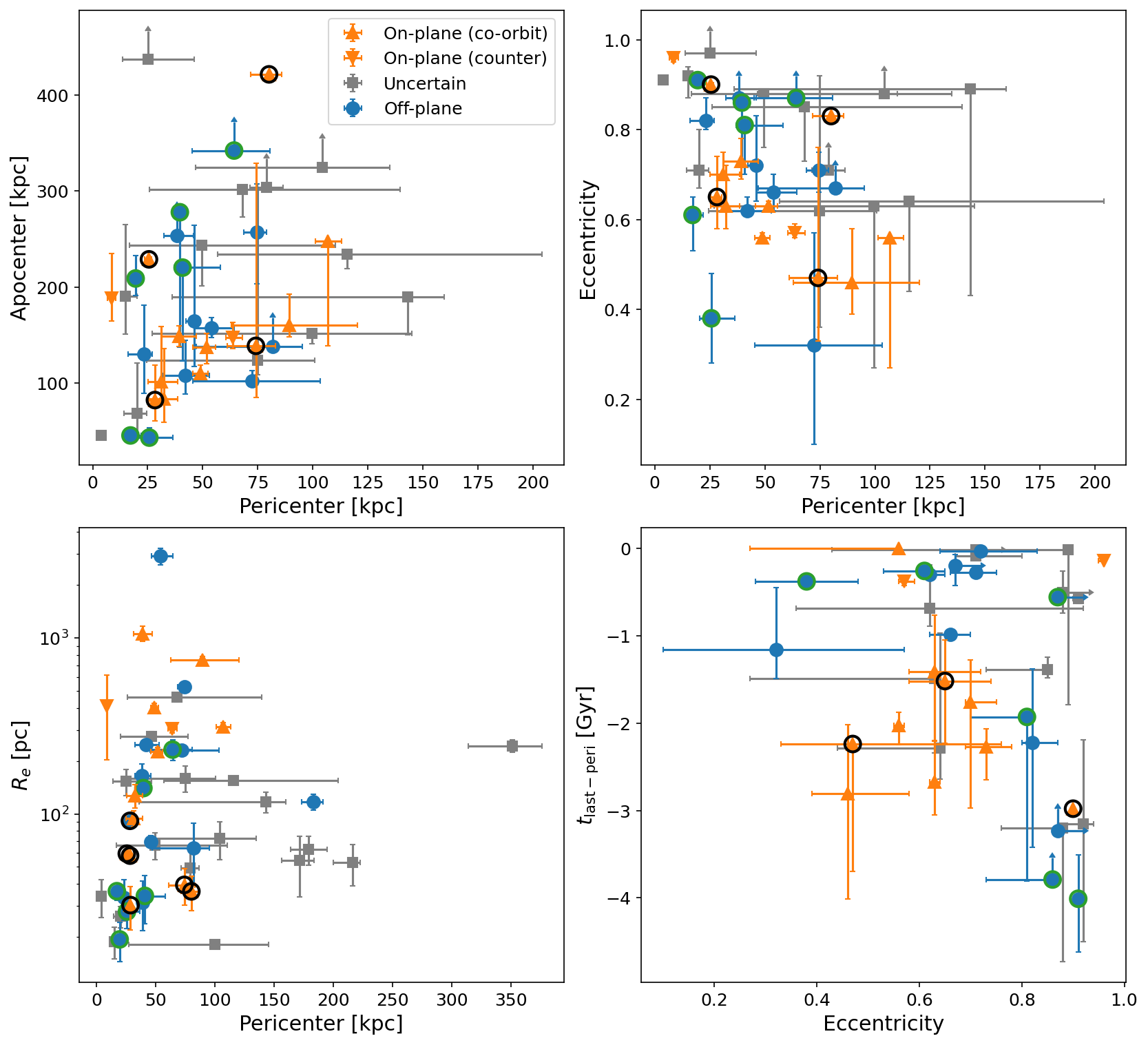}
    \caption{Comparison of the orbital properties for our sample of dwarf galaxies, assuming a low-mass potential for the MW. \textit{Top left:} pericentre versus apocentre (in kpc). \textit{Top right:} pericentre (in kpc) versus eccentricity. \textit{Bottom left:} pericentre (in kpc) versus the half-light radius $R_e$ (in pc). \textit{Bottom right:} eccentricity versus the orbital phase (i.e., the time since the last pericentric passage normalised by the orbital period). In all panels, symbols and colors are as described in Fig.~\ref{fig:VPOS_ph}. An arrow in the error bars indicates that a target lack a complete orbital solution and therefore the reported value is the calculated 16th quantile.}
    \label{fig:VPOS_orb_iso}
\end{figure*}

\begin{figure*}
    \centering
    \includegraphics[width=.95\textwidth]{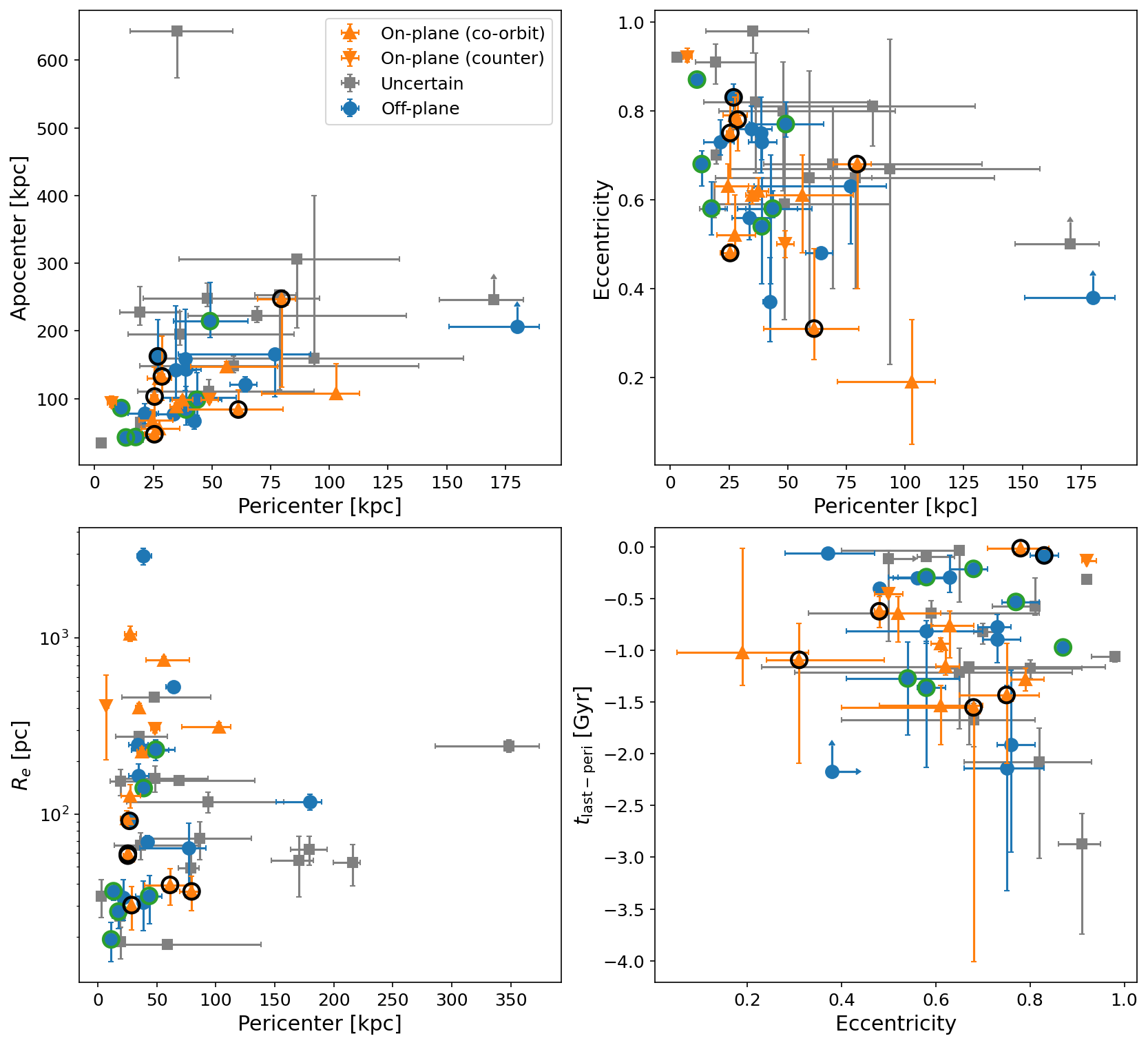}
    \caption{As in Fig.~\ref{fig:VPOS_orb_iso}, but with the orbital parameters derived assuming a high-mass potential for the MW.}
    \label{fig:VPOS_orb_heavy}
\end{figure*}

In Figs.~\ref{fig:VPOS_orb_iso} and \ref{fig:VPOS_orb_heavy}, we show instead the comparison between orbital parameters (i.e., pericentre, apocentre, eccentricity, time since the last pericentre) calculated using a low-mass and high-mass MW potential, respectively. We remind to Sect.~\ref{sec:comp_orb_props} for more details and Table~\ref{tab:params} for the results of the statistical tests performed on the considered parameter distributions.

\begin{figure*}
    \centering
    \includegraphics[width=.49\textwidth]{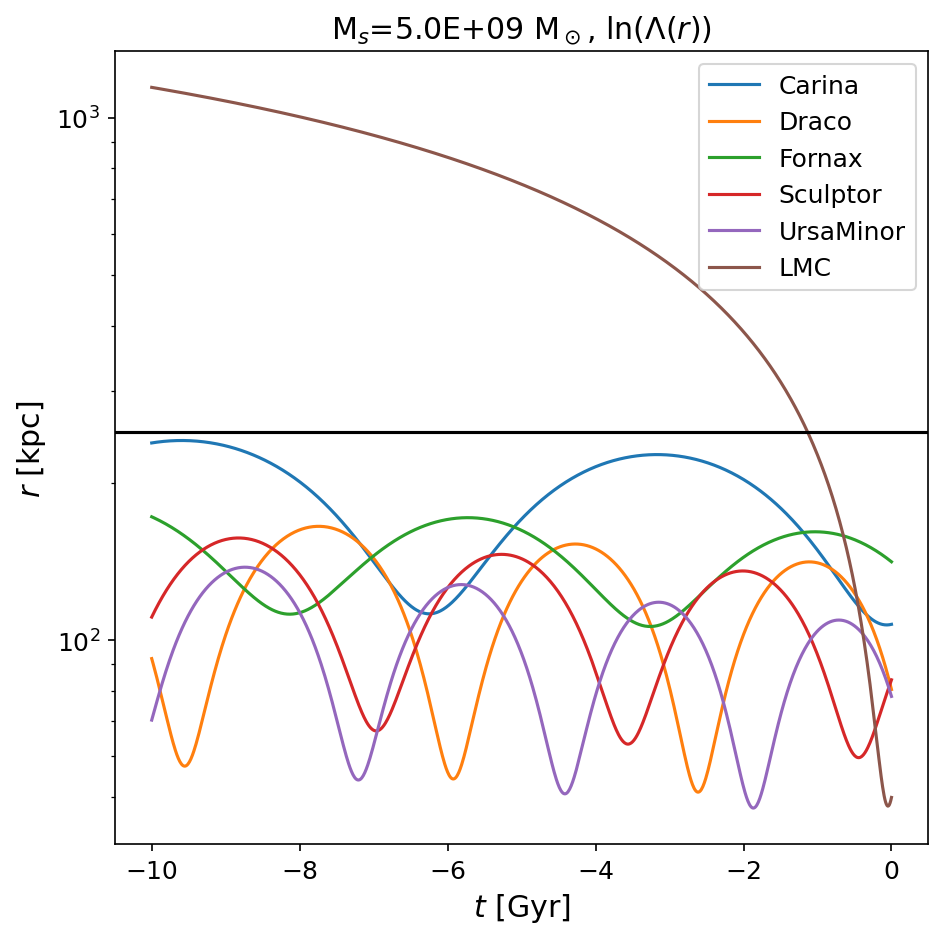}
    \includegraphics[width=.49\textwidth]{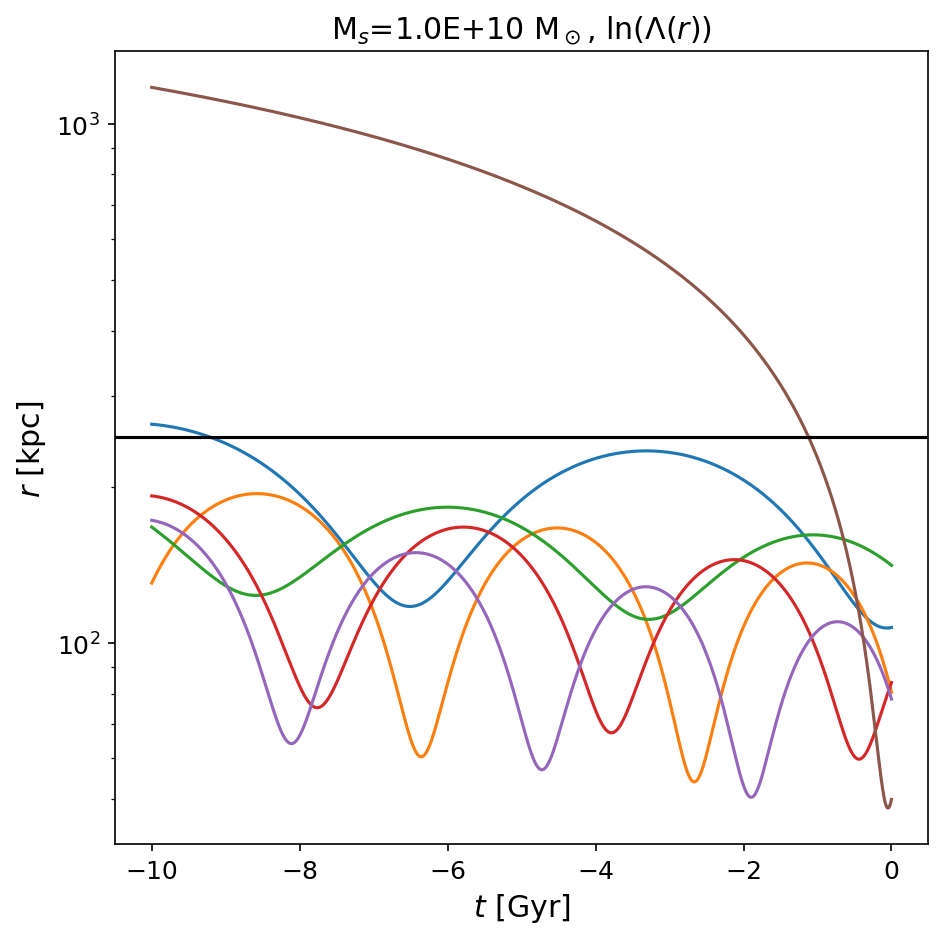}\\
    \includegraphics[width=.49\textwidth]{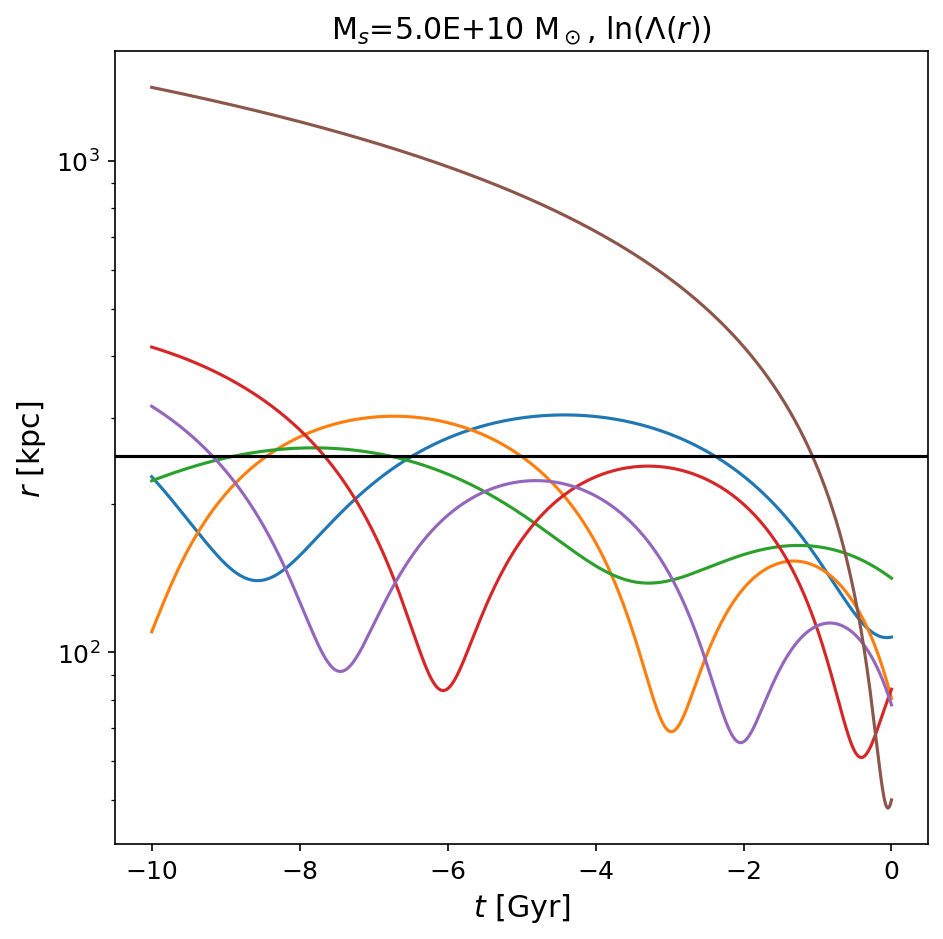}
    \includegraphics[width=.49\textwidth]{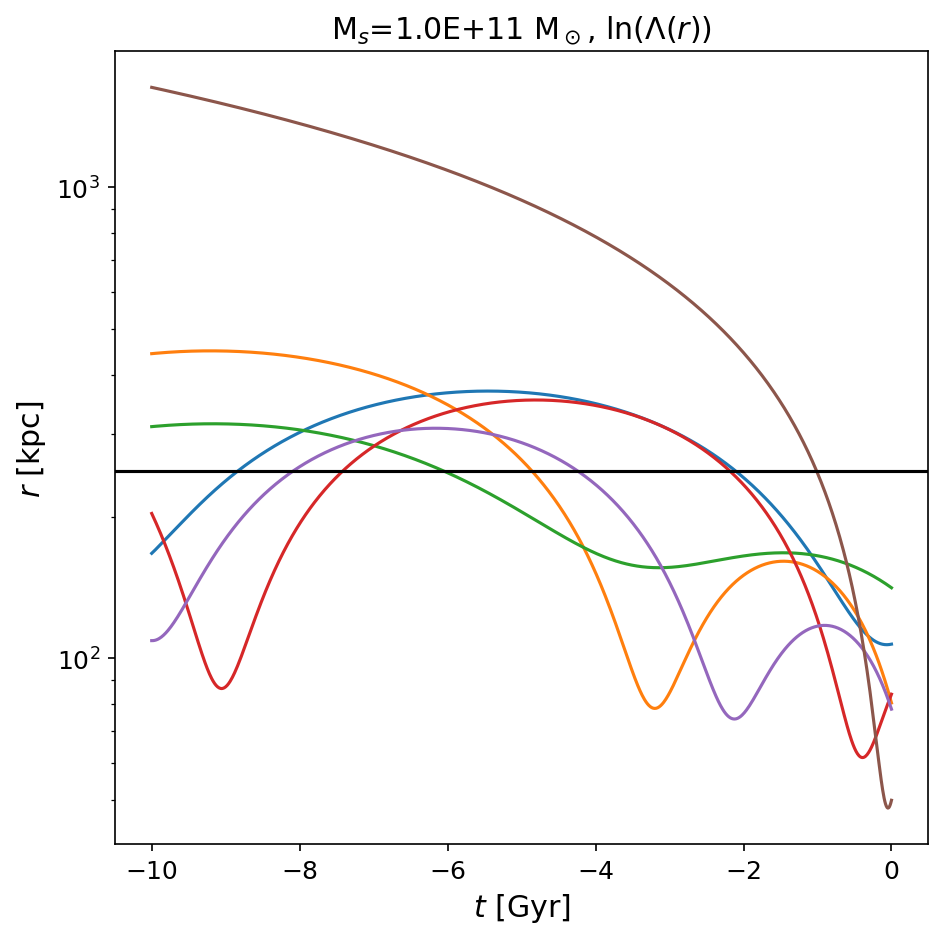}
    \caption{Orbital evolution of the bright dwarfs of the VPOS integrating backwards in time for 10~Gyr assuming the \texttt{MWPotential2014} \citep{Bovy2015} and taking into account the effects of dynamical friction, assuming that the satellites have all a total mass of $M_S=5\times10^{9}\,M_\odot$ (\textit{top left}), $M_S=10^{10}\,M_\odot$ (\textit{top right}), $M_S=5\times10^{10}\,M_\odot$ (\textit{bottom left}), and $M_S=10^{11}\,M_\odot$ (\textit{bottom right}); the black solid line in all panels represents the MW virial radius $r_{\rm vir}=245$~kpc.}
    \label{fig:VPOS_orb_dyn_frict}
\end{figure*}

In Sect.~\ref{subsec:comp_orb_props_infall}, we discussed the effect of dynamical friction on the orbital properties of the brightest satellites of the MW on the VPOS.
In Fig.~\ref{fig:VPOS_orb_dyn_frict}, we show their orbital evolution integrating backwards in time for 10~Gyr assuming the \texttt{MWPotential2014} \citep{Bovy2015} and taking into account the effects of dynamical friction (assuming a varying ${\rm ln}(\Lambda(r))$ and a scale radius for the dwarfs of 5~kpc, \citealp[see e.g.][]{Iorio2019}). For the satellites total mass we assumed four values of $M_S=(5\times10^{9};10^{10};5\times10^{10};10^{11})\,M_\odot$, compatible with abundance matching expectations for such luminous systems \citep[e.g.,][]{Bullock+Boylan-Kolchin2017}. We see that the effects of dynamical friction start to be significant at $M_S\geq5\times10^{10}\,M_\odot$. 

\begin{figure*}
    \centering
    \includegraphics[width=.49\textwidth]{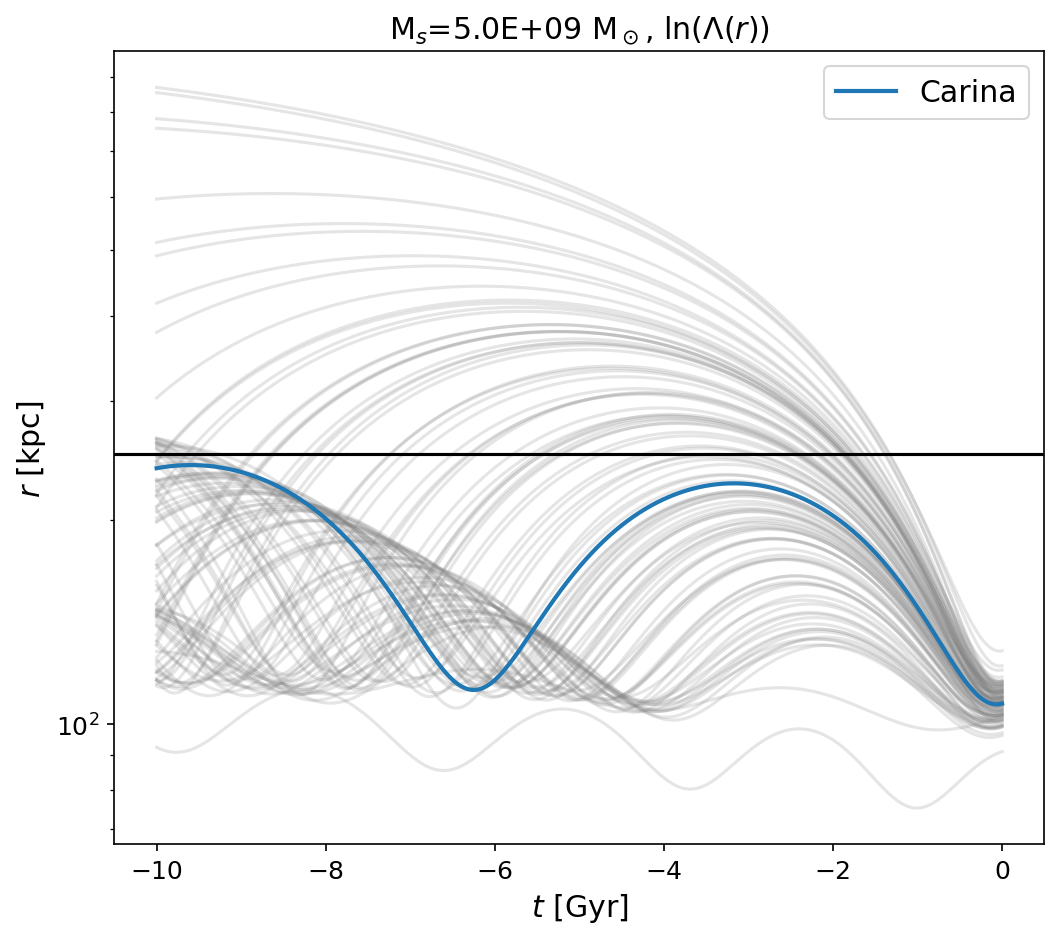}
    \includegraphics[width=.49\textwidth]{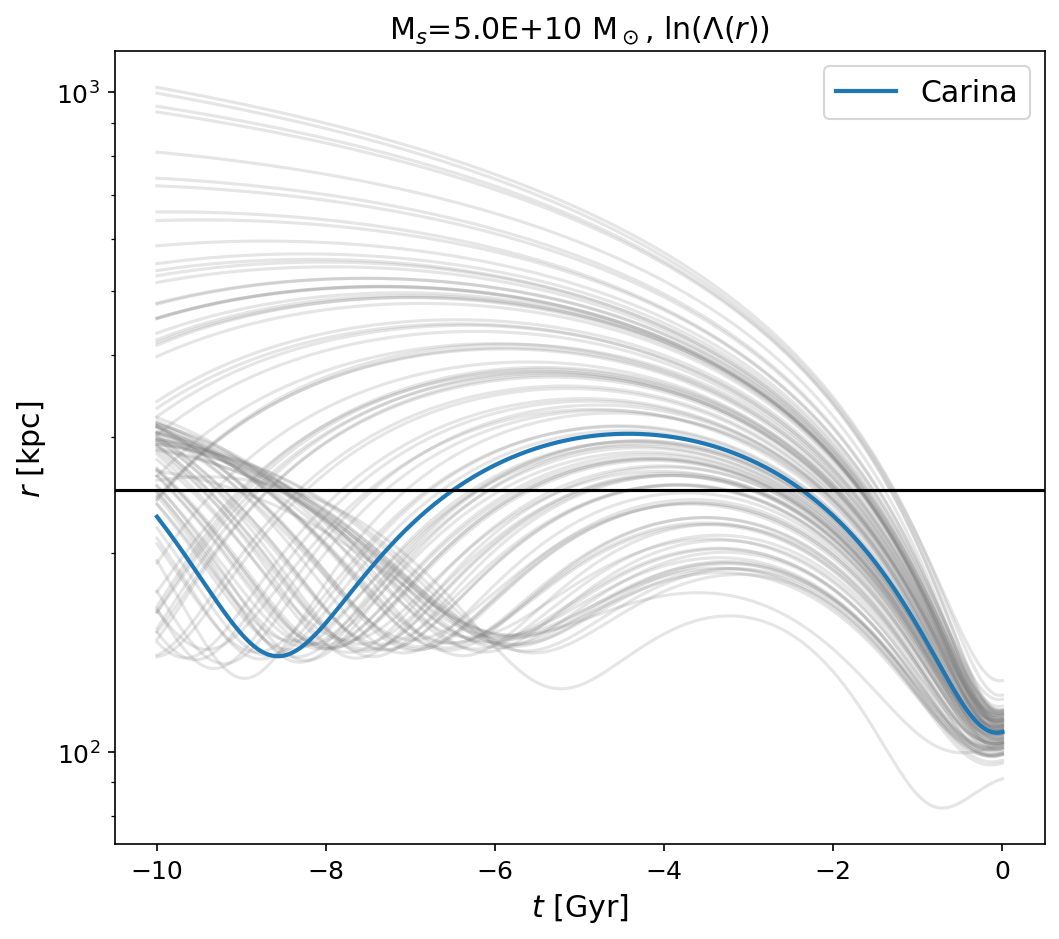}\\
    \includegraphics[width=.49\textwidth]{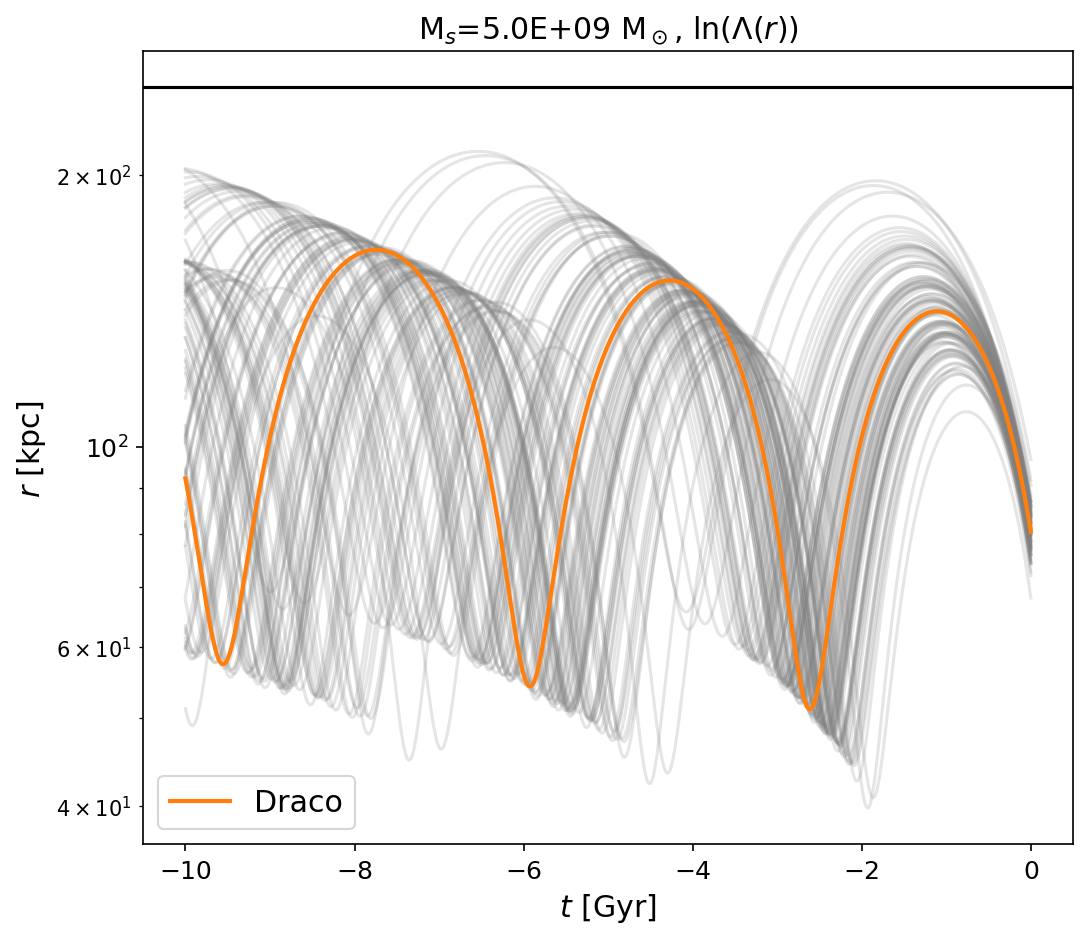}
    \includegraphics[width=.49\textwidth]{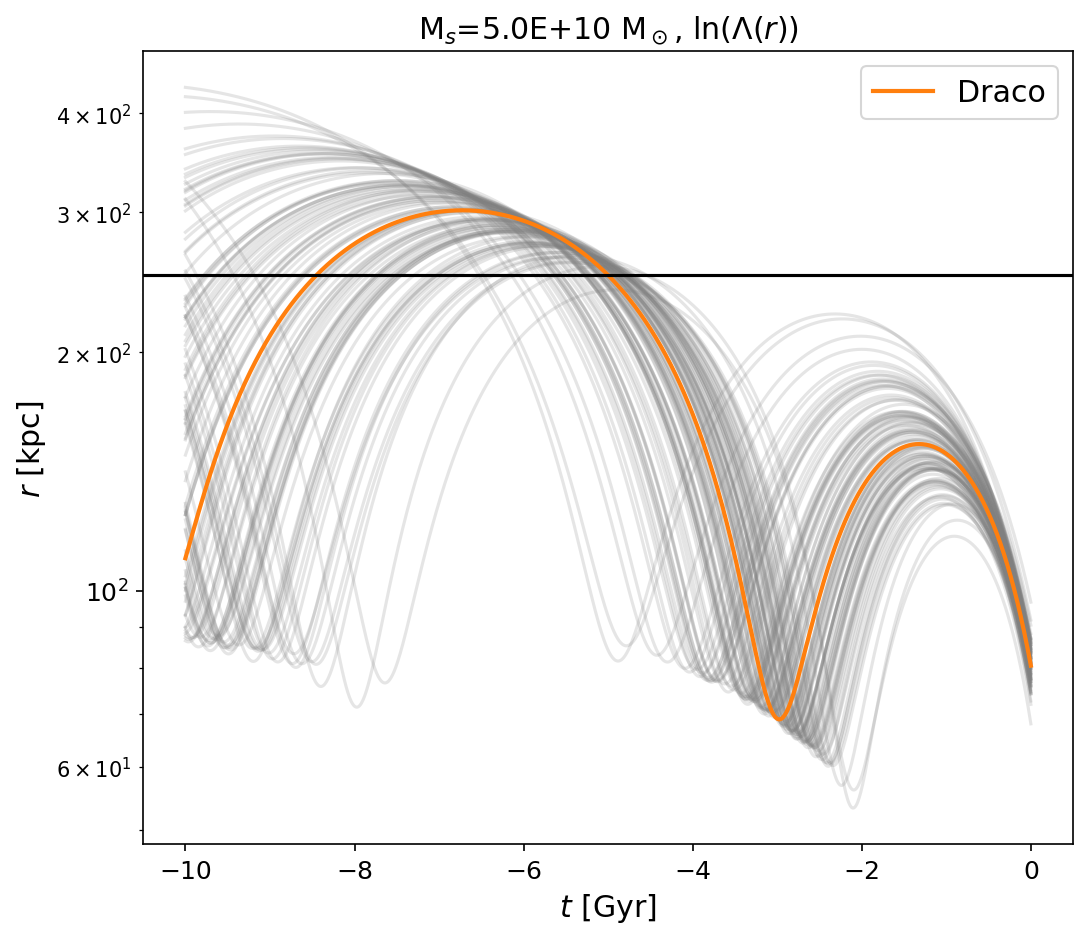}\\
    \caption{As in Fig.~\ref{fig:VPOS_orb_dyn_frict}, but here showing the impact of uncertainties on the orbital evolution of the individual systems. For each galaxy, we show as grey lines 100 realisations made by a Monte-Carlo sampling of the position and velocity vector, taking into account errors on distance, proper motion and radial velocity. We show results assuming that the satellites have a total mass of $M_S=5\times10^{9}\,M_\odot$ (\textit{left panels}) and $M_S=5\times10^{10}\,M_\odot$ (\textit{right panels}); the black solid line in all panels represents the MW virial radius $r_{\rm vir}=245$~kpc.}
    \label{fig:VPOS_orb_dyn_frict_sats_1}
\end{figure*}

\begin{figure*}
    \centering
    \includegraphics[width=.49\textwidth]{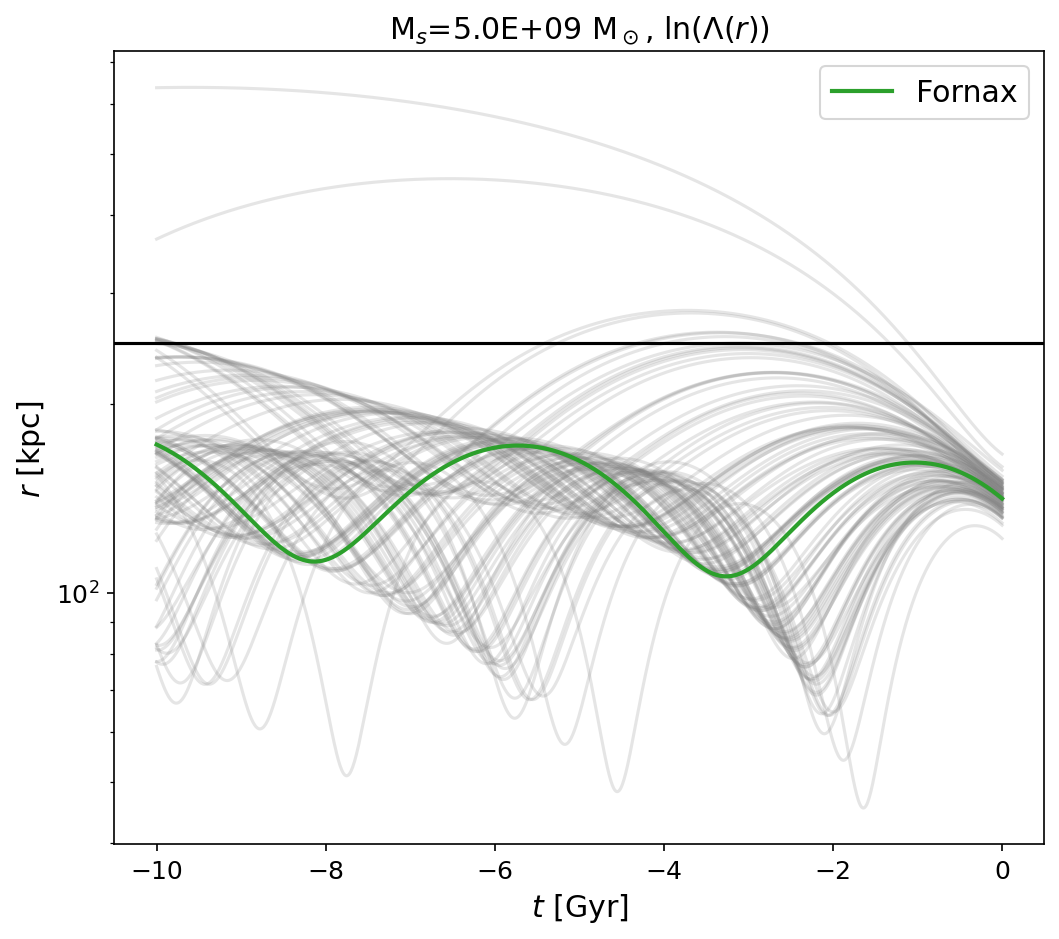}
    \includegraphics[width=.49\textwidth]{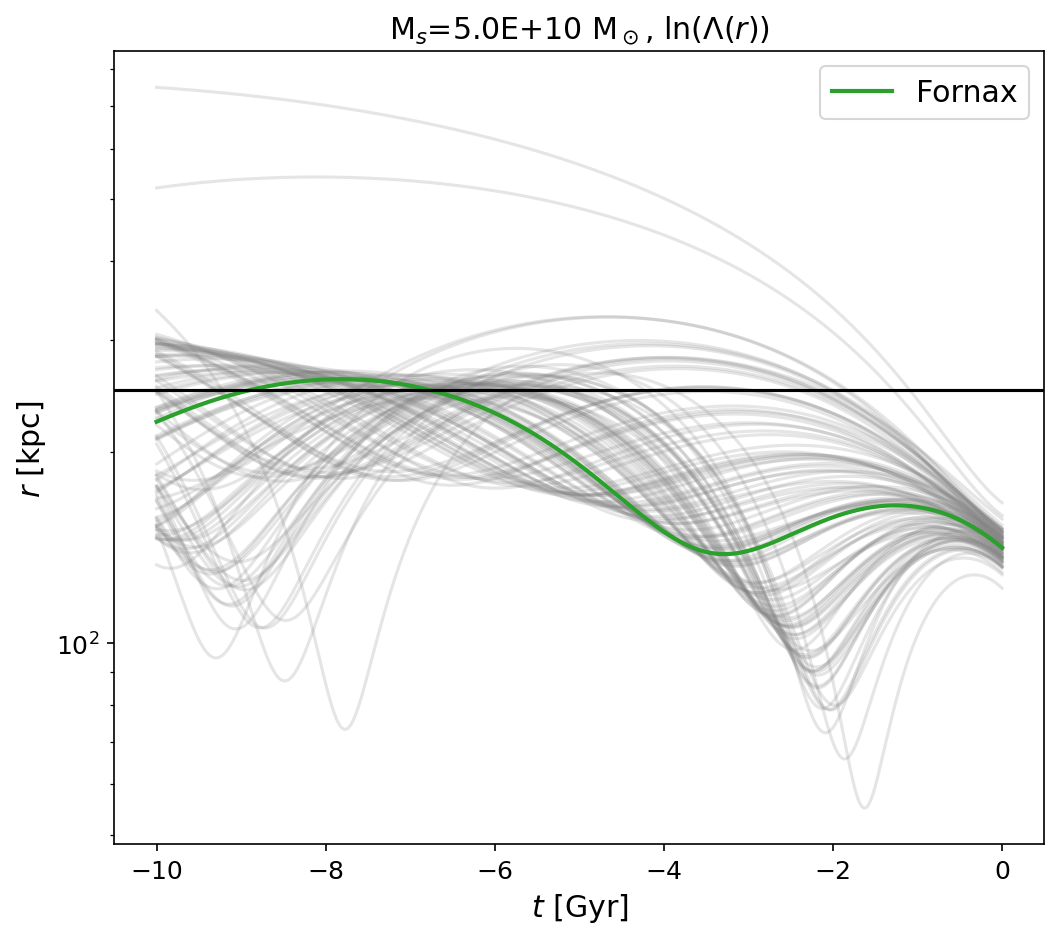}\\
    \includegraphics[width=.49\textwidth]{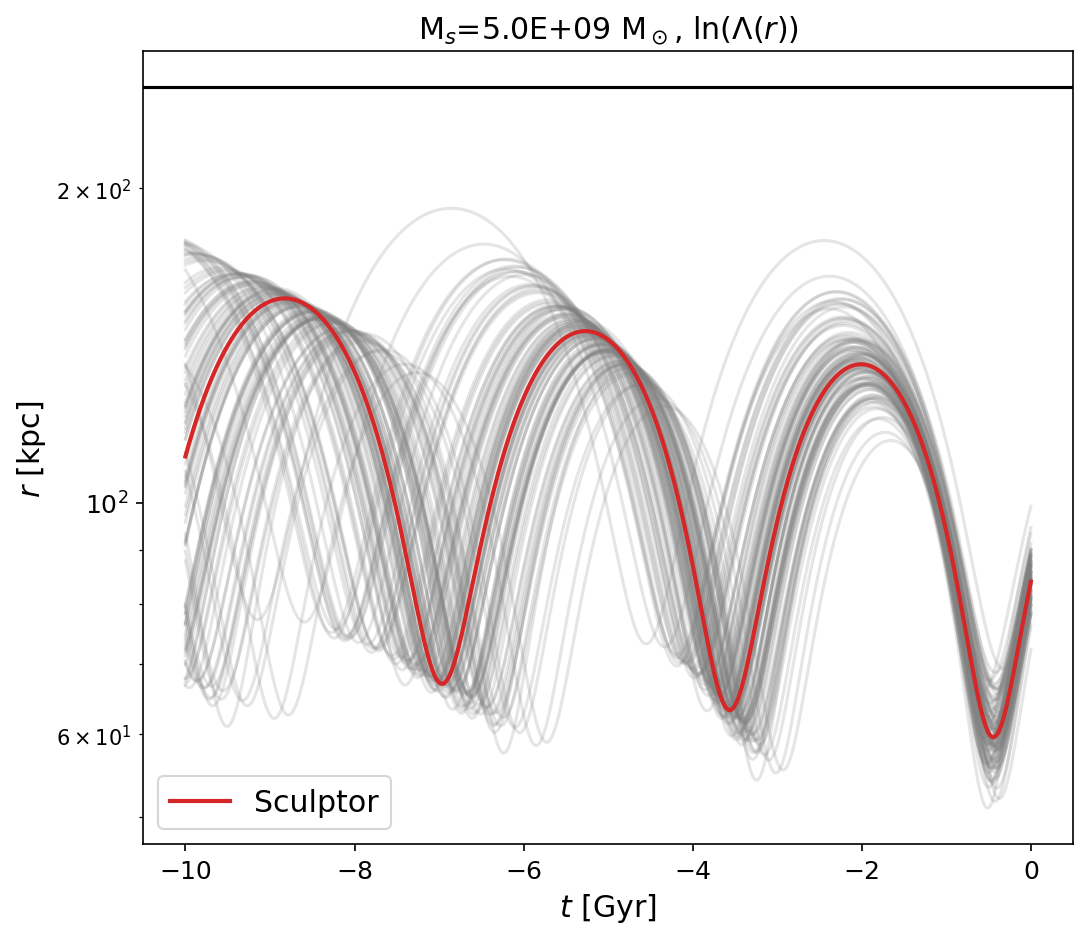}
    \includegraphics[width=.49\textwidth]{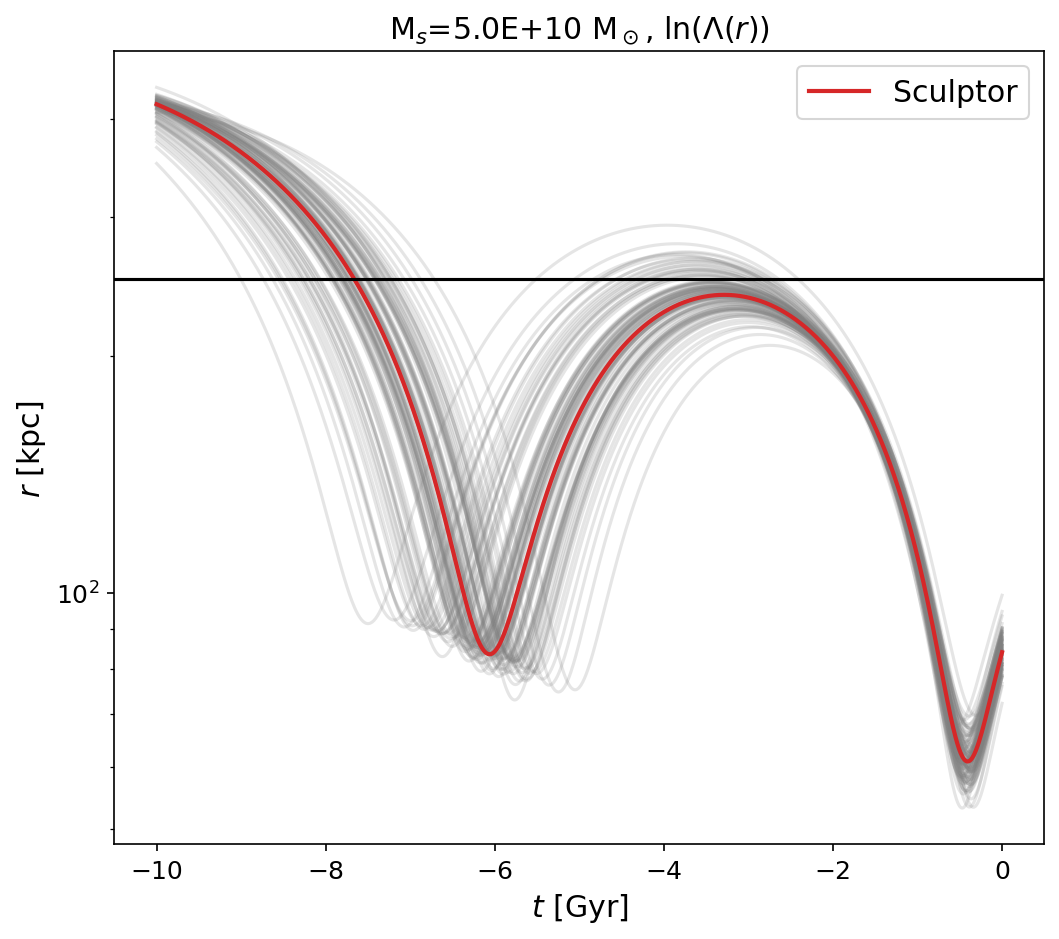}\\
    \includegraphics[width=.49\textwidth]{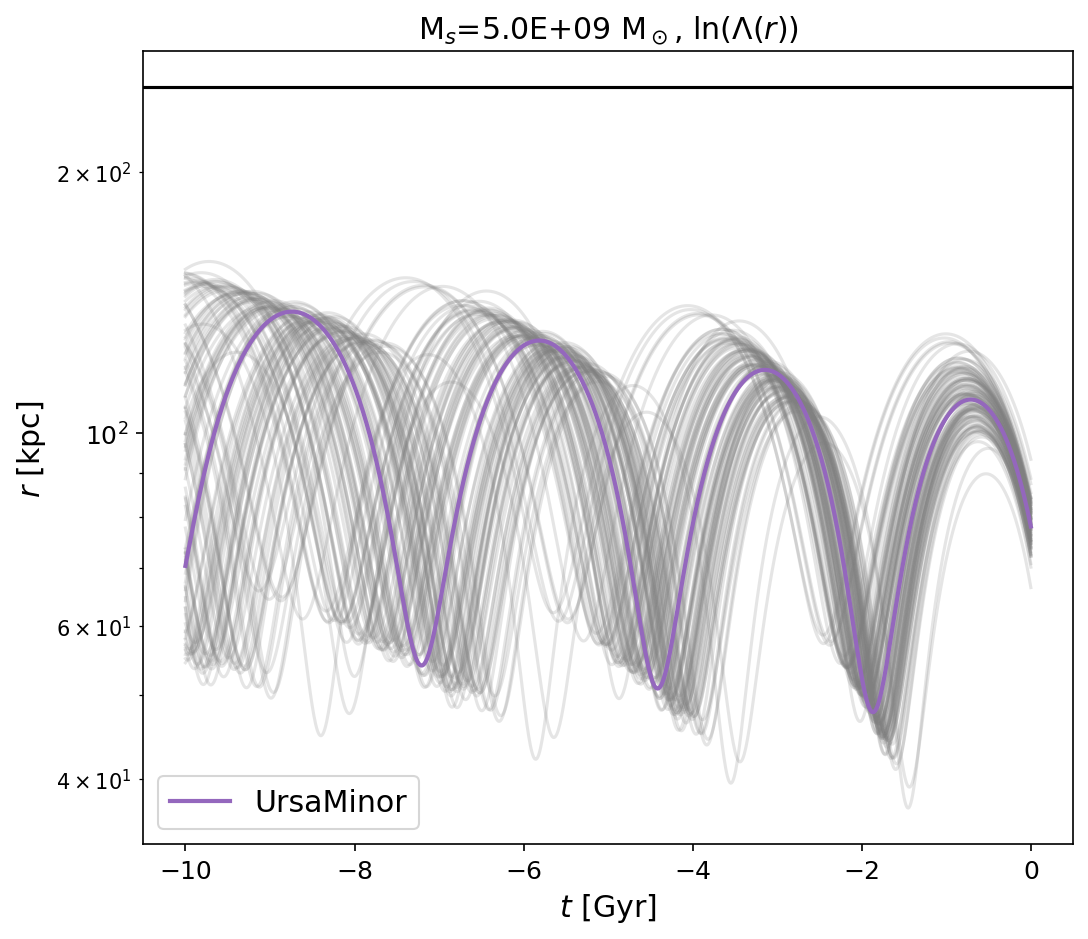}
    \includegraphics[width=.49\textwidth]{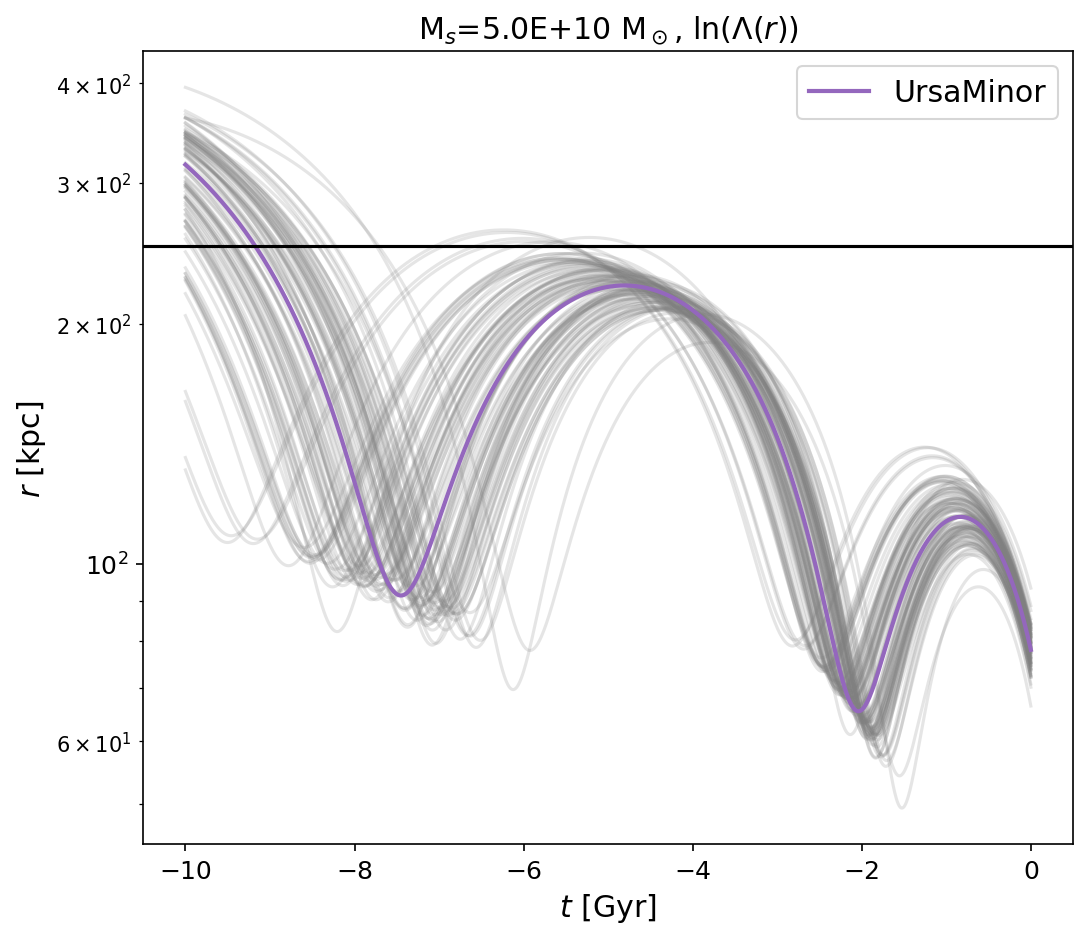}\\
    \caption{Figure~\ref{fig:VPOS_orb_dyn_frict_sats_1}. Continued.}
    \label{fig:VPOS_orb_dyn_frict_sats_2}
\end{figure*}

In Figs.~\ref{fig:VPOS_orb_dyn_frict_sats_1} and~\ref{fig:VPOS_orb_dyn_frict_sats_2}, we also show the impact of uncertainties on the orbit recovery of each considered galaxy by performing 100 random realisations sampling from their position-velocity vector. We show in figures the results for two mass limits indicating on one side a typical value for these dwarf galaxies and on the other the minimum mass at which the effect of dynamic friction starts to be significant. We see that in most cases the errors mainly affect the orbital period rather than the amplitude, although for Carina and Fornax a fraction of the recovered orbits are unbound. This means that, in general, the number of pericentric passages tends to be conserved, even if the time at which they occur is the more uncertain the further back in time one integrates (particularly for times prior to 6 Gyr ago).

\begin{figure*}
    \centering
    \includegraphics[width=.49\textwidth]{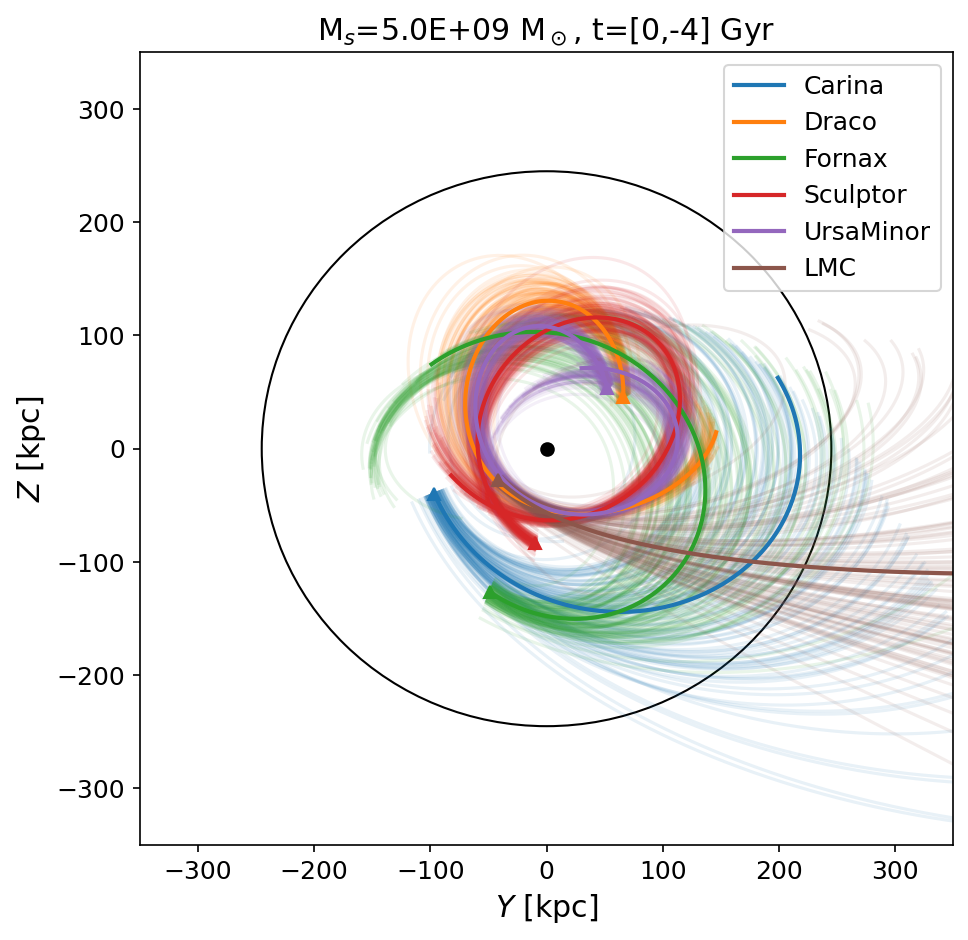}
    \includegraphics[width=.49\textwidth]{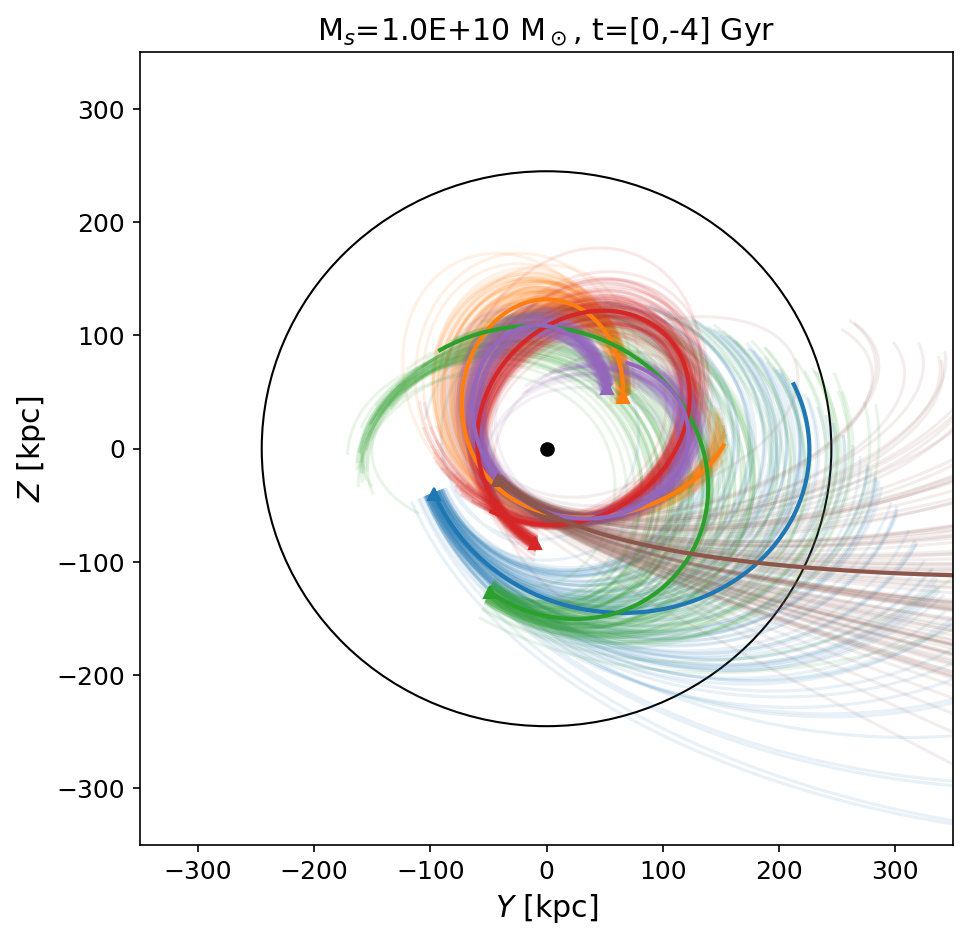}\\
    \includegraphics[width=.49\textwidth]{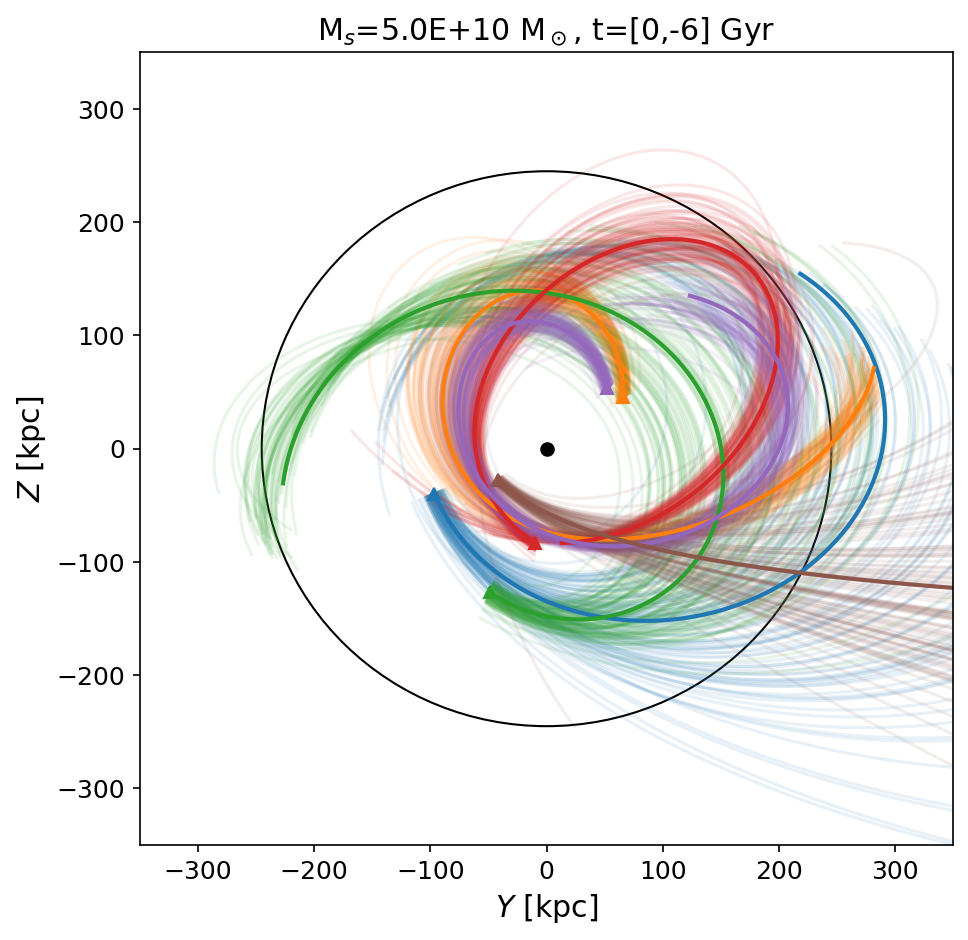}
    \includegraphics[width=.49\textwidth]{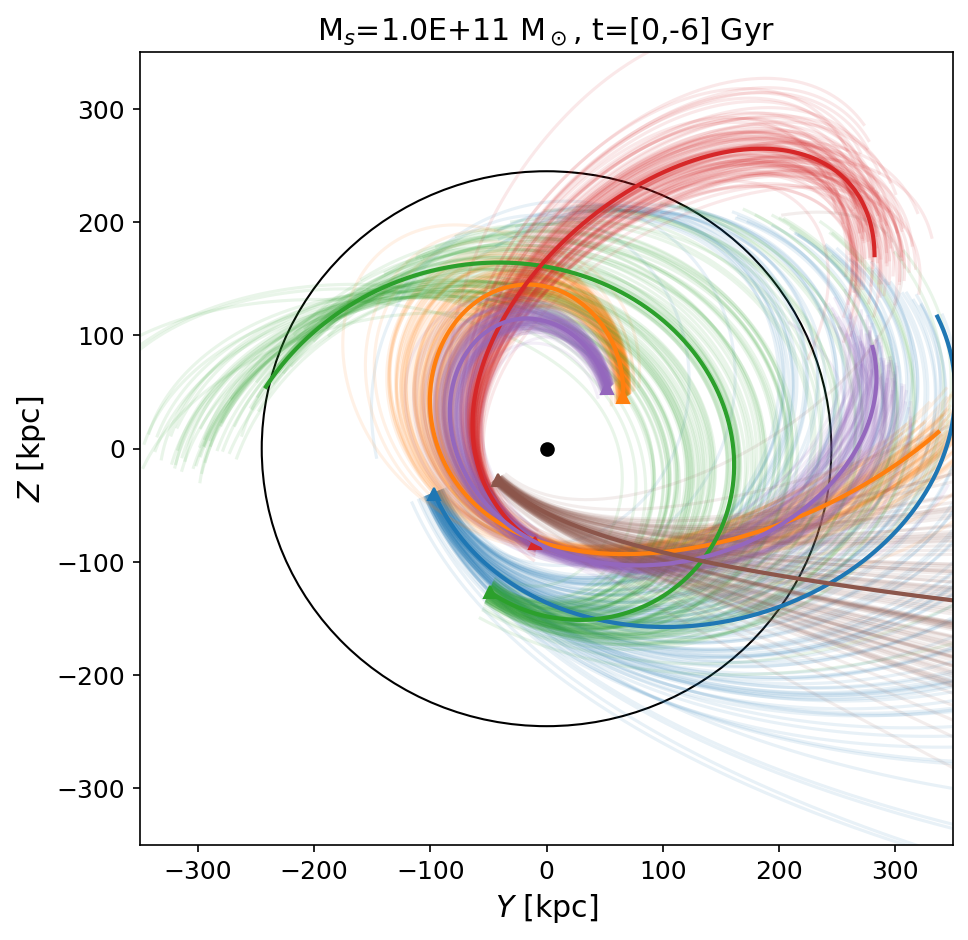}
    \caption{Panels as in Fig.~\ref{fig:VPOS_orb_dyn_frict}, but here showing the orbital evolution of the selected satellites up to 4~Gyr ago in the Galactocentric Cartesian $Y-Z$ plane, which roughly coincides with the one defined by the VPOS. For each galaxy, we also show the orbits obtained from the 100 realisations made by Monte-Carlo sampling of the position and velocity vector, as in Fig.~\ref{fig:VPOS_orb_dyn_frict_sats_1}. The circle in each panel has the same radius as the virial radius of the MW, $r_{\rm vir}=245$~kpc.}
    \label{fig:VPOS_orb_dyn_frict_YZ}
\end{figure*}

Finally, in Figs.~\ref{fig:VPOS_orb_dyn_frict_YZ} we show the orbital evolution of the selected satellites in the Galactocentric Cartesian $Y-Z$, approximately coinciding with the one defined by the VPOS. 
We show orbits up to 4~Gyr ago for $M_S\leq1\times10^{10}\,M_\odot$ (roughly the time to complete an orbital period) and up to 6~Gyr ago for $M_S\ge5\times10^{10}\,M_\odot$ (roughly the time of entry into the MW potential). 
We see that within the errors, and especially at high masses ($M_S\ge5\times10^{10}$), it is possible for the co-orbiting systems to share a common location in the past, however, the errors are too large to reach any significant conclusion. As wrote in the main text, a dedicated simulation is necessary to prove the hypothesis of a group infall origin for the VPOS.

\end{appendix}

\end{document}